\documentclass[12pt]{article}
\usepackage{microtype}
\usepackage[normalem]{ulem}
\usepackage{slashed}
\usepackage{blkarray}
\usepackage[greek,english]{babel}
\usepackage{cancel}
\usepackage{upgreek}
\usepackage{float}
\usepackage{url}
\usepackage{latexsym}
\usepackage{amsmath}
\usepackage{amsfonts}
\usepackage{amssymb}
\usepackage{epsfig}
\usepackage{isomath}
\usepackage{amsthm}
\usepackage{lmodern}
\usepackage{blkarray, bigstrut}
\usepackage{booktabs}
\usepackage{xcolor}
\usepackage{graphicx}
\usepackage[toc,page]{appendix}
\usepackage{multirow}%
 \usepackage{hyperref}

\definecolor{blue-violet}{rgb}{0.54, 0.17, 0.89}
\definecolor{PineGreen}{cmyk}{0.92, 0, 0.59, 0.25}
\definecolor{YellowOrange}{cmyk}{0, 0.42, 1, 0}
\definecolor{orange}{rgb}{0.95, 0.5, 0.1}

\newcommand{\mathsym}[1]{{}}
\newcommand{\unicode}[1]{{}}
\DeclareFontFamily{OMS}{rsfs}{\skewchar\font'60}
\DeclareFontShape{OMS}{rsfs}{m}{n}{<-5>rsfs5 <5-7>rsfs7 <7->rsfs10 }{}
\DeclareSymbolFont{rsfs}{OMS}{rsfs}{m}{n}
\DeclareSymbolFontAlphabet{\Scr}{rsfs}
\def\IC{\relax\,\hbox{$\inbar\kern-.3em{\rm C}$}}
\def\IG{\relax\,\hbox{$\inbar\kern-.3em{\rm G}$}}
\def\IB{\relax{\rm I\kern-.18em B}}
\def\ID{\relax{\rm I\kern-.18em D}}
\def\IL{\relax{\rm I\kern-.18em L}}
\def\IF{\relax{\rm I\kern-.18em F}}
\def\IH{\relax{\rm I\kern-.18em H}}
\def\II{\relax{\rm I\kern-.17em I}}
\def\IN{\relax{\rm I\kern-.18em N}}
\def\IP{\relax{\rm I\kern-.18em P}}
\def\IQ{\relax\,\hbox{$\inbar\kern-.3em{\rm Q}$}}
\def\bfzero{\relax\,\hbox{$\inbar\kern-.3em{\rm 0}$}}
\def\IK{\relax{\rm I\kern-.18em K}}
\def\IG{\relax\,\hbox{$\inbar\kern-.3em{\rm G}$}}
 \font\cmss=cmss10 \font\cmsss=cmss10 at 7pt
\def\IR{\relax{\rm I\kern-.18em R}}
\def\ZZ{\relax\ifmmode\mathchoice
{\hbox{\cmss Z\kern-.4em Z}}{\hbox{\cmss Z\kern-.4em Z}}
{\lower.9pt\hbox{\cmsss Z\kern-.4em Z}} {\lower1.2pt\hbox{\cmsss
Z\kern-.4em Z}}\else{\cmss Z\kern-.4em Z}\fi}
\def\bfone{\relax{\rm 1\kern-.35em 1}}

\def\n010{\mathrm{N^{0,1,0}}}
\def\inbar{\vrule height1.5ex width.4pt depth0pt}
\def\bfzero{\relax{\rm I\kern-.18em 0}}
\def\bfone{\relax{\rm 1\kern-.35em 1}}

\DeclareFontFamily{U}{rsf}{} \DeclareFontShape{U}{rsf}{m}{n}{
  <5> <6> rsfs5 <7> <8> <9> rsfs7 <10-> rsfs10}{}
\DeclareMathAlphabet\Scr{U}{rsf}{m}{n}

\def\tilde{\widetilde}

\def\1bar{1\hskip -.275cm -}
\def\2bar{2\hskip -.275cm -}
\def\3bar{3\hskip -.275cm -}

\newsavebox{\uuunit}
\sbox{\uuunit}
                 {\setlength{\unitlength}{0.825em}
                      \begin{picture}(0.6,0.7)
                                      \thinlines
                                      \put(0,0){\line(1,0){0.5}}
                                      \put(0.15,0){\line(0,1){0.7}}
                                      \put(0.35,0){\line(0,1){0.8}}
                                     \multiput(0.3,0.8)(-0.04,-0.02){10}{\rule{0.5pt}{0.5pt}}
                      \end {picture}}

\makeatletter \@addtoreset{equation}{section} \makeatother

\def\bfone{\relax{\rm 1\kern-.35em 1}}

\def\bfone{\relax{\rm 1\kern-.35em 1}}
\font\cmss=cmss10 \font\cmsss=cmss10 at 7pt

\def\bfone{\relax{\rm 1\kern-.35em 1}}
\def\inbar{\vrule height1.5ex width.4pt depth0pt}
\def\IC{\relax\,\hbox{$\inbar\kern-.3em{\rm C}$}}
\def\ID{\relax{\rm I\kern-.18em D}}
\def\IF{\relax{\rm I\kern-.18em F}}
\def\IH{\relax{\rm I\kern-.18em H}}
\def\II{\relax{\rm I\kern-.17em I}}
\def\IN{\relax{\rm I\kern-.18em N}}
\def\IP{\relax{\rm I\kern-.18em P}}
\def\IQ{\relax\,\hbox{$\inbar\kern-.3em{\rm Q}$}}
\def\IR{\relax{\rm I\kern-.18em R}}
\font\cmss=cmss10 \font\cmsss=cmss10 at 7pt
\def\ZZ{\relax\ifmmode\mathchoice
{\hbox{\cmss Z\kern-.4em Z}}{\hbox{\cmss Z\kern-.4em Z}}
{\lower.9pt\hbox{\cmsss Z\kern-.4em Z}} {\lower1.2pt\hbox{\cmsss
Z\kern-.4em Z}}\else{\cmss Z\kern-.4em Z}\fi}

\def\ni{\noindent}
\def\tilde{\widetilde}
\def\bar{\overline}

\def\hat{\widehat}

\def\Coe#1.#2.{{#1\over #2}}

\def\coe#1.#2.{\relax{\textstyle {#1 \over #2}}\displaystyle}

\def\to{\rightarrow}
\def\notin{\hbox{{$\in$}\kern-.51em\hbox{/}}}

\def\IE{\relax{{\rm I\kern-.18em E}}}

\def\IGam{\relax{{\rm I}\kern-.18em \Gamma}}

\def\IA{\relax{\hbox{{\rm A}\kern-.82em {\rm A}}}}

\def\ben{\begin{equation}}
\def\een{\end{equation}}
\def\bea{\begin{eqnarray}}
\def\eea{\end{eqnarray}}
\def\bea{\begin{eqnarray}}
\def\eea{\end{eqnarray}}

\begin{document}
\begin{titlepage}
\allowdisplaybreaks
\begin{center}
\vskip 0.2cm
\renewcommand{\thefootnote}{\fnsymbol{footnote}}
{\Large 
{Gauged $\mathcal{N}=3,D=4$ Supergravity: a New Web of Marginally Connected Vacua}}\\[1cm]
Pietro~Fr\'e${}^{\; a,b,d}$\footnote{pietro.fre@unito.it}, Alfredo Giambrone\,$^{c,b,d}$\footnote{alfredo.giambrone@polito.it}, Daniele
Ruggeri\footnote{daniele.rug@gmail.com},  Mario Trigiante${}^{\; c,b,d}$\footnote{mario.trigiante@polito.it} and  Petr Va\v{s}ko${}^{\; e}$\footnote{vasko@ipnp.mff.cuni.cz} \\[10pt]
\vspace{4pt}  {${}^a$\sl\small Dipartimento di Fisica, Universit\'a
di Torino}\\
\vspace{2pt} {{\em $^{b}$\sl\small INFN --
 Sezione di Torino \\
via P. Giuria 1, \ 10125 Torino \ Italy}}\\
\vspace{2pt}
{${}^c${{\sl\small  Dipartimento di Fisica Politecnico di Torino,}}\\
{\em C.so Duca degli Abruzzi, 24, I-10129 Torino, Italy}\\
\vspace{2pt}
 \centerline{$^{d}$ \it Arnold-Regge Center,}
 \centerline{\it via P. Giuria 1,  10125 Torino,
Italy}}
\vspace{2pt}
 \centerline{$^{e}$ \it Institute of Particle and Nuclear Physics, Charles University,}
 \centerline{\it V Hole\v{s}ovi\v{c}k\'{a}ch 2, 180 00 Prague 8, Czech Republic}

\vspace{10pt}
\begin{abstract}

We analyze the vacuum structure of $\mathcal{N}=3,D=4$ supergravity coupled to 9 vector multiplets with gauge group ${\rm SO}(3)\times {\rm SU}(3)$. Aside from the central $\mathcal{N}=3$ AdS$_4$ vacuum at the origin, on which the supermultiplet structure reproduces the massless sector of  M-theory  compactified on $\mathrm{N^{0,1,0}}$, we find a rich structure of AdS$_4$ vacua preserving $\mathcal{N}=0,1,2,3$ supersymmetry. These new vacua are arranged in a manifold spanned by scalar fields corresponding to exactly marginal deformations of the dual CFT. This manifold has the form $T^3/K$, where $K$ is a discrete subgroup of the gauge group: $\mathcal{N}=3,2$ and $1$ vacua correspond, respectively, to a point, a line and a surface in the three-dimensional vacuum manifold. We study RG flows from the central $\mathcal{N}=3$ vacuum and elaborate on the possible higher dimensional origin of the new vacua. For the reader's convenience we also provide a review of the embedding tensor formulation of $D = 4,\,\mathcal{N}=3$ gauged
supergravities. In particular we provide formulas involving the fermion shift tensors and mass matrices in $\mathcal{N}=3$ theories, which can be applied to a generic gauging.
\end{abstract}
\end{center}
\end{titlepage}
\tableofcontents \noindent {}
\newpage
\renewcommand{\thefootnote}{\arabic{footnote}}
\setcounter{footnote}{0}
\section{Introduction}
Lower-dimensional gauged supergravities have provided a valuable framework for consistently studying the dynamics of a subset of physical degrees of freedom associated with type II superstring theories in $D=10$ or $D=11$ supergravity. Defining such  models amounts to constructing \emph{consistent truncations} of the higher dimensional theories on specific backgrounds. 
Of particular interest are solutions of $D=10$ type II supergravities or of $D=11$ supergravity whose geometry is a warped product $AdS_4 \times_w M_d$, $d=D-4$, of   a 4-dimensional anti-de Sitter spacetime and an internal compact manifold $M_d$.  Once this is achieved, one can try to explore relevant properties such as perturbative and non-perturbative stability of the solutions. As far as the construction of consistent truncations within maximal lower-dimensional supergravities is concerned, Exceptional Field Theory \cite{Hohm:2013uia, Hohm:2014qga} provides an efficient framework for embedding certain lower dimensional models into superstring or M-theories and for studying perturbative stability of their solutions \cite{Malek:2020yue}. In the more general case important progress has been made towards a systematic construction of lower-dimensional consistent truncations  \cite{Cassani:2019vcl,Josse:2021put}. In this case the set-up is the one of Generalised Geometry in which a wide class of consistent truncations can be described by exploiting the concept of generalised $G_S$-structure manifolds with singlet intrinsic torsion. Earlier results related to the construction of consistent truncations of $D=11$ supergravity compactified on manifolds $M_d$ with tri-sasakian geometry were obtained in \cite{Cassani:2011fu}.\par 
Our present work is inspired by one of these spontaneous compactifications, which has the form
\begin{equation}\label{ads4n010}
    \mathrm{AdS_4} \times \mathrm{N^{0,1,0}}\,,
\end{equation}
where, within the infinite class of sasakian homogeneous spaces
$\mathrm{N^{p,q,r}}$ introduced by Castellani and Romans in
\cite{Castellani:1983tc}, the case $\{p,q,r\}=\{0,1,0\}$ defines the unique instance of a 7-dimensional homogeneous tri-sasakian manifold. \par
As shown in the original paper \cite{Castellani:1983tc} and
systematically reviewed in \cite{castdauriafre}, the vacuum
(\ref{ads4n010}) admits three anti-de Sitter Killing spinors and
correspondingly the whole spectrum of Kaluza-Klein states is
arranged into supermultiplets of the following supergroup:
\begin{equation}\label{supercamillo}
    \mathcal{G}_{iso} \, = \, \mathrm{OSp(3|4)} \times \mathrm{SU(3)}\,\,\,.
\end{equation}
Such organization of the Kaluza-Klein states was achieved in
\cite{Fre1999xp}, which also provides the general form of the
$\mathrm{OSp(3|4)}$ supermultiplets. Later  in \cite{ringoni} this
result was compared  with the spectrum of primary conformal fields
pertaining to a candidate $D=3$ superconformal field theory
suggested by the HyperK\"ahler quotient construction of the metric
cone $\mathcal{C}\left(\n010\right)$. All the Kaluza-Klein towers
are perfectly reproduced but there are also additional ones that the
superpotential of the candidate theory does not suppress, as it was
remarked in \cite{ringoni}. A precise comparison of these very early
results with those much later derived in the framework of quiver
theories associated with orbifolds $\n010/\mathbb{Z}_k$,
\cite{Gaiotto:2009tk},\cite{Hosseini:2016ume} is still missing in
the literature.
\par
As for the massless supermultiplets the above mentioned spectrum is
very simple, it just contains the massless graviton multiplet and
the 9 massless vector multiplets. The massless graviton multiplet
includes the graviton $g_{\mu\nu}$, three gravitinos
$\psi_{A\mu}$, three gauge fields ${A}^{AB}_\mu$ gauging the
R-symmetry $\mathrm{SO(3)}$ and one spin one-half field $\chi_{\bullet}$. Each
massless $\mathcal{N}=3$ vector multiplet has a field content equal
to that of an $\mathcal{N}=4$ multiplet, namely one vector
${A}_\mu$, four spin one half spinors $(\lambda_A,\lambda)$,
organized into a triplet and a singlet of $\mathrm{SO(3)}$ and six
scalars organized into two triplets of $\mathrm{SO(3)}$. The nine
vector multiplets are divided into eight in the adjoint of
$\mathrm{SU(3)}$, the bosonic group factor in (\ref{supercamillo}),
and one in a singlet (the so called Betti multiplet originating from
the non trivial cohomology group of $\n010$ in degree two).\par

In the present work we start considering a four-dimensional $\mathcal{N}=3$ supergravity coupled to $9$ vector multiplets, whose gauge group $\mathcal{G}={\rm SO}(3)\times {\rm SU}(3)$, coincides with the isometry group of the internal manifold. We are aware, however, that this model does not fit the $\mathcal{N}=4$ consistent truncation defined in \cite{Cassani:2011fu}. Nevertheless the vacua we shall analyze are described within a smaller truncation of the original $\mathcal{N}=3$ model, with scalar manifold $({\rm SU}(1,1)/{\rm U}(1))^3$. The study of the possible embedding of these vacua within the consistent truncation of \cite{Cassani:2011fu} and thus their actual relation with the compactification \eqref{ads4n010} will be the subject of future investigation.
As for the full $\mathcal{N}=3$ model with nine vector multiplets, it certainly reproduces, around the central $\mathcal{N}=3$ vacuum at the origin, properties of the linearized theory on the  \eqref{ads4n010} background, in particular the massless AdS-supermultiplets, though possibly not their complete non-linear interactions.

In this first work, we focus on this $\mathcal{N}=3$ gauged supergravity and its vacuum structure independently of its possible relation with M-theory compactifications. In particular we find, besides the central  $\mathrm{AdS_4}$ $\mathcal{N}=3$-supersymmetric vacuum, naturally associated with the  compactification \eqref{ads4n010}, a rich structure of new vacua, with different supersymmetries.

This model has also been recently studied in \cite{Karndumri:2016miq,Karndumri:2016fix}. Here we present a broad analysis of the vacuum structure of the theory that is not contained in the above research. Aside from the $\mathcal{N}=3$ AdS$_4$ vacua with ${\rm SU}(2)\times {\rm U}(1)$ and ${\rm SO}(3)$ symmetries, which were already found in \cite{Karndumri:2016miq,Karndumri:2016fix}, our analysis unveils new compact loci of $\mathcal{N}=1$ and $\mathcal{N}=2$ vacua, besides perturbatively stable $\mathcal{N}=0$ ones. These new vacua, to our knowledge, were overlooked in the literature. We provide for the $\mathcal{N}=3,2,1$
vacua the corresponding supermultiplet structures and study the relevant RG flows. All these vacua form a compact
manifold $T^3/K$, where $K$ is a discrete subgroup of the gauge group, isomorphic to the symmetric group $S_4$. The $\mathcal{N}=3,2$ and $1$ vacua correspond, respectively, to a point, a line and a surface in the three-dimensional vacuum manifold, the remaining points define perturbatively stable $\mathcal{N}=0$ anti-de Sitter vacua. The compact vacuum manifold, with geometry $T^3/K$, is spanned by three angular variables which define flat directions of the scalar potential and which are thus natural candidates to correspond to exactly marginal deformations of the dual CFT. In this latter theory these deformations would therefore realize a pattern of supersymmetry breaking, when moving from a supersymmetric vacuum to a less supersymmetric one in the same moduli space, by marginal deformations.\par
Let us end this Introduction with few more details abut the model under consideration and our results.
As first derived in \cite{Castellani:1985wk,Castellani:1985ka} the
$6n$ scalars of a matter coupled supergravity theory with $n$ vector
multiplets are organized into the complex coordinates of the
non-compact K\"ahlerian manifold:
\begin{equation}\label{labirintone}
    \mathcal{M}_{\mathrm{scalar}} \, = \,\frac{\mathrm{SU(3,n)}}{\mathrm{SU(3)\times SU(n)
    \times U(1)}}\,\,\,.
\end{equation}
Hence, a choice of $n$, in our case $n=9$, defines a unique ungauged supergravity
theory.
\par
Using the embedding tensor formalism
\cite{embeddone,Nicolai:2000sc,deWit:2002vt,deWit:2005ub} (for reviews see \cite{Samtleben:2008pe,Trigiante:2016mnt}) we study the gauging of the group:
\begin{equation}\label{zerowat}
   \mathcal{G}=\, \mathrm{SO(3) \times SU(3)}\,,
\end{equation}
and search for extrema of the corresponding scalar potential.
As mentioned above, such a gauge theory has, at the origin of the coset
manifold
\begin{equation}\label{m39}
    \mathcal{M}_{S} \, \equiv \mathcal{M}_{\mathrm{scalar}}^{n=9}=
\,\frac{\mathrm{SU(3,9)}}{\mathrm{SU(3)\times SU(9)
    \times U(1)}}
\end{equation}
an anti-de Sitter vacuum preserving $\mathcal{N}=3$ supersymmetries that
naturally corresponds to the original M-theory compactification
(\ref{ads4n010}). Obviously there are other extrema whose
geometrical interpretation in higher dimensions is yet to be
understood.
\par
In particular, by means of a consistent truncation of our theory
to singlets of certain specified subgroups of the gauge
group (\ref{zerowat}) we find two other vacua with $\mathcal{N}=3$
supersymmetry in $D=4$. Specifically
\begin{description}
  \item[A)] Truncating to the singlets under the subgroup:
  \begin{eqnarray}\label{rappanuiaA}
    \mathrm{SO_{diag}^{A}(3)} & = &\mbox{diag} \left( \mathrm{SO(3) \times
    SO_I(3)}\right)\subset \mathcal{G}
\end{eqnarray}
where $\mathrm{SO(3)} \subset \mathrm{OSp(3|4)}$ is the R-symmetry
group, while $\mathrm{SO_I(3)}\subset \mathrm{SU(3)}$ is the real
restriction of the complex group under which the fundamental
representation remains irreducible
$$\mathbf{3}_{\mathrm{SU(3)}}\stackrel{\mathrm{SO_I(3)}}{\longrightarrow}
\mathbf{3}\,,$$ we find a second $\mathcal{N}=3$ vacuum whose isometry
is simply $\mathrm{OSp(3|4)}$ and all the  other fields arrange
themselves into $\mathcal{N}=3$ massive vector multiplets (for details see Section \ref{sec:supermult}).
  \item[B)] Truncating to the singlets under the subgroup:
  \begin{eqnarray}\label{rappanuiaB}
    \mathrm{SO_{diag}^{B}(3)} & = &\mbox{diag} \left( \mathrm{SO(3) \times
    SO_{II}(3)}\right)\subset \mathcal{G}
\end{eqnarray}
where $\mathrm{SO(3)} \subset \mathrm{OSp(3|4)}$ is once again the
R-symmetry group, while $\mathrm{SO_{II}(3)}\sim
\mathrm{SU_{II}(2)}\subset \mathrm{SU(3)}$ is locally isomorphic to the
natural $\mathrm{SU_{II}(2)}$  subgroup of $\mathrm{SU(3)}$  under
which the fundamental representation splits  into a singlet plus a
doublet
$$\mathbf{3}_{\mathrm{SU(3)}}\stackrel{\mathrm{SU_{II}(2)}}{\longrightarrow}
\mathbf{2}\oplus \mathbf{1}$$
we find a third $\mathcal{N}=3$ vacuum whose isometry is simply
$\mathrm{OSp(3|4)}$ and all the  other fields arrange themselves
into $\mathcal{N}=3$ massive vector multiplets but with different
energy (scaling dimension) eigenvalues than in the previous case (for details see Section \ref{sec:supermult}).
\end{description}
Each of the above $\mathcal{N}=3$ vacua is connected to loci of $\mathcal{N}=2,1$ and $0$ vacua through three angular flat-directions of the scalar potential. In fact they are part of vacuum manifolds with geometry $T^3/K$, as mentioned above.
Our paper is organized as follows:
\par
In Section~\ref{barengo} we review $\mathcal{N}=3$ supergravity in four dimensions. Starting from the ungauged theory, we illustrate the general procedure  to construct the gauged one using the embedding tensor formalism. Of particular relevance to our analysis is the derivation of the fermion-shift tensors, the mass matrices and the scalar potential from the ${\rm S}[{\rm U}(3)\times {\rm U}(n)]$-irreducible components of the so-called $T$-tensor.

In Section~\ref{sec:gauged_model} we specialize to the model with nine vector multiplets (one of them, corresponding to the the Betti multiplet, being completely decoupled). Supplemented by three vectors from the gravity multiplet, these $3+8$ vector fields gauge the $(3+8)$-dimensional compact subgroup $\mathcal{G}=\mathrm{SO}(3)\times\mathrm{SU}(3)$ of the isometry group ${\rm G}=\mathrm{SU}(3,9)$ associated with the scalar manifold of the ungauged theory. 

Section~\ref{sec:trunc} sets up the stage for analysis of the vacuum structure of the above model. Since the scalar manifold is $54$-dimensional, it is a daunting task to extremize the scalar potential in general. Thus, we restrict the study to two different consistent truncations of the theory, each associated with a $6$-dimensional scalar manifold that is embedded into the full scalar manifold inequivalently. On these subspaces the scalar potential can be extremized and we find that in both cases the vacuum manifold has the topology of an orbifold: a $3$-torus quotiented by a particular discrete subgroup of the gauge group. We then describe loci of different co-dimensions in the vacuum manifold, based on the amount of preserved supersymmetry as well as on breaking patterns of the gauge group.

Section~\ref{sec:supermult} is devoted to decomposing the mass spectra on vacua preserving $\mathcal{N}$ of the three original supersymmetries, into unitary irreducible representations of the supergroup $\mathrm{OSp}(\mathcal{N}|4)$, which represents the superconformal group of the holographically dual $\mathrm{SCFT}_3$.

In Section~\ref{sec:DW} we present Domain-Wall solutions dual to RG-flows between the maximally symmetric vacuum at the origin of the scalar manifold and other less symmetric vacua, whose holographic meaning remains to be uncovered. We verify the $a$-theorem for these flows.

Finally, we summarize the content of this paper and offer a brief outlook in the Conclusions.

Technical details are given in the Appendices. Appendices~\ref{ward}-~\ref{D} deal with various aspects of the embedding tensor formalism. Appendix~\ref{E} fixes conventions for gauge group generators. In Appendix \ref{DWapp} we provide the details of the Domain-Wall solutions of Section \ref{sec:DW}. Appendix~\ref{app:supermult} presents relevant unitary irreducible representations of $\mathrm{OSp}(\mathcal{N}|4)$. Finally, Appendix~\ref{app:spectra} provides detailed tables of mass spectra of the gauged supergravity in various vacua. 

\section{Gauged \texorpdfstring{$\mathcal{N}=3$ Supergravity}{N=3}}
\label{barengo}
In this section we define the general field theoretical setting of our analysis by reviewing the main facts about $\mathcal{N}=3$ supergravity and its gaugings \cite{Castellani:1985ka}.
\subsection{The Ungauged Model}
For the sake of fixing the relevant notations, let us start with reviewing the general features of an ungauged $\mathcal{N}=3$ supergravity, namely of the version of the theory in which the vector fields are not minimally coupled to any other field.\par
A generic model of this kind features, besides the supergravity multiplet, a number $n$ of vector multiplets.
In particular the gravity multiplet consists of the graviton $g_{\mu \nu}$, $\mu,\nu=\,0,...,3$ being the space-time index, three gravitinos $\psi_{A\,\mu}$, $A=1,\dots,3$, three vector fields (graviphotons) $A^{AB}_{\mu}$, and one dilatino $\chi_{ABC}=\chi_\bullet\,\epsilon_{ABC}$. Each of the $n$ vector multiplets (labeled by $I=1,\dots, { n}$), contains a vector field $A_{I\mu}$, four gauginos $\lambda_{IA},\,\lambda_I$, and three complex scalar fields $\phi_{IAB}$.\par
Therefore the model features $n_v=3+n$ vector fields and $3\times n$ complex scalar fields spanning a complex scalar manifold of the form:
\begin{equation}\label{labirinto}
    \mathcal{M}_{scalar} \, = \,\frac{{\rm G}}{H}\,=\,\frac{\mathrm{SU}(3,n)}{{\rm S}[{\rm U}(3)\times {\rm U}(n)]}\,,
\end{equation}
the isotropy group being locally isomorphic to the product of the R-symmetry group $H_R={\rm U(3)}$ and the group $H_{{\rm matter}}= {\rm SU}(n)$ acting on the vector multiplets only.
\paragraph{The electric-magnetic duality symmetry.} The global on-shell symmetry group of the ungauged model is the isometry group ${\rm G}=\mathrm{SU}(3,n)$ of the scalar manifold, provided its non-linear action on the scalar fields is combined with a symplectic, electric-magnetic duality action on the vector field strengths and their magnetic duals. \par
 The symplectic duality action of ${\rm G}$ on the electric and magnetic charges is defined by the representation:
 \begin{equation}
  {\Scr R}_\eta={\bf (3+ n)}\oplus\overline{{\bf (3+ n)}}\,.
 \end{equation}
 The representation ${\bf (3+ n)}$, in turn, branches with respect to the subgroup $H$ as follows:
 \begin{equation}
 {\bf (3+ n)}\,\rightarrow \,{\bf (3,1)}_{-1}\oplus {\bf (1,n)}_{\frac{3}{n}} \,.
 \end{equation}
 There is an obvious complex basis of the representation space of ${\Scr R}_\eta$, in which the action of the group ${\rm G}$ is block-diagonal. A vector in this basis is
denoted by
 \begin{equation}
V^{\underline{M}}=(V^{\underline{\Lambda}},\,V_{\underline{\Lambda}})\,\,,\,\,\,\,
V^{\underline{\Lambda}}=
 (V^{AB},\,V_{I})\,\,,\,\,\,V_{\underline{\Lambda}}=(V^{\underline{\Lambda}})^*\,.
 \end{equation}
 where $V^{AB}$ is a complex  vector in the representation ${\bf (3,1)}_{-1}$ of $H$ while $V_I$ transforms in the ${\bf (1,n)}_{+\frac{3}{n}}$ of
the same group. In this basis a representation of a generic element  $T=(T_{\underline{\Lambda}}{}^{\underline{\Sigma}})$ of ${\rm SU}(3,n)$ in its fundamental representation ${\bf (3+n)}$ has the form:
 \begin{equation}
 T\in {\rm G}\,\,\rightarrow\,\, {\Scr R}_\eta[T]_{\underline{M}}{}^{\underline{N}} \equiv
  \left(
  \begin{matrix}
    T &  {\bf 0} \\
  {\bf 0}  & T^*
  \end{matrix} \right) \,,\label{cbasis}
\end{equation}
where $T$ satisfies the defining condition $T^\dagger \eta T=\eta$, $\eta={\rm diag}(+1,+1,+1,-1,\dots,-1)$.
The structure of the matrix $T$ in terms of $H$-covariant blocks is:
\begin{equation}
T=\left(\begin{matrix}T_{AB}{}^{CD} & T_{AB\,J}\cr T^{I\, CD} & T^{I}{}_J\end{matrix}\right)\,.
\end{equation}
The drawback of this basis is that the matrix ${\Scr R}_\eta[T]_{\underline{M}}{}^{\underline{N}} $ is not symplectic. \footnote{Notice that in this complex basis the matrix ${\Scr R}_\eta[T]$ is symplectic with respect to an antisymmetric $2(n+3)\times 2(n+3)$ matrix $\mathbb{C}_\eta$ of the form:
\begin{equation}
\mathbb{C}_\eta\equiv \left(\begin{matrix}{\bf 0} & \eta\cr -\eta & {\bf 0}\end{matrix}\right)\,.
\end{equation}
Indeed the reader can verify, using the property $T^\dagger \eta T=\eta$, that ${\Scr R}_\eta[T]^t\,\mathbb{C}_\eta\,{\Scr R}_\eta[T]=\mathbb{C}_\eta$.} The real symplectic representation of ${\rm G}$ in terms of matrices in ${\rm Sp}(2(3+n),\mathbb{R})$, is obtained through the following change of basis:
\begin{equation}
V^M=({\Scr A}^{\dagger}{\Scr O})^{M}{}_{\underline{N}}\,V^{\underline{N}}\,,
\end{equation}
where we have denoted a vector in the real symplectic basis by $V^M=(V^\Lambda,\,V_\Lambda)$ and the matrices ${\Scr O}$ and ${\Scr A}$ are given by:
\begin{equation}
{\Scr O} = \begin{array}{c c}
    & \begin{array} {@{} c c c c @{}}
      \mbox{{\scriptsize 3}} & \mbox{{\scriptsize n}} & \mbox{{\scriptsize 3}} & \mbox{{\scriptsize n}}
    \end{array} \\
    \begin{array}{c}
      \mbox{{\scriptsize 3}} \\ \mbox{{\scriptsize n}} \\ \mbox{{\scriptsize 3}} \\\mbox{{\scriptsize n}}
    \end{array}\hspace{-1em} &
    \left(
      \begin{array}{@{} c | c|| c | c @{}}
        {\bf 1} & {\bf 0} & {\bf 0} & {\bf 0}   \\ \hline
        {\bf 0} & {\bf 0} & {\bf 0} & {\bf 1}   \\ \hline\hline
        {\bf 0} & {\bf 0} & {\bf 1} & {\bf 0}\\ \hline
        {\bf 0} & {\bf 1} & {\bf 0} & {\bf 0}\\
      \end{array}
    \right) \\ \nonumber
    \mbox{} 
  \end{array}\,;\,\, {\Scr A} = \frac{1}{\sqrt{2}}\hskip -0.5cm \begin{array}{c c}
    & \begin{array} {@{} c c  @{}}
      \mbox{{\scriptsize 3 + n}} & \mbox{{\scriptsize 3 + n}}
    \end{array} \\
    \begin{array}{c}
      \mbox{{\scriptsize 3 + n}} \\ \mbox{{\scriptsize 3 + n}}
    \end{array}\hspace{-17em} &
    \left(
      \begin{array}{@{}  c|| c @{}}
        \phantom{3} {\bf 1}  \phantom{3}  & \phantom{3} i {\bf 1}  \phantom{3} \\ \hline\hline
         \phantom{3} {\bf 1} \phantom{3}   &  -i {\bf 1} \phantom{3}
      \end{array}
    \right) \\ \nonumber
    \mbox{} 
  \end{array}\quad \quad\quad .
\end{equation}
${\Scr A}$ being the Cayley matrix and each block in its matrix representation has dimension $(3+n)\times (3+n)$. We denote  by ${\Scr R}[T]_M{}^N$ the representation of a generic element $T$ of ${\rm G}$ is the new real basis. It defines an embedding of ${\rm G}$ into the group ${\rm Sp}(2(3+n),\mathbb{R})$:
\begin{equation}
{\Scr R}\,:\,\,\,{\rm G}\,\longrightarrow\,\,\,{\rm Sp}(2(3+n),\mathbb{R})\,\,\Leftrightarrow\,\,\,\forall T\in {\rm G}\,:\,\,\,{\Scr R}[T]^T\cdot\mathbb{C}\cdot{\Scr R}[T]=\mathbb{C}\,,
\end{equation}
where $$\mathbb{C}\equiv \left(\begin{matrix}{\bf 0} & {\bf 1}\cr -{\bf 1} & {\bf 0} \end{matrix}\right)\,.$$
The real symplectic basis is the one in which the vector field strengths $F^\Lambda_{\mu\nu}=\partial_\mu A^\Lambda_\nu-\partial_\nu A^\Lambda_\mu$, together with their magnetic duals $G_{\Lambda\,\mu\nu}$, transform,
as components of a single symplectic vector:
\begin{equation}
{ G}^M_{\mu\nu}\equiv \left(\begin{matrix}F^\Lambda_{\mu\nu}\cr G_{\Lambda\,\mu\nu}\end{matrix}\right)\,,
\end{equation}
the dual field strengths $G_{\Lambda\,\mu\nu}$ being defined, as usual, in the following way:
\begin{equation}
G_{\Lambda\,\mu\nu}=-\epsilon_{\mu\nu\rho\sigma}\,\frac{\delta {\Scr L}}{\delta F^\Lambda_{\rho\sigma}}\,.
\end{equation}
The electric-magnetic duality action of an element $T$ in ${\rm SU}(3,n)$
is effected as follows:
\begin{equation}
{G}^M_{\mu\nu}\,\,\rightarrow\,\,\,\,{ G}^{\prime M}_{\mu\nu}={\Scr R}[T]^{-1}{}_N{}^M\,{ G}^N_{\mu\nu}\,.\label{dualiT}
\end{equation}
We shall collectively denote by $A^M_\mu$ the vector of electric gauge fields and their magnetic duals, so that, locally, ${ G}^M=dA^M$. The representations of the various fields with respect to ${\rm G}$ and $H\sim H_R\times H_{{\rm matter}}$ are given in Table \ref{N3content}.
\begin{table}
\begin{center}
\begin{tabular}{|c||c|c|c|c|c|c|c|}
  \hline
    & $A^M_\mu$ &  $\psi_{A\mu}$& $\chi_\bullet$& $\lambda_{I A}$ &$\lambda_{I}$ & $F^{AB}_{\mu\nu}$ & $F^{I}_{\mu\nu}$\\
    \hline \hline
 $G$ & ${\bf (3+n)}+\overline{{\bf (3+n)}}$ & ${\bf 1}$ & ${\bf 1}$ & ${\bf 1}$ & ${\bf 1}$& ${\bf 1}$& ${\bf 1}$\\
  \hline
 $H$ & ${\bf (1,1)}_0$ & ${\bf (3,1)}_{+\frac{1}{2}}$ & ${\bf (1,1)}_{+\frac{3}{2}}$ & ${\bf (3,n)}_{\frac{n+6}{2n}}$ &${\bf (1,n)}_{\frac{3(n+2)}{2n}}$ & ${\bf (3,1)}_{-1}$ & ${\bf (1,\bar{n})}_{-\frac{3}{n}}$\\
  \hline
\end{tabular}
  \caption{Relevant representations with respect to ${\rm G}$ and $H$. The fermions $\lambda_{IA}$ and $\lambda_I$ have opposite chirality.}\label{N3content}
\end{center}
\end{table}
\paragraph{The coset geometry.} The scalar fields $\phi=(\phi^s)$ are described in the theory by a coset representative ${L}(\phi)\in {\rm SU}(3,n)$ so that the action on $\phi^s$, by an element $T\in {\rm SU}(3,n)$ of the isometry group of the scalar manifold:
\begin{equation}
T:\,\,\,\phi^s\,\rightarrow\,\,\phi^{\prime s}=\phi^{\prime s}(\phi^r)\,,
\end{equation}
is defined by the left action of $T$ on the coset representative, modulo the right action of $H$, namely by the equation:
\begin{equation}
T\cdot {L}(\phi)={L}(\phi')\cdot h(\phi,T)\,,\label{Tphi}
\end{equation}
where $h(\phi,T)$ is a compensator in $H$.
The Lie algebra $\mathfrak{g}=\mathfrak{su}(3,n)$ of ${\rm G}$ can be written, according to the Cartan decomposition, as the direct sum of its maximal
compact subalgebra $\mathfrak{H}=\mathfrak{u}(3)\oplus \mathfrak{su}(n)$, generating $H$, and the subspace $\mathfrak{K}$ of non-compact generators:
\begin{equation}
\mathfrak{g}=\mathfrak{H}\oplus \mathfrak{K}\,.\label{Cartandec}
\end{equation}
We shall find it convenient to choose for the scalar manifold an $H$-covariant parametrization, which amounts to choosing the coset representative
as follows:
\begin{equation}
{L}(\phi)\in e^\mathfrak{K}\,,
\end{equation}
namely the scalar fields $\phi^s$ to be parameters of the non-compact generators $\mathfrak{K}\subset\mathfrak{g}$. Being $[\mathfrak{H},\, \mathfrak{K}]\subset \mathfrak{K}$, $\mathfrak{K}$ supports a representation of $H$ and, in the
chosen parametrization, the scalar fields transform under $H$ in the same
representation. This representation is the ${\bf (\bar{3},{n})}_{k}+{\bf (3,\bar{n})}_{-k}$, where $k=1+3/n$ and the scalar fields have the following index structure:
\begin{equation}
\phi=(\phi_{AB\,J},\,\phi^{AB\,J})\,\,,\,\,\,\,\,\phi^{AB\,J}=(\phi_{AB\,J})^*\,.
\end{equation}
According to the general theory of coset-spaces, in terms of ${L}(\phi)$ we can construct the left-invariant 1-form $\Omega(\phi)$ with values in $\mathfrak{g}$
\begin{equation}
\Omega(\phi)\equiv L^{-1}dL(\phi)={\Scr Q}(\phi)+{\Scr P}(\phi)\,,\label{OPQ}
\end{equation}
where ${\Scr Q},\,{\Scr P}$ are the projections of $\Omega$ on the subspaces $\mathfrak{H}$ and $\mathfrak{K}$, respectively. The
$\mathfrak{su}(3,n)$-Maurer-Cartan equations $d\Omega+\Omega\wedge \Omega=0$ imply the following relations:
\begin{align}
\mathcal{ R}[{\Scr Q}]\equiv d{\Scr Q}+{\Scr Q}\wedge {\Scr Q}=-{\Scr P}\wedge {\Scr P}\,,\nonumber\\
\mathcal{D}{\Scr P}\equiv d{\Scr P}+{\Scr Q}\wedge {\Scr P}+{\Scr P}\wedge {\Scr Q}=0\,,
\end{align}
where $\mathcal{R}[{\Scr Q}]$ is the curvature 2-form, with values in $\mathfrak{H}$, while $\mathcal{D}$ defines the exterior $H$-covariant derivative acting on ${\Scr P}$. The $H$-irreducible components of ${\Scr Q}$ and ${\Scr P}$ can be read from the matrix form of $\Omega$ in the fundamental representation of ${\rm G}$:
\begin{equation}
\Omega=\left(\begin{matrix}{\Scr Q}_{AB}{}^{CD} & {\Scr P}_{AB\,J}\cr {\Scr P}^{I\, CD} & {\Scr Q}^{I}{}_J\end{matrix}\right)\,,
\end{equation}
where ${\Scr P}_{AB\,I}=({\Scr P}^{I\,AB})^*$. We further define ${\Scr
P}^{AB\,I}=({\Scr P}_{AB\,I})^*$ and ${\Scr P}_{I\,AB}=({\Scr P}^{I\,AB})^*$.
The  Riemannian metric on the  scalar manifold can be computed as follows:
\begin{equation}
{\Scr G}_{st}(\phi)\,d\phi^s\otimes d\phi^t={\Scr P}_s^{AB\,I}{\Scr P}_{AB\,I\vert t}\,d\phi^s\otimes d\phi^t\,.
\end{equation}
The coset representative can be evaluated as a symplectic matrix in the ${\Scr R}$-representation:
We shall also equivalently describe ${\Scr P}_s^{AB\,I}$ by the tensor ${\Scr P}_{s A}^{~~~I}=\frac{1}{2}\epsilon_{ABC}{\Scr P}_s^{BC\,I}$.
\begin{equation}
{\Scr R}[L(\phi)]=(L(\phi)_M{}^N)\,.
\end{equation}
Similarly equation (\ref{Tphi}) can be written in terms of matrices in the same representation. We choose the symplectic basis so that the compensator in (\ref{Tphi}) is represented by an orthogonal $(6+2n)\times (6+2n)$ matrix ${\Scr R}[h(\phi,\,T)]$. Since the coset representative is acted
on from the left and from the right by two different groups, namely ${\rm G}$ and $H$, respectively, we can refer the corresponding indices to two different bases. We choose in particular the real basis for the left index and
the complex one for the right index, so as to define the matrix $\tilde{L}(\phi) \equiv (L(\phi)_M{}^{\underline{N}})$.\par
One can define on the scalar manifold  the following $(6+2n)\times (6+2n)$ symmetric, symplectic, negative-definite matrix (summation over $\underline{N}$ being understood)
\begin{equation}
\mathcal{M}(\phi)=-\tilde{L}(\phi)\cdot \tilde{L}(\phi)^\dagger=-\,L(\phi)_M{}^{\underline{N}}(L(\phi)_N{}^{\underline{N}})^*\,,\label{Mdef}
\end{equation}
which encodes the non-minimal couplings of the scalar fields to the vector ones.
Under an isometry $T\in {\rm SU}(3,n)$ which maps $\phi$ into $\phi'(\phi)$, the matrix $\mathcal{M}(\phi)$ transforms as follows:
\begin{equation}
\mathcal{M}(\phi')={\Scr R}[T]\cdot \mathcal{M}(\phi)\cdot{\Scr R}[T]^T\,,
\end{equation}
as it can be verified by applying eq. (\ref{Tphi}) in the relevant representation, together with the property that ${\Scr R}_\eta[h(\phi,T)]$ is unitary (orthogonal in the real basis).\par
The definition of the dual field strengths $G_{\Lambda\mu\nu}$ can be encoded in a symplectic covariant condition on ${ G}^M_{\mu\nu}$
known as \emph{twisted self-duality condition} \cite{Cremmer:1978ds} (we suppress the spacetime indices for convenience):
\begin{equation}
{}^*{ G}^M=-(\mathbb{C}\cdot\mathcal{M}(\phi))^M{}_N\, { G}^N\,.\label{tsdc}
\end{equation}
One can verify that the group ${\rm G}$ is a global symmetry of the field equations and Bianchi identities provided the action of a generic isometry $T\in {\rm G}$ on the scalar fields is combined with a symplectic duality action  (\ref{dualiT}) on the vector field strengths and their magnetic duals and
with the action of the compensating transformation $h(\phi,T)$ on the fermionic fields in the appropriate $H$-representation \cite{Gaillard:1981rj}.
\subsection{The Gauged Model}
So far we have been dealing with the ungauged $\mathcal{N}=3,\,D=4$ models, focussing on their main features and in particular on their on-shell global symmetry properties, encoded in the group ${\rm G}={\rm SU}(3,\,n)$. Supersymmetry requires these models to have no scalar potential and thus the only vacuum is a Minkowski spacetime with $3n$ complex scalar moduli. Non-trivial dynamics for the scalar fields, encoded in a scalar potential, can be introduced, without manifestly breaking supersymmetry, through the \emph{gauging procedure}, which amounts to introducing an internal gauge group $\mathcal{G}$ and corresponding minimal couplings of the vector fields to the other fields (see \cite{Castellani:1985ka} for the original construction of $\mathcal{N}=3,\,D=4$ models with electric gaugings).
Although in the present work we shall focus on a specific gauging and study the corresponding vacuum structure, we describe in this section the general duality covariant formulation of the gauging procedure based on the
embedding tensor \cite{Cordaro:1998tx,Nicolai:2000sc,deWit:2002vt,deWit:2005ub}, see \cite{Samtleben:2008pe,Trigiante:2016mnt} for reviews.\par
The gauging procedure consists in promoting a suitable subgroup $\mathcal{G}$ of the global symmetry group ${\rm G}$ of the ungauged theory to local symmetry and in modifying the Lagrangian and the supersymmetry transformation laws in order for the resulting theory to feature the same amount of  supersymmetry as the original one ($\mathcal{N}=3$ in our case). Gauging
a group $\mathcal{G}$ requires  the introduction of minimal couplings of the vector fields to the other fields. This in general would break the original electric-magnetic duality symmetry ${\rm G}$ of the ungauged model. In the embedding tensor formulation of the gauging procedure, we keep a formal ${\rm G}$-covariance of the field equations by encoding all the information about the local embedding of $\mathcal{G}$ inside ${\rm G}$ in a ${\rm G}$-covariant tensor $\Theta$. This formalism requires a certain level of redundancy in the description of the theory by introducing, aside from the electric gauge fields $A^\Lambda_\mu$ corresponding to gauge generators $X_\Lambda$, also magnetic ones $A_{\Lambda\mu}$ gauging the generators $X^\Lambda$ and two- forms $B_{\alpha}=(B_{\alpha\mu\nu})$, $\alpha=1,\,\dots,\,{\rm
dim}({\rm G})$, in the adjoint representation of the global symmetry group ${\rm G}$.
Grouping the electric and magnetic vectors, as well as the corresponding gauge generators, in symplectic vectors $A^M_\mu,\,X_M$, respectively, we
can write the gauge connection as follows:
\begin{equation}
\Omega_{g\,\mu}\equiv g\,A^M_\mu\,X_M\,, \label{gaugeconn}
\end{equation}
$g$ being the coupling constant.
The condition that the gauge algebra of $\mathcal{G}$ be a subalgebra of ${\rm G}$ in turn requires that the gauge generators $X_M$ be linear combinations of the global symmetry group generators $t_\alpha$:\footnote{The
electric-magnetic duality representation ${\Scr R}$ being symplectic, ${\Scr R}[t_\alpha]$ are symplectic generators and thus satisfy the following condition: ${\Scr R}[t_\alpha]_M{}^P\mathbb{C}_{NP}={\Scr R}[t_\alpha]_N{}^P\mathbb{C}_{MP}$.}
\begin{equation}
X_M=\Theta_M{}^\alpha\,t_\alpha\,.\label{Theta}
\end{equation}
This defines the embedding tensor $\Theta_M{}^\alpha$ which encodes all the information about the choice of $\mathcal{G}$ inside ${\rm G}$. This object formally transforms in the product ${\Scr R}\times {\rm Adj}({\rm G})$ of the symplectic (electric-magnetic) duality representation ${\Scr R}$ of ${\rm G}$ times its adjoint representation. Consistency of the gauging procedure, namely the possibility of constructing a locally $\mathcal{G}$-invariant, $\mathcal{N}$-supersymmetric action, requires $\Theta$ to satisfy linear and quadratic constraints. These are best expressed in terms of the tensor $$X_{MN}{}^P\equiv \Theta_M{}^\alpha\,{\Scr R}[t_\alpha]_N{}^P\,.$$
The linear constraint reads:
\begin{equation}
X_{(MN}{}^R\mathbb{C}_{P)R}=0\,.\label{conl}
\end{equation}
The quadratic constraints are two:
\begin{align}
a)&:\,\,[X_M,\,X_N]+X_{MN}{}^P\,X_P=0\,,\label{cona}\\
b)&:\,\,\mathbb{C}^{MN}\,\Theta_{M}{}^\alpha\,\Theta_{N}{}^\beta=0\,.\label{conb}
\end{align}
The former expresses the property of $\Theta$ being gauge invariant and implies that the gauge fields transform under the duality action of $\mathcal{G}$, as a global symmetry group, in its co-adjoint representation. The latter condition (\ref{conb}) guarantees that no more than the number $n_v$ of the vector fields of the model are involved in the gauging, namely that there are no more than $n_v$ linearly independent gauge generators $X_M$. This condition, in particular, implies the existence of a symplectic frame (related to the original one by a symplectic transformation) in which the ``magnetic'' components $\Theta^{\Lambda\,\alpha}$ of $\Theta$ are zero (``electric'' frame). It can be shown that, for $\mathcal{N}\ge 3$, condition (\ref{cona}) implies (\ref{conb}), while in the maximal theory, $\mathcal{N}=8$, they are equivalent.\par
Spacetime derivatives in the action are then replaced by covariant ones:
\begin{equation}
\partial_\mu\rightarrow \partial_\mu- \Omega_{g\,\mu}\,,
\end{equation}
and the abelian field strengths by non-abelian ones:
\begin{equation}
\partial_\mu A^M_\nu-\partial_\nu A^M_\mu\,\rightarrow\,\partial_\mu A^M_\nu-\partial_\nu A^M_\mu+g X_{NP}{}^M\,A^N_{[\mu} A^P_{\nu ]}\,.
\end{equation}
This, as far as the scalar part of the action is concerned, requires considering the \emph{gauged Maurer-Cartan vielbein and connection matrices} $\hat{{\Scr P}},\,\hat{{\Scr Q}}$ on the scalar manifold.
The latter are constructed out of the gauged Maurer-Cartan left-invariant 1-form, in the complex basis (\ref{cbasis}), as follows:
\begin{equation}
\hat{\Omega} ={\Scr R}[L^{-1}(d-\Omega_g)L]=\hat{{\Scr P}}+\hat{{\Scr Q}}\,,
\end{equation}
where $2 \hat{{\Scr P}}=\hat{\Omega}+\hat{\Omega}^\dagger$ and $2\hat{{\Scr Q}} = \hat{\Omega}-\hat{\Omega}^\dagger$ are the non-compact and compact components of $\hat{\Omega}$ respectively. In other words, $\hat{{\Scr P}}$ is the gauged vielbein
and $\hat{{\Scr Q}}$ is the gauged $H$-connection in the real symplectic representation.\\
The gauged scalar kinetic term reads as
\begin{equation}
{\rm e}^{-1} {\Scr L}_{{\rm scal. kin}}= \frac{1}{2} {\Scr G}_{rs} {\Scr D}_\mu \phi^r {\Scr D}^\mu \phi^s=\frac{1}{2}\,{\rm Tr}( \hat{{\Scr P}}_\mu\cdot \hat{{\Scr P}}^\mu)=\hat{{\Scr P}}_{\mu}^{\,AB\,I}\,\hat{{\Scr P}}^{\,\mu}_{AB\,I}\,, \label{sympscal}
\end{equation}
where ${\rm e}=\sqrt{|{\rm det}(g_{\mu\nu})|}$ and ${\Scr D}_\mu \phi^s = \partial_\mu \phi^s -g A^M_\mu 
\Theta_M^\alpha k_\alpha^s$ is the gauge covariant derivative of the scalar fields, $k_\alpha^s$ being the Killing vectors of the scalar manifold isometries generatoed by $t_\alpha$. \par
The vector kinetic terms in the Lagrangian read:
\begin{equation}
  {\rm e}^{-1} {\Scr L}_{{\rm v. kin}}= \frac{1}{4}\mathcal{I}_{\Lambda \Sigma} \mathcal{H}^\Lambda_{\mu \nu} \mathcal{H}^{\Sigma ~ \mu \nu} + \frac{1}{8 {\rm e}}\mathcal{R}_{\Lambda \Sigma} \epsilon^{\mu \nu \tau \gamma} \mathcal{H}^\Lambda_{\mu \nu} \mathcal{H}^\Sigma_{\tau \gamma}\,,\label{Lvkin}
\end{equation}
The symmetric matrices $\mathcal{I}_{\Lambda \Sigma} $ and $\mathcal{R}_{\Lambda \Sigma} $ are derived from the symplectic matrix $\mathcal{M}(\phi)$ defined in (\ref{Mdef}) as follows\footnote{Recall that we have chosen the symplectic basis so that ${\Scr R}[H] \in {\rm SO}(6+2n)$.}
\begin{equation}
  \mathcal{M}= -\tilde{L}\,\tilde{L}^\dagger= \left(
    \begin{matrix}
      \mathcal{R}\mathcal{I}^{-1}\mathcal{R}+\mathcal{I} & -\mathcal{R}\mathcal{I}^{-1} \\
      -\mathcal{I}^{-1}\mathcal{R} & \mathcal{I}^{-1}
  \end{matrix}  \right)\,.
\end{equation}
$\mathcal{H}^\Lambda$ is the electric component of the symplectic field strength $\mathcal{H}^M=F^M+\frac{g}{2}\mathbb{C}^{MN} \Theta_M^\alpha B_\alpha$, $F^M=dA^M+\frac{g}{2} X_{NP}^{~~~M} A^N \wedge A^P$ being the symplectic non-abelian field strength of $A^M$. In general, this latter is not gauge covariant since the generalized gauge structure constants $X_{MN}^{~~~~P}=X_{[MN]}^{~~~~~P}+X_{(MN)}^{~~~~~P}$ do not satisfy the Jacobi identity whenever the symmetric component  $X_{(MN)}^{~~~~~~P}$ is non-vanishing. The auxiliary fields $B_{\alpha\,\mu\nu}$, and their suitably defined gauge variation, must be introduced in order to define the gauge covariant field strength ${ G}^M = (\mathcal{H}^\Lambda, G_\Lambda$). $G_\Lambda=-\epsilon_{\mu \nu \rho \sigma} \frac{\delta {\Scr L}}{\delta \mathcal{H}_{\rho \sigma}}$ is the dual of $\mathcal{H}^\Lambda$\footnote{This is true once the field equations for $B_{\alpha\,\mu \nu}$ are implemented. The latter imply the identification of  $\mathcal{H}_\Lambda$ and $G_\Lambda$.}. In terms of ${ G}^M$ we obtain the equations of motion for $A^M_\mu$ (which comprise the field equations for the electric vector fields and the Bianchi identities):
\begin{equation}
  \epsilon^{\mu \nu \rho \sigma} {\Scr D}_\nu { G}^M_{\rho \sigma}
= 2\, \mathbb{C}^{MN} \frac{\delta {\Scr L}_{{\rm matter}}}{\delta A^M_\mu}\,, \label{Amotion}
\end{equation}
where ${\Scr L}_{{\rm matter}}$ denotes the part of the Lagrangian describing the coupling of the vectors to the scalar and fermion fields.
\par
Gauge invariance of the action and supersymmetry require the addition of order-$g$ topological terms and Yukawa terms
to the action as well as an order-$g^2$ scalar potential
$V(\phi)$.
The Yukawa terms have the following general form (we use the notation of \cite{Trigiante:2016mnt}):
\begin{align}
{\rm e}^{-1}\,\mathcal{L}_{{\rm Yukawa}}=&g\,\left(2\,\bar{\psi}^A_\mu\gamma^{\mu\nu}\,\psi_\nu^B\,\mathbb{S}_{AB}+{\rm i}\,\bar{\lambda}^{\mathcal{I}}\gamma^\mu\psi_{A\mu}\,\mathbb{N}_{\mathcal{I}}{}^A+\bar{\lambda}^{\mathcal{I}}\lambda^{\mathcal{I}}\,
\mathbb{M}_{\mathcal{I}\mathcal{J}}\right)+{\rm h.c.}\,.\label{Yukawa}
\end{align}
where $\lambda_{\mathcal{I}}$ is a collective symbol to describe the positive-chirality spin-$1/2$ fermions
$$\lambda_{\mathcal{I}}\equiv \{\lambda_{IA},\,\lambda^I,\,\chi_\bullet\}\,.$$
As usual, in the Weyl representation, the positive-chirality spinor fields $\lambda^{\mathcal{I}}$ are the charge-conjugate of the negative-chirality ones.
The quantities $\mathbb{S}_{AB}=\mathbb{S}_{BA}$, $\mathbb{N}_{\mathcal{I}}{}^A$ and $\mathbb{M}_{\mathcal{I}\mathcal{J}}$, as well as their complex conjugates $\mathbb{S}^{AB}\equiv (\mathbb{S}_{AB})^*$, $\mathbb{N}^{\mathcal{I}}{}_A\equiv (\mathbb{N}_{\mathcal{I}}{}^A)^*$, $\mathbb{M}^{\mathcal{I}\mathcal{J}}\equiv (\mathbb{M}_{\mathcal{I}\mathcal{J}})^*$, are $H$-covariant tensors which depend on the scalar fields and (linearly) on the embedding tensor. The same quantities uniquely define the scalar potential which satisfies the so-called \emph{potential Ward identity}:
\begin{equation}
\delta_B^A\,V(\phi)=
g^2\left(\mathbb{N}_{\mathcal{I}}{}^A\mathbb{N}^{\mathcal{I}}{}_B-12\,\mathbb{S}^{AC}\,
\mathbb{S}_{BC}\right)\,.\label{PWard}
\end{equation}
Besides order-$g$ and $g^2$ modifications to the action,
the gauging procedure also requires additional order-$g$ terms in the supersymmetry transformation laws of the fermion fields:
\begin{align}
\delta\psi_{A\mu}&=\mathcal{D}_\mu\epsilon_A+{\rm i}\,g\,\mathbb{S}_{AB}\gamma_\mu\epsilon^B+\dots\,,\nonumber\\
\delta \lambda_{\mathcal{I}}&=g\,\mathbb{N}_{\mathcal{I}}{}^A\,\epsilon_A+\dots\,.
\end{align}
Note that all the modifications of the action and the fermion supersymmetry transformation laws implied by the gauging procedure are defined in terms of the composite fields $\mathbb{S}_{AB}$, $\mathbb{N}_{\mathcal{I}}{}^A$ and $\mathbb{M}_{\mathcal{I}\mathcal{J}}$. These $H$-covariant tensors in $\mathcal{N}=3$ supergravities are
\begin{align}
\mathbb{S}_{AB}=\mathbb{S}_{BA}\,\,,\,\,\,\mathbb{N}^{IA}{}_B\,\,,\,\,\,\mathbb{N}^{AI}\,\,,\,\,\,
\mathbb{N}_A\,\,,\,\,\,\mathbb{M}_{IA}{}^J\,\,,\,\,\,\mathbb{M}^{\bullet,IA}\,\,,\,\,\,\mathbb{M}^{\bullet,I}\,\,,\,\,\,\mathbb{M}^{IA,JB}\,,\label{NsMs}
\end{align}
and are components of a single $H$-tensor calledd \emph{T-tensor}, defined
in terms of $X_M$ and $L$ as:
\begin{equation}
  \mathbb{T}_{\underline{MN}}{}^{\underline{P}} \equiv (\tilde{L}^{-1})_{\underline{M}}{}^Q \left(\tilde{L}^{-1}\,{\Scr R}[X_Q] \,\tilde{L} \right)_{\underline{N}}{}^{\underline{P}}\,. \label{eq:T_tensor}
\end{equation}
The $T$-tensor $\mathbb{T}_{\underline{MN}}{}^{\underline{P}}$ being the transform of $X_{MN}{}^P$ via $\tilde{L}^{-1}$, it satisfies the same linear and quadratic constraints as the latter:
\begin{eqnarray}
  \mathbb{T}_{(\underline{MN}}{}^{\underline{P}} \mathbb{C}_{\underline{Q})\underline{P}} \equiv ~~ -\mathbb{T}_{(\underline{MNQ})}
& = & 0 \label{linea} \\
  \left[ \mathbb{T}_{\underline{M}} , \mathbb{T}_{\underline{N}} \right]_{\underline{R}}{}^{\underline{T}} + \mathbb{T}_{\underline{MN}}{}^{\underline{P}} \mathbb{T}_{\underline{PR}}{}^{\underline{T}} & = & 0 \label{quadrato}
\end{eqnarray}
 The former selects, within the product of $\Scr{R}\times {\rm Adj}({\rm G})$, the representation (using the Dynking label notation):
 \begin{equation}
 (0,1,0,\dots,0,1)\oplus (1,0,\dots,0,1,0)
 \end{equation}
 corresponding to the tensors:
 $$\mathbb{T}_{\underline{\Lambda\Sigma}}{}^{\underline{\Gamma}}=
\mathbb{T}_{[\underline{\Lambda\Sigma}]}{}^{\underline{\Gamma}}\,\,,\,\,\,\,\,
\mathbb{T}^{\underline{\Lambda\Sigma}}{}_{\underline{\Gamma}}=
\left(\mathbb{T}_{\underline{\Lambda\Sigma}}{}^{\underline{\Gamma}}\right)^*=
\mathbb{T}^{[\underline{\Lambda\Sigma}]}{}_{\underline{\Gamma}}\,,$$
respectively. The underlined indices refer to the complex basis (\ref{cbasis}) and run from 1 to $3+n$.
The fermion shift tensors and the mass matrices (\ref{NsMs}) are identified with the $H={\rm S}[{{\rm U}(3)\times {\rm U}(n)}]$-irreducible components of $\mathbb{T}_{\underline{\Lambda\Sigma}}{}^{\underline{\Gamma}}$, $\mathbb{T}^{\underline{\Lambda\Sigma}}{}_{\underline{\Gamma}}$.
 The precise relations are given in Appendices \ref{SM} and \ref{D}.
 In particular the fermion-shift tensors entering are expressed in terms of the components of the $T$-tensor as follows:
\begin{eqnarray}
  \mathbb{S}_{A B} &=& - \frac{1}{2} \epsilon_{(A | C D} \mathbb{T}^{C D}_{~~~~B)}\,, \nonumber\\
  \mathbb{N}^B &=& \mathbb{T}^{E B}_{~~~~E}\,, \nonumber\\
  \mathbb{N}_{C I} &=& \, \epsilon_{A B C} \mathbb{T}^{A B}_{~~~~I} \,,\nonumber\\
  \mathbb{N}^{I A}_{~~~~B} &=& \,- 2 \, \mathbb{T}^{I A}_{~~~~B} + \, \mathbb{T}^{I C}_{~~~~C} \delta^{A}_{~B}\,. \label{eq:fermionic_shifts}
\end{eqnarray}\par
There are differential relations among these $H$-tensors named \emph{gradient flow} equations \cite{DAuria:2001rlt,Trigiante:2016mnt}. They are found by decomposing in irreducible $H$-components the general relation:
\begin{equation}
\mathcal{D}\mathbb{T}_{MN}^{~~~~P}=-{\Scr R}[{\Scr P}]_M^{~~Q}\mathbb{T}_{QN}^{~~~~P}+\left[ \mathbb{T}, {\Scr R}[{\Scr P}]\right]_N^{~~P}\,,\label{GF}
\end{equation}
where $\mathcal{D}$ is the $H$-covariant derivative. The above equation is obtained from the definition of the $T$-tensor and eq. (\ref{OPQ}) in the ${\Scr R}$-representation. \par
Finally the quadratic constraints (\ref{quadrato}) also imply the potential Ward identity (\ref{PWard}) which, specialized to the $\mathcal{N}=3$ models under consideration, reads:
\begin{equation}
\mathbb{N}_A \mathbb{N}^B + \mathbb{N}_{A I} \mathbb{N}^{B I} + \mathbb{N}_{IC}^{~~~~B} \mathbb{N}^{IC}_{~~~~A} - \, 12 \,
\mathbb{S}_{AC} \mathbb{S}^{BC} = \delta_{A}^{B} V\,, \label{wardid}
\end{equation}
where, for the sake of notational convenience, we have absorbed the coupling constant $g$ in the definition of the embedding tensor and thus in the fermion-shift tensors. This identity is necessary in order to preserve $\mathcal{N}=3$ supersymmetry of the gauged action to quadratic order in the embedding tensor.
For a derivation of the potential Ward identity from the quadratic constraints see Appendix \ref{ward}.
\subsection{General Mass Formulae}
We give below the general mass formulae for the fermionic and bosonic fields in a given vacuum of the model. On this background the vector and fermion fields vanish, while the scalar fields take constant values $\phi_0=(\phi_0^s)$ which extremize the scalar potential:
\begin{equation}
    \left.\frac{\partial V}{\partial\phi^s}\right\vert_{\phi=\phi_0}=0\,.
\end{equation}
The value $V_0=V(\phi_0)$ of the scalar potential in $\phi_0$ defines the cosmological constant: $\Lambda=V_0$. We shall study vacua of anti-de Sitter (AdS) type, for which $V_0<0$. In this case the AdS radius $L$ is given by $L=\sqrt{-\frac{3}{V_0}}$.
\subsubsection{Scalar Masses}
The scalar masses can be computed on the vacuum $\phi_0$ by expanding, up to second order terms in the scalar fluctuations around $\phi_0$, the scalar-field part of the gauged  action
\begin{equation}
e^{-1} {\Scr L}_{{\rm scal}}=\frac{1}{2}{\Scr G}_{rs}\,{\Scr D}_\mu \phi^r {\Scr D}^\mu \phi^s -V(\phi)\,,
\end{equation}
where the kinetic part was defined in (\ref{sympscal}).
The square-mass matrix for the scalar fields then reads:
\begin{equation}
M^{({\rm scal})}{}_r{}^{t}=\left.{\Scr G}^{ts}\frac{\partial^2 V}{\partial \phi^s \partial \phi^r} \right\vert_{\phi=\phi_0}\,.
\end{equation}
The squared-mass spectrum of the scalar fields on $\phi_0$ is then given by the eigenvalues of $M^{({\rm scal})}{}_r{}^{t}$.

\subsubsection{Vector Mass Matrix}
The masses for the vector fields originate from their minimal couplings to the scalars. 
By using the twisted self-duality condition (\ref{tsdc}) 
\begin{equation}
{}^*{G}=-\mathbb{C}\cdot\mathcal{M}\cdot{G}
\end{equation}
which holds also for the gauged field strengths ${G}^M$,
and restricting only to the couplings of the vector fields to the scalar ones,
one can rewrite eqs. (\ref{Amotion}) in the following form
\begin{equation}
\mathcal{M}_{MN}{}^*{\Scr D}{}^*{G}^N\,=\,g^2 \Theta_M^\alpha\,k_\alpha^r\, {\Scr G}_{rs}\,
k_\beta^s\, \Theta_N^\beta A^N\,.
\end{equation}
From this we obtain the vector squared-mass matrix on the vacuum:
\begin{equation}
  M^{({\rm vector})}{}^P{}_M=g^2\, {\Scr R}[{L}^{-1}]^{ ~~P}_Q{\Scr R}[{L}^{-1}]^{ ~~~N}_Q {\Scr K}_{NM}=-g^2\,(\mathcal{M}^{-1}\cdot{\Scr K})_{~~~~M}^{~P}
\end{equation}
where
\begin{equation}
{\Scr K}_{MN} \equiv \left.\Theta_M^\alpha\,k_\alpha^r\, {\Scr G}_{rs}\,
k_\beta^s\, \Theta_N^\beta\right\vert_{\phi =\phi_0}\,.
\end{equation}
The eigenvalues of $M^{({\rm vector})}{}^P{}_M$ will correspond to the vector squared-mass
spectrum\footnote{Note that this spectrum does not depend on the symplectic frame. Since, by virtue of the quadratic constraint on the embedding tensor, we can always rotate the latter to the electric frame in which $\Theta^{\Lambda\,\alpha}=0$. Because of the quadratic constraint, half of the eigenvalues of the matrix $M^{({\rm vector})}$ vanish.}. Let us observe that ${\rm
det}\left(M^{({\rm vector})}\right) \propto \,{\rm det}\left({\Scr R}[{L}^{-1}]\cdot {\Scr K}\cdot {\Scr R}[{L}^{-T}]\right)$. This allows us to compute the vector mass spectrum as the eigenvalues of
\begin{equation}
  \mathbb{M}^{({\rm vector})}_{PN}=\frac{g^2}{4}{\rm Tr}\left(\mathbb{T}_{  P}\cdot\mathbb{T}_{N}+\mathbb{T}_{
P}\cdot(\mathbb{T}_{ N})^\dagger\right)\,.
\end{equation}
Indeed \footnote{See also eq.(\ref{sympscal}). Naively, we trade the scalar product on $\frac{{\rm G}}{H}$ with the trace on the ${\Scr R}$ symplectic
representation.}
\begin{equation}
  {\Scr K}_{MN}=\frac{1}{2}{\rm Tr}\left(K_M K_N\right)\,,
\end{equation}
where $$K_M\equiv \frac{1}{2}\,\left({\Scr R}[{L}]^{-1} \cdot X_M \cdot {\Scr R}[{L}]+({\Scr R}[{L}]^{-1} \cdot X_M \cdot {\Scr R}[{L}])^\dagger\right)$$ 
is
the projection of the adjoint action of ${L}$ on $X_M$ along its non-compact component.

\subsection{Fermionic masses}
The masses for the fermion fields originate from the Yukawa terms (\ref{Yukawa}).
As mentioned earlier, the tensors 
$\mathbb{S}_{AB}$, $\mathbb{N}_{\mathcal{I}}^A$ and  $\mathbb{M}_{\mathcal{I J}}$ are defined as components of the $\mathbb{T}$-tensor.

\subsubsection{Gravitinos masses and Supersymmetry breaking}

Let us choose as fermionic vacuum $\langle\psi^A_\mu\rangle=\langle\lambda_{\mathcal{I}}\rangle=0$. In order to preserve the supersymmetry generated by $\epsilon_A Q^A$ we must have
\begin{eqnarray}
\langle\delta \psi_{A \mu}\rangle &=& \nabla_\mu \epsilon_A + {\rm i} \mathbb{S}_{AB}|_{\phi=\phi_0} \gamma_\mu \epsilon^B = 0 \label{killingspinors} \\
\langle\delta \lambda_{\mathcal{I}}\rangle&=&\mathbb{N}_{\mathcal{I}}^A|_{\phi=\phi_0}\epsilon_A =0 \label{goldstini}
\end{eqnarray}
Let us assume that the vacuum is $\mathcal{N}^{'}$ supersymmetric, $0 \leq \mathcal{N}^{'} \leq 3$. Then, we have $\mathcal{N}^{'}$ Killing spinors $\epsilon_a$, $a:1,...,\mathcal{N}^{'}$. Integrability of eq.(\ref{killingspinors}) implies
\begin{equation}
\mathbb{S}_{aA}\mathbb{S}^{bA}|_{\phi=\phi_0}=-\frac{V_0}{12} \delta_a^b
\end{equation}
While from eq.(\ref{goldstini}) we obtain $\mathbb{N}_{\mathcal{I}}^a=0$.
If the vacuum is $\mathcal{N}^{'}=3$ supersymmetric, then the gravitinos mass matrix $\mathbb{S}\mathbb{S}^\star|_{\phi=\phi_0}$ will be proportional to the identity with eigenvalues $m_\psi^2=-\frac{V_0}{12}$. When $V_0<0$, this will correspond to AdS-massless gravitinos, as expected in the case of fully preserved supersymmetry. Indeed, the goldstinos $\eta_A \propto \mathbb{N}^{\mathcal{I}}_A \lambda_{\mathcal{I}}$ vanish in that case.

\subsubsection{Fermionic matter masses}

Upon the redefinition \footnote{In other words, we reabsorb the massless goldstinos in the gravitinos. The sum is intended over the non vanishing goldstinos components, the one corresponding to a non singular sub-block of $\mathbb{S}\mathbb{S}^\star + \frac{V_0}{12}{\bf 1}_{3\times 3}$.}
\begin{equation}
  \psi^A_\mu \rightarrow \psi^A_\mu + \frac{{\rm i}}{12} \sum_{C} \left( \frac{\mathbb{S}}{\mathbb{S}\mathbb{S}^\star + \frac{V_0}{12}{\bf 1}_{3\times
3}} \right)^{A}_{C} \gamma_\mu \eta^C\end{equation}
we obtain the linearized equations
\begin{eqnarray}
  {\rm i} \gamma^{\mu} {\Scr D}_\mu \lambda_{\mathcal{I}} = \left(2\mathbb{M}_{\mathcal{IJ}}-\frac{1}{3} \sum_{AB} \left( \frac{\mathbb{S}}{\mathbb{S}\mathbb{S}^\star + \frac{V_0}{12}{\bf 1}_{3\times 3}} \right)_{AB}\mathbb{N}^A_{\mathcal{I}} \mathbb{N}_{\mathcal{J}}^{B}\right)\lambda^{\mathcal{J}} \equiv {\rm
M}_{\mathcal{IJ}}\lambda^{\mathcal{J}}
\end{eqnarray}
So we compute the fermionic matter mass spectrum as the eigenvalues of ${\rm M M^{\dagger \mathcal{J}}_{\mathcal{I}}}$.

\section{The Model with Gauge Group \texorpdfstring{${\rm SO}(3)\times {\rm SU}(3)$}{SO(3) x SU(3)}}\label{sec:gauged_model}
After having given, in the previous Section, a general review of the $D=4,\,\mathcal{N}=3$ gauged supergravity in the duality-covariant formulation, we focus here on the special choice of the gauge group $\mathcal{G}={\rm SO}(3)\times {\rm SU}(3)$ which, as discussed in the Introduction, is the natural candidate to describes the ${\rm AdS}_4$ vacuum resulting from a Freund Rubin compactification of eleven dimensional supergravity on $N^{0,1,0}$. As we shall see, from inspection of the vacuum structure of the model, besides the latter vacuum, a rich web of new vacua arises.
\subsection{The Model}
We shall restrict ourselves to electric gaugings, namely to an embedding tensor with only electric components $\Theta_{\Lambda}{}^\alpha$ ($\Theta^{\Lambda \alpha}=0$), since we have verified that a \emph{dyonic} gauging of the same group does not lead to new physical properties of the model. \par
The quadratic constraints (\ref{cona}) require the branching of the representation ${\Scr R}$ of ${\rm G}$ with respect to the subgroup $\mathcal{G}$ to contain the adjoint representation of the latter, which defines the gauge vector fields among the $3+n$ vectors. As pointed out in the Introduction, we choose the model with $n=9$ vector multiplets so that the fundamental representation of ${\rm G}={\rm SU}(3,9)$ branches with respect to $\mathcal{G}$ as follows:
\begin{equation}
{\bf 12}\,\xrightarrow{{\rm SO}(3)\times {\rm SU}(3)}\,\,{\bf (3,1)}\oplus {\bf (1,8) }\oplus {\bf (1,1) }\,.
\end{equation}
The last singlet on the right-hand-side represents the Betti multiplet.\par
The $54$ real scalar  fields span the manifold:
\begin{align}
\mathcal{M}_S=\frac{\mathrm{SU}(3,9)}{{\rm S}[\mathrm{U}(3)\times\mathrm{U}(9)]}\,,\label{MS39}
\end{align}
and all belong to the vector multiplets.\par
The gauge generators $X_M$ are expressed in terms of the isometry ones $t_\alpha$ through the embedding tensor, as in (\ref{Theta}).
Denoting by $\hat{t}_\ell$, $\ell=1,2,3$, and $\hat{t}_m$, $m=1,\dots, 8$, the infinitesimal isometry generators of the groups ${\rm SO}(3)$ and ${\rm SU}(3)$, respectively (see  Appendix \ref{E} for the matrix form of these generators in the fundamental representation of the corresponding groups), we can define the  gauge generators $X_\ell,\,X_m$ as \footnote{With respect to \ref{gaugeconn}, $\mathcal{G}$ is the direct product of two simple groups so that we can introduce two different coupling constants, one for each factor.}
\begin{equation}
X_\ell=g_1\,\hat{t}_\ell\,\,,\,\,\,\,X_m=g_2\,\hat{t}_m\,,
\end{equation}
where we have denoted by $g_1,\,g_2$ the coupling constants associated with the two groups (in other words, in the chosen basis of the isometry generators, the embedding tensor is diagonal with entries $g_1$ and $g_2$). These are the only non-vanishing components of the symplectic vector of generators $X_M$:
 $$X^\Lambda={\bf 0},\,\{X_\Lambda\}=\{X_\ell,\,X_m,\,X_{\Lambda=12}={\bf 0}\}\,.$$
 The representation ${\Scr R}_\eta$  of $\hat{t}_\ell$ and $\hat{t}_m$ in the complex basis (\ref{cbasis}) reads
 \begin{equation}
  {\Scr R}_\eta[\hat{t}] = \left(
  \begin{matrix}
    {\rm adj}(\hat{t}) & 0_{\scalebox{.5}{$3 \times 9$}} & 0_{\scalebox{.5}{$3 \times 3$}} & 0_{\scalebox{.5}{$3 \times 9$}} \\
    0_{\scalebox{.5}{$9 \times 3$}} & 0_{\scalebox{.5}{$9 \times 9$}} & 0_{\scalebox{.5}{$9 \times 3$}} & 0_{\scalebox{.5}{$9 \times 9$}} \\
    0_{\scalebox{.5}{$3 \times 9$}} & 0_{\scalebox{.5}{$3 \times 9$}} & {\rm adj}(\hat{t})^* & 0_{\scalebox{.5}{$3 \times 9$}} \\
    0_{\scalebox{.5}{$9 \times 3$}} & 0_{\scalebox{.5}{$9 \times 9$}} & 0_{\scalebox{.5}{$9 \times 3$}} & 0_{\scalebox{.5}{$9 \times 9$}}
  \end{matrix} \right) ~~ \hat{t} \in \mathfrak{so}(3)
\end{equation}
\begin{equation}
  {\Scr R}_\eta[\hat{t}] = \left(
  \begin{matrix}
    ~~~0_{\scalebox{.5}{$3 \times 3$}} & ~~~0_{\scalebox{.5}{$3 \times 8$}} & 0 & ~~~0_{\scalebox{.5}{$3 \times 3$}} & ~~~0_{\scalebox{.5}{$3 \times 8$}} & 0 \\
    ~~~0_{\scalebox{.5}{$8 \times 3$}} & {\rm adj}(\hat{t}) & 0 & ~~~0_{\scalebox{.5}{$8 \times 3$}} & ~~~0_{\scalebox{.5}{$8 \times 8$}} & 0 \\
    0 & 0 & 0 & 0 & 0 & 0\\
    ~~~0_{\scalebox{.5}{$3 \times 3$}} & ~~~0_{\scalebox{.5}{$3 \times 8$}} & 0 &~~~0_{\scalebox{.5}{$ 3 \times 3 $}} & ~~~0_{\scalebox{.5}{$3 \times 8$}} & 0\\
    ~~~0_{\scalebox{.5}{$8 \times 3$}} & ~~~0_{\scalebox{.5}{$8 \times 8$}} & 0 &  ~~~0_{\scalebox{.5}{$8 \times 3$}} & {\rm adj}(\hat{t}) & 0\\
    0 & 0 & 0 & 0 & 0 & 0
  \end{matrix} \right) ~~ \hat{t} \in \mathfrak{su}(3)\,.
\end{equation}
The 12 vector fields transform, with respect to $\mathcal{G}=\mathrm{SO}(3)\times \mathrm{SU}(3)$ in the representation
\begin{equation}
A^\Lambda_\mu\,\,:\,\,\,\,\,\,(\mathbf{3},\mathbf{1})\oplus(\mathbf{1},\mathbf{8}+\mathbf{1})\,,
\end{equation}
while we can choose an $H$-covariant parametrization of the scalar manifold (\ref{MS39})
in which the scalar fields have the index structure  $\{\phi^s\}=\{\phi^{\ell,m},\,\phi^\ell\}$ and transform under $\mathcal{G}$ in the following representations:
$$\phi^s\,\,:\,\,\,\,\,\,(\mathbf{3},\mathbf{8})\,[\phi^{\ell,m}]\oplus (\mathbf{3},\mathbf{1})\,[\phi^{\ell}]+c.c.$$

\section{Consistent Truncations and Two Classes of Vacua \label{vacuasect}}\label{sec:trunc}

The scalar potential is a complicated non-linear function of the 54 scalar fields and is therefore very hard to extremize in general. Thus it is often useful to restrict to consistent truncations of the model characterized by a lower number of scalar fields. Consistency of the truncation then guarantees that the extrema of the scalar potential found in the smaller model, are vacua of the full theory. A consistent truncation can be defined by all the fields which are singlets with respect to a subgroup of the gauge group (or, in general, a subgroup of the duality group which leaves the embedding tensor invariant). We shall consider two consistent truncations characterized by three complex scalar fields each, spanning a manifold of the form $\left[\tfrac{\mathrm{SU}(1,1)}{\mathrm{U}(1)}\right]^3$. Then we study extrema of the potential restricted to these subspaces and find two compact hypersurfaces of vacua (not systematically discussed in the literature so far). They are defined by two different
embeddings of the $\left[\tfrac{\mathrm{SU}(1,1)}{\mathrm{U}(1)}\right]^3$ submanifold inside $\mathcal{M}_S$. To characterize them we consider the Cartan decomposition (\ref{Cartandec}). The generic element ${\bf k}\in \mathfrak{K}$ of the coset space $\mathfrak{K}$ has the following block-form in the fundamental representation of the $\mathfrak{su}(3,9)$ Lie algebra
\begin{align}\label{eq:H_fixed_points}
  {\bf k}=\left(\begin{array}{c|c} \mathbf{0}_{3\times3} & \mathbf{X}_{3\times9} \\ \hline \mathbf{X}^\dagger_{9\times3} & \mathbf{0}_{9\times9}\end{array}\right),\quad\mathbf{X}\in\mathrm{Mat}_{3\times9}(\mathbb{C})
\end{align}
Then the two embeddings are defined as
\begin{align}\label{eq:scal_embedd}
  \left[ \frac{\mathrm{SU}(1,1)}{\mathrm{U}(1)}\right]^3\hookrightarrow\mathcal{M}_S:\;(z_1,z_2,z_3)\mapsto\exp(\mathbf{k})
\end{align}
with $\mathbf{X}$ for the two embeddings given by
\begin{align}
  &\text{Type (i):}&\mathbf{X}=\begin{pmatrix}\label{eq:coset1}
    z_1 & 0 & 0 & 0 & 0 & 0 & 0 & 0 & 0\\
    0 & z_2 & 0 & 0 & 0 & 0 & 0 & 0 & 0\\
    0 & 0 & z_3 & 0 & 0 & 0 & 0 & 0 & 0
  \end{pmatrix}
\\
  &\text{Type (ii):}&\mathbf{X}=\begin{pmatrix}\label{eq:coset2}
                        0 & z_1 & 0 & 0 & 0 & 0 & 0 & 0 & 0\\
                        0 & 0 & 0 & 0 & z_2 & 0 & 0 & 0 & 0\\
                        0 & 0 & 0 & 0 & 0 & 0 & z_3 & 0 & 0
                      \end{pmatrix}
\end{align}
The above two choices of ${\bf X}$ define two 3-dimensional complex subspaces of $\mathfrak{K}$ defined by the singlets with respect to two discrete subgroups (\emph{stabilizers}) of ${\rm G}=\mathrm{SU}(3,9)$ which leave the embedding tensor invariant. The stabilizer is defined as the subgroup of ${\rm G}$, whose elements leave ${\bf k}$ invariant:
\begin{align}
{\bf g}^{-1}\, {\bf k}\,{\bf g}={\bf k}, \;\;{\bf g}\in {\rm G}\,\,\,.
\end{align}
The $12$-dimensional (including the decoupled Betti multiplet) adjoint representation $\mathrm{Ad}(\mathcal{G})$ of the gauge group has a homomorphism into the fundamental representation of ${\rm G}$. The image of the generators in the fundamental representation $J_i\in\mathfrak{so}(3)$, $i\lambda_I/2\in\mathfrak{su}(3)$ (see Appendix~\ref{E} for their definition) under this homomorphism will be denoted $\hat{J}_i$ and $\hat{\lambda}_I$, respectively. Then the stabilizer subgroup fixing $\mathbf{k}$ corresponding to Type (i) embedding reads~\footnote{In the expressions below we slightly abuse notation, the $\mathrm{SU/SO}$ groups are meant to represent the image of their adjoint representations under the homomorphism.}
\begin{align}
\text{Type (i):}\;\;&{\bf g}_1=\exp(\pi(\hat{J}_1+\hat{\lambda}_1))\in\mathrm{SU}(2)_D\subset \mathrm{Ad}(\mathcal{G}) \subset{{\rm G}}  \nonumber \\
&{\bf g}_2=\exp(\pi(\hat{J}_2+\hat{\lambda}_2)) \in\mathrm{SU}(2)_D\subset \mathrm{Ad}(\mathcal{G})\subset {\rm G}\,\,\,, 
\end{align}
while the stabilizer subgroup fixing $\mathbf{k}$ associated with Type (ii) embedding takes the form
\begin{align}
&\text{Type (ii):} \nonumber\\
&{\bf g}_1=\text{diag}(1,-1,-1,1,1,1,-1,-1,-1,-1,1,1)=\exp(\pi(-\hat{J}_1+2\hat{\lambda}_2))\in\mathrm{SO}(3)_D\subset \mathrm{Ad}(\mathcal{G})\subset {\rm G}  \nonumber \\
&{\bf g}_2=\text{diag}(-1,1,-1,-1,-1,1,1,1,-1,-1,1,1)=\exp(\pi(-\hat{J}_2+2\hat{\lambda}_5))\in\mathrm{SO}(3)_D\subset \mathrm{Ad}(\mathcal{G})\subset {\rm G}  \nonumber \\
&{\bf g}_3=\text{diag}(1,-1,-1,-1,1,-1,1,-1,1,-1,-1,-1)\not\subset \mathrm{Ad}(\mathcal{G})\subset {\rm G}\,.
\end{align}
The fact that the Lie algebra generators $\mathbf{k}$ are unique singlets under these discrete transformations allows us to restrict to a minimal truncation of the
theory, described just by the three complex scalar fields $z_i$.

Let us briefly comment on the structure of these two discrete groups. For Type (i) the two generators form the quaternionic group $\mathcal{Q}$,
where the map to the usual notation is (${\bf g}_1\rightarrow i,~{\bf g}_2\rightarrow j) $.
For Type (ii) each generator forms a $\mathbb{Z}_2$ group, hence the full discrete group is $\mathbb{Z}_2\times\mathbb{Z}_2\times\mathbb{Z}_2$. Let us observe that in both cases the order of the discrete group is $8$, however there is a crucial difference: in Type (ii) case we need to use an element ${\bf g}_3$ outside (the adjoint representation of) the gauge group. This
is fine because the adjoint ${\bf g}_3$ action leaves the $X$-tensor invariant.~\footnote{One can understand this in the following way: in both cases the $SO(3)$ factor of the generators represents a rotation by $\pi$ around different axes, the $SU(3)$ factor involves the spinorial representation of $SO(3)$ in the Type (i) case, while this is not true in the Type (ii) case. Hence, the two generators in the Type (i) case really represent a rotation by $\pi/2$ around two different axes. On the other hand, in the Type (ii) case ${\bf g}_1$ and ${\bf g}_2$ are not enough to obtain the necessary discrete group, since they square to the identity.}
\par
The second point of view is that the two types of vacua (derived within Type (i) and (ii) consistent truncations) have very specific
gauge group breaking patterns. As it will be explained later, in some sense they correspond to precisely two different
preserved non-abelian subgroups of the $\mathrm{SU}(3)$ factor of the gauge group
\begin{align}
\mathrm{SU}(3)\supset
\begin{cases}\label{eq:SU3_nAb_embedd}
\mathrm{SU}(2)\text{ generated by }\{\lambda_1,\lambda_2,\lambda_3\} \\ \mathrm{SO}(3)\text{ generated by }\{\lambda_2,\lambda_5,\lambda_7\}
\end{cases}
\end{align}

In order to compute the relevant quantities (in particular the scalar
potential), we associate with $\mathbf{k}$ the coset representative $L$
\begin{align}
L=\exp{(\mathbf{k})}\,.
\end{align}
From $L$ we can in turn compute the T-tensor (see~\eqref{eq:T_tensor})-- the fundamental object which contains the
fermionic shifts (which are the building blocks of the scalar potential) together with mass
matrices of fermions. Once we have the T-tensor we project to the fermionic
shifts using formulae~\eqref{eq:fermionic_shifts}. The scalar potential is
finally obtained from the Ward identity~\eqref{wardid}
\begin{align}
  \left.V\right\vert_{\mathrm{singlet}}= \left.\frac{1}{3}\mathrm{Tr}\left( \mathbb{N}_A \mathbb{N}^A + \mathbb{N}_{A I} \mathbb{N}^{A I} + \mathbb{N}_{IC}^{~~~~A} \mathbb{N}^{IC}_{~~~~A} - \, 12 \, \mathbb{S}_{AC} \mathbb{S}^{AC}\right)\right\vert_{\mathrm{singlet}}\,.
\end{align}
It is useful to write the three complex scalar fields of the two truncations, appearing in (\ref{eq:coset1}) or (\ref{eq:coset2}), as follows:
\begin{align}
z_j=r_j\exp(i\alpha_j),\,\,j=1,2,3,\,\,\text{ where }r_j\in\mathbb{R}_{\geq0}\text{ and }\alpha_j\in[0,2\pi)\,.
\end{align}
In this parametrization, denoting by ${\phi}^r$ the six real scalar fields $\{r_1,r_2,r_3,\alpha_1,\alpha_2,\alpha_3\}$, the coset metric reads
\begin{align}
ds^2\,=\,{\Scr G}_{rs}(\phi)d {\phi}^r d {\phi}^s\,&=\,\sum_i^3 \left(\,2 dr_i{}^2\,+\,\frac{1}{2}\sinh^2\left(2r_i\right)d\alpha_i{}^2\,\right) \,\,.\label{metrixSTU}
\end{align}
For the Type (i) model the potential is computed to be
\begin{eqnarray}
V(r_i,\alpha_i)&=& g_1^2\left(\,-3\,-2\cosh\left(2r_3\right)-\cosh\left(2r_1\right)\left(2+\cosh\left(2r_2\right)+\cosh\left(2r_3\right)\right)- \right. \nonumber \\
 &&\left. \cosh\left(2r_2\right)\left(2+\cosh\left(2r_3\right)\right)\right)\,+\,g_2^2\left(\,3\,+\cosh\left(2r_2\right)\left(-2+\cosh\left(2r_3\right)\right) \right. \nonumber\\
 && \left. -2\cosh\left(2r_3\right)\,+\,\cosh\left(2r_1\right)\left(-2+\cosh\left(2r_2\right)+\cosh\left(2r_3\right)\right)\,\right)\,\,\label{potgen},
\end{eqnarray}
while for the Type (ii) model it takes the following form
\begin{eqnarray}
V(r_i,\alpha_i)&=&g_1^2\left(\,-3\,-2\cosh\left(2r_3\right)\,-\,\cosh\left(2r_1\right)\left(2+\cosh\left(2r_2\right)+\cosh\left(2r_3\right)\right)-\right.\nonumber \\ 
&&\left. \cosh\left(2r_2\right)\left(2+\cosh\left(2r_3\right)\right)\right)\,+\frac{g_2^2}{4}\left(\,3\,+\cosh\left(2r_2\right)\left(-2+\cosh\left(2r_3\right)\right)\,\right. \nonumber \\
&&\left. \,-2\cosh\left(2r_3\right)\,+\,\cosh\left(2r_1\right)\left(-2+\cosh\left(2r_2\right)+\cosh\left(2r_3\right)\right)\,\right)\,\,\label{potgen2}.
\end{eqnarray}
Note that the above expressions do not depend on $\alpha_i$. Since these angular variables are not Goldstone bosons, they correspond to genuine flat directions.\\
Now, by virtue of the Gradient-Flow equations, the potential in (\ref{potgen},\ref{potgen2}) can be re-interpreted in terms of a ``\emph{superpotential}''; such a superpotential, $\mathcal{W}$, is strictly dependent on the eigenvalues of the fermionic shift $\mathbb{S}_{AB}$ , which are given by
\begin{align}
&\text{Type (i):}&&\mathbb{S}_{AB}=\delta_{AB}\left(g_1 \prod_{j=1}^{3} \cosh \left(r_j\right) - g_2 e^{i\left(- \alpha_{B}+ \alpha_C+\alpha_D\right)}\prod_{j=1}^{3} \sinh \left(r_j\right)\,  
  \right)\,\,\,,  \label{eq:fshift1} \\ 
  &\text{Type (ii):}&&\mathbb{S}_{AB}=\delta_{AB}\left(g_1 \prod_{j=1}^{3} \cosh \left(r_j\right) - \frac{g_2}{2}e^{i\left(- \alpha_{B}+ \alpha_C + \alpha_D \right)} \prod_{j=1}^{3} \sinh \left(r_j\right) \,   
  \right)\,\,\,, \label{eq:fshift2}
\end{align}
with $\alpha_B\neq \alpha_C\neq \alpha_D$.
In both type (i) and (ii) truncations, we can construct the ``\emph{superpotential}''  $\mathcal{W}(r_i,\alpha_i)$ in terms of the modulus of any of the diagonal entries of $\mathbb{S}_{AB}$ (e.g. $\mathbb{S}_{11}$):
\begin{equation}
    \mathcal{W}(r_i,\alpha_i)=2\,|\mathbb{S}_{AA}|\,.
\end{equation}
The scalar potential is defined through the ''superpotential equation''
\begin{equation}
V\left(r_j\right)\,=\,2\,{\Scr G}^{rs}\,\frac{\partial }{\partial \phi^r}\mathcal{W}(r_j,\alpha_j)\,\frac{\partial }{\partial \phi^s}\mathcal{W}(r_j,\alpha_j)\,-\,3\,\mathcal{W}(r_j,\alpha_j)^2\,\,,
\end{equation}
which holds both for Type (i) and Type (ii) vacuum. Notice that the dependence on $\alpha_i$ drops out in the expression of the potential. 
For this reason we can define an $\alpha_i$-independent  superpotential as follows:
\begin{equation}
     \mathcal{W}_0(r_i)\equiv  \mathcal{W}(r_i,\alpha_i=0)\,,\label{Wsup0}
\end{equation}
in terms of which the potential reads:
\begin{equation}
    V(r_k)=\sum_{i=1}^3 \left(\frac{\partial}{\partial r_i}\mathcal{W}_0\right)^2-3\,\mathcal{W}_0^2\,.\label{VWr}
\end{equation}
We shall use this function to derive the domain wall solution in section \ref{sec:DW}. In contrast to the analogous results in the literature, we find a scalar potential with three flat directions (i.e. not Goldstone bosons) when restricted to the above defined truncations. In the dual CFT, these flat directions are natural candidates for \emph{exactly marginal deformations}. In fact the three angles will parametrize two 3-tori ($T^3_{(i)},\,T^3_{(ii)}$) of vacua, to be discussed below.
Although the potential at these extrema does not depend on $\alpha_i$, the amount of preserved supersymmetry does, thus realizing a phenomenon of spontaneous supersymmetry breaking through marginal deformations. To our knowledge, these manifolds of vacua of the $\mathcal{N}=3$ model under consideration, preserving different amounts of supersymmetry, have not been discussed in the literature so far. Let us discuss them in detail.
\par
Inspection of the gradient of the potential shows that one can consistently set $r_1=r_2=r_3=r$\footnote{ The other vacua of the truncations have $r_2=r_3=0$ (when $g_1=0$) or $r_1=r_2 $ and $ r_3=0$ (modulo permutations of the radii). They correspond to supersymmetric Minkowski vacua or to non-supersymmetric and perturbatively unstable AdS vauca respectively. The first case corresponds to a model with ungauged graviphotons. Here we shall focus on perturbatively stable AdS vacua.}. 
This allows us to write a
more
compact formula for the scalar potential to be extremized
\begin{align}
&\text{Type (i):}&&V(r,\alpha_1,\alpha_2,\alpha_3)=V(r)=-12\left[ \, g_1^2\cosh^4(r) - g_2^2\sinh^4(r)\, \right]\,,\label{vacuumi} \\
  &\text{Type (ii):}&&V(r,\alpha_1,\alpha_2,\alpha_3)=V(r)=-12\left[\, g_1^2\cosh^4(r) - \frac{g_2^2}{4}\sinh^4(r) \, \right]\,.\label{vacuumii}
\end{align}
The extremality condition $\frac{\partial V}{\partial r}=0$ determines the following three distinct values $r=r_{\mathrm{vac}}$ for $r$ at the extrema:
\begin{align}
  &\text{Type (i):}&& r_{\mathrm{vac}}=\frac{1}{2}\log\left( \frac{g_2+g_1}{g_2-g_|} \right) &&\implies T_{(i)}^3\text{ of extrema:}~\exists\;g_2>g_1\,, \label{eq:rstar} \\
  &\text{Type (ii):}&&r_{\mathrm{vac}}=\frac{1}{2}\log\left( \frac{g_2+2g_1}{g_2-2g_1} \right) &&\implies T_{(ii)}^3\text{ of extrema:}~\exists\;g_2>2g_1\,, \\
&\text{Origin:}&&r_{\mathrm{vac}}=0 &&\implies \text{isolated extremum:}~\forall\;g_1,g_2\,.
\end{align}
We see that we have one isolated vacuum that exists for all values of the
couplings $g_1,g_2$. It is located at the origin $\mathcal{O}$ of the scalar manifold as
expected. Aside from it there are two types of non-trivial vacuum manifolds: both of them
are three-tori $T^3$ parameterized by $(\alpha_1,\alpha_2,\alpha_3)$, though
embedded differently into the scalar manifold $\mathcal{M}_S$. The
Type (i) and type (ii) $T^3$-vacua only exist for $g_2>g_1$ and $g_2>2g_1$, respectively. The corresponding values of the scalar potential $V$ (i.e. the cosmological constants at the extrema) are:
\begin{align}
&\text{Type (i):}&& \Lambda=\left.V\right\vert_{r_{\mathrm{vac}}}=-12\frac{g_1^2g_2^2}{g_2^2-g_1^2} \,,\label{eq:cosmi} \\
&\text{Type (ii):}&& \Lambda=\left.V\right\vert_{r_{\mathrm{vac}}}=-12\frac{g_1^2\,g_2^2}{g_2^2-4g_1^2}\,,\\
&\text{Origin:}&&\left.V\right\vert_{r_{\mathrm{vac}}}=-12\,g_1^2\,.
\end{align}
Thus all vacua have a negative constant scalar curvature, as expected for $AdS_4$ spacetime geometries.

We still need to introduce one more refinement since the discussion above
was
slightly imprecise. The points of the tori $T^3$ of Type (i) or (ii) are not all
gauge inequivalent. There is a discrete subgroup $\Gamma\subset \mathcal{G}$ of the
gauge group that identifies them. It acts on the $(z_1,z_2,z_3)$ coordinates
introduced in~\eqref{eq:scal_embedd} in terms of a 3-dimensional irreducible
representation
\begin{align}
\begin{array}{cccc}
\text{Inversions:}&\boxed{\begin{pmatrix} -1&0&0\\0&-1&0\\0&0&1\end{pmatrix}}&\begin{pmatrix}-1&0&0\\0&1&0\\0&0&-1\end{pmatrix}&\begin{pmatrix}1&0&0\\0&-1&0\\0&0&-1\end{pmatrix} \\
\text{Permutations:}&\begin{pmatrix}0&1&0\\1&0&0\\0&0&1\end{pmatrix}&\boxed{\begin{pmatrix}1&0&0\\0&0&1\\0&1&0\end{pmatrix}}&\boxed{\begin{pmatrix}0&0&1\\0&1&0\\1&0&0\end{pmatrix}}
\end{array}\label{eq:discrete_group}
\end{align}
The first line represents inversions of all possible pairs of the
$z$-coordinates (shifts of their $\alpha$-phases by $\pi$), while the second line acts
by permutations. These matrices generate the discrete group
\begin{align}
\Gamma\simeq S_4 \simeq S_3 \ltimes K_4 \simeq T^{h}_{24}\simeq
\mathrm{O}_{24}
\end{align}
where $S_n$ is the symmetric group of $n$ objects,
$K_4\simeq\mathbb{Z}_2\times\mathbb{Z}_2$ is the Kleinian
four-group, $T^{h}_{24}\subset \mathrm{O(3})$ is the full
tetrahedral group (including inversions) and finally
$\mathrm{O}_{24} \subset \mathrm{SO(3)}$ is the rotational
(orientation preserving) octahedral group. The discrete group $\Gamma\simeq S_4$ can be presented by $3$ generators and
relations among them. A possible choice of these generators (in the
3-dimensional irrep) consists of the $3$
boxed matrices in~\eqref{eq:discrete_group}.
So the conclusion of this analysis is that the vacuum manifold
$\mathcal{M}_{\mathrm{vac}}$ depends on the couplings $g_1,g_2$ and takes
the
form
\begin{align}
\mathcal{M}_{\mathrm{vac}}=\left\{ \begin{array}{ll} g_2\leq g_1: &
\mathcal{O} \\ g_1<g_2\leq 2g_1: & \mathcal{O}\cup T^3_{(i)}/S_4 \\ g_2>2g_1: & \mathcal{O}\cup T^3_{(i)}/S_4\cup T^3_{(ii)}/S_4 \end{array}\right.
\end{align}
We may interpret the appearance of new vacua for the above ranges of the coupling constants in terms of the occurrence of phase transitions. As it will be discussed in the sequel, according to the specific phases, different RG-flows between the above vacua can exist.  
Next, we will characterize interesting submanifolds of the vacuum manifold
according to supersymmetry or gauge symmetry breaking patterns.

\begin{figure}[tb]
\centering
\includegraphics[width=1\columnwidth]{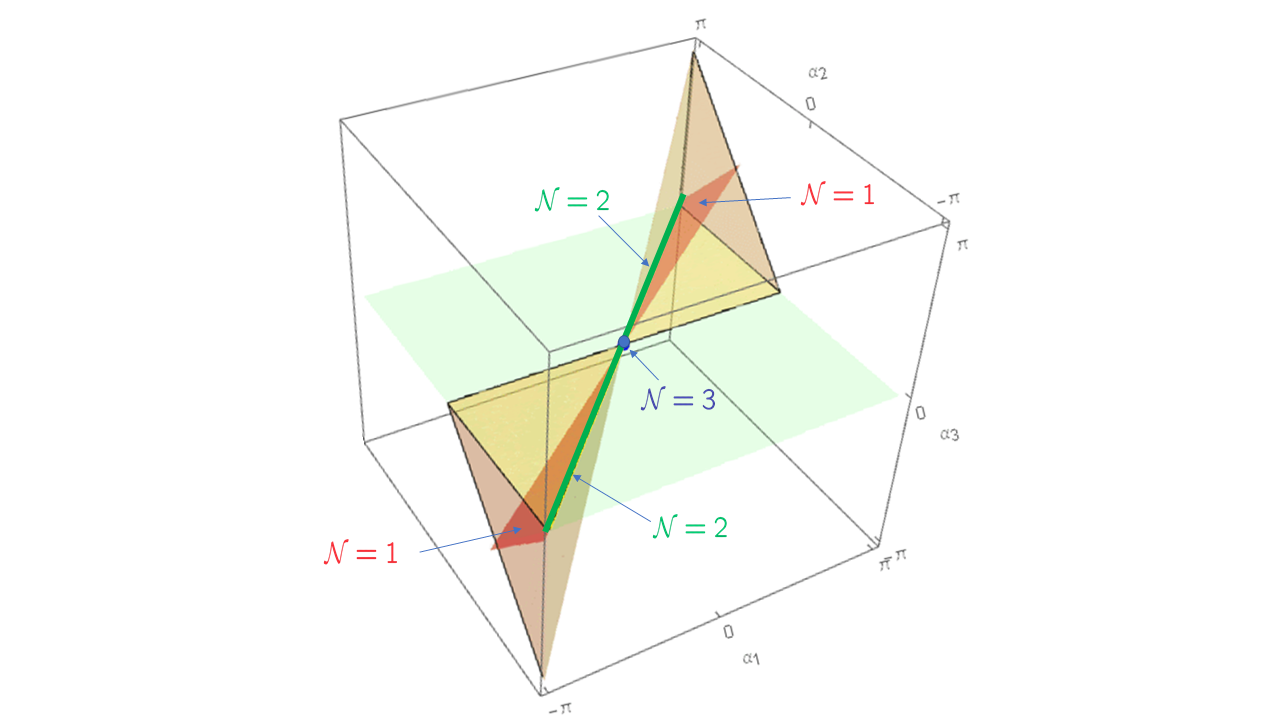}
\caption{Representation of one of the two  manifolds of vacua parametrized by $\alpha_1,\,\alpha_2,\,\alpha_3$. There is a residual identification \eqref{resident} among the points on the plane $\alpha_3=0$. The vertices $(-\pi,-\pi,-\pi)$ and $(\pi,\pi,\pi)$ are also identified.} 
\label{fig:vacstructure} 
\end{figure}

In order to analyze supersymmetry breaking it is sufficient to study
the kernel (or equivalently image) of the generalized fermionic shift tensor
$\mathbb{N}_{\mathcal{I}}^A$ of spin--$\tfrac{1}{2}$ fields. The index
$\mathcal{I}$ runs over all spin--$\tfrac{1}{2}$ fields in the theory. In
the
case of an $\mathcal{N}=3$ supergravity in $d=4$ dimensions under consideration it
means $\mathcal{I}=1,\ldots,37$ in the following order: $\mathcal{I}\in$\{1
dilatino, $9\times1$ gaugino R--symmetry singlets, $9\times3$ gaugino R--symmetry
triplets\}. Then the number of unbroken supersymmetries in a given vacuum
is determined as
\begin{align}\label{eq:susy_vac}
  \mathcal{N}_{\mathrm{vac}}=\mathrm{dim}\left( \mathrm{Ker}\mathbb{N}_{\mathcal{I}}^A\Big\rvert_{\mathrm{vac}} \right)
\end{align}
In light of the potential Ward identity \eqref{wardid}, the number of preserved supersymmetries is equal to the number of eigenvalues $\mathbb{S}_{AA}$ of the diagonal matrix $\mathbb{S}_{AB}$ (see~\eqref{eq:fshift1} and~\eqref{eq:fshift2}) satisfying
\begin{align}
\lvert\mathbb{S}_{AA}\rvert\,=\frac{1}{2L}=\sqrt{-V_0/12},
\end{align}    
where $L=\sqrt{-3/V_0}$ is the AdS radius.
Both for type $i)$ and $ii)$, the above condition is met (modulo permutations in angles $\alpha_i$) for one, two and three eigenvalues when:
\begin{align}
    \mathcal{N}=1\,\,&\,\,\,\,\alpha_1=\alpha_2+\alpha_3\,,\nonumber\\
    \mathcal{N}=2\,\,&\,\,\,\,\alpha_1=\alpha_2\,\,,\,\,\,\alpha_3=0\,,\nonumber\\
     \mathcal{N}=3\,\,&\,\,\,\,\alpha_1=\alpha_2=\alpha_3=0\,.
\end{align}
All other points break supersymmetry completely. In Figure~\ref{fig:vacstructure} we graphically illustrate the structure of both type $i)$ and $ii)$ vacua, parametrized by $\alpha_1,\,\alpha_2\,\alpha_3$, where the identifications implemented by the group $\Gamma$ are taken into account. The inversions in this group amount to shifting two angles by $\pm \pi$, leaving the third unaltered. We can fix these symmetries, as well as the permutations in $\Gamma$, by restricting the values of the angles to the following domains:
\begin{align}
D_1:&\,-\pi\le \alpha_1\le \alpha_2\le \alpha_3\le 0\,,\nonumber\\
D_2:&\,0\le \alpha_3\le \alpha_2\le \alpha_1\le \pi\,.
\end{align}
which are represented in Figure~\ref{fig:vacstructure} by the colored tetrahedra. There is still an identification to be considered among the points in the shaded region of the graph. It identifies the two triangular faces of the tetrahedra at $\alpha_3=0$ and acts as follows: 
\begin{equation}(\alpha_1,\,\alpha_2)\in D_1\,\,\sim \,\,\,(\alpha_2+\pi,\,\alpha_1+\pi)\in D_2\,.\label{resident}
\end{equation}
Hence we can describe the independent $\mathcal{N}=2$ vacua ($\alpha_1=\alpha_2,\,\alpha_3=0$) by the segment belonging to $D_1$ only.

Let us now describe the gauge group breaking patterns in various vacua. To determine the subgroup $H_0\subset \mathcal{G}$ of the gauge group that
remains unbroken in the vacuum, one solves for the centralizer
$h_0\in\mathrm{Lie}(H_0)\subset\mathfrak{su}(3,9)$ of the coset generator $\mathbf{k}$
in~\eqref{eq:scal_embedd} evaluated at the given vacuum
\begin{align}
\left[\mathbf{k}\big|_{\mathrm{vac}},h_0\right]=0
\end{align}
Equipped with this knowledge let us classify the submanifolds of
$\mathcal{M}_{\mathrm{vac}}$ based on the residual gauge symmetry. We
systematize the discussion starting from most generic submanifolds with least
residual gauge symmetry, going to more restricted submanifolds with bigger gauge
symmetry according to the following chain of subgroups
\begin{align}
\label{eq:gauge_subgroups}
\mathbf{1}\subset\cdots\subseteq H_0^{(k)}\subseteq\cdots\subseteq H_0^{(1)}\subset \mathcal{G}
\end{align}
Below we give the list of special submanifolds of $\mathcal{M}_{\mathrm{vac}}$,
together with their properties, i.e. topology, preserved supersymmetry and
residual gauge symmetry~\footnote{In the following diagrams, the upper inclusion sign captures the relation
between various submanifolds, while the lower one represents relations among
unbroken gauge groups $H_0$. The inclusion between gauge groups is regular,
but this is not always the case for the vacuum manifolds. For instance the two
circles (with antipodal identification) are disjoint up to one point that
they
share. The first circle is $\mathcal{N}=2$, the second one is $\mathcal{N}=0$
and the single common point is actually $\mathcal{N}=3$.}
\subsubsection*{Type (i): $g_2>g_1$}
\begin{align}\label{eq:T1_vac}
&\boxed{\begin{array}{c}(\alpha_1,\alpha_2,\alpha_3)~\text{generic}\\\mathcal{M}_{\mathrm{vac}}=T^3/S_4\\\mathcal{N}=0\\H_0=\mathrm{U}(1)\end{array}}\begin{array}{c}\\ \supset\\ \\\subseteq\end{array}\boxed{\begin{array}{c}(\alpha_2+\alpha_3,\alpha_2,\alpha_3)\\\mathcal{M}_{\mathrm{vac}}=T^2/K_4\\\mathcal{N}=1;~ (\alpha_2,\alpha_3\neq 0)\\H_0=\mathrm{U}(1)\end{array}}\begin{array}{c}\\\supset\\ \\\subset\end{array}\boxed{\begin{array}{c}(\alpha_2,\alpha_2,0)\\\mathcal{M}_{\mathrm{vac}}=S^1/\mathbb{Z}_2\\\mathcal{N}=2;~(\alpha_2\neq 0)\\H_0=\mathrm{U}(1)_D\times \mathrm{U}(1)\end{array}}\begin{array}{c}\\\not\supseteq\\\\\subset\end{array}
\notag \\
&\boxed{\begin{array}{c}(\alpha_1,\alpha_1,\alpha_1)\\\mathcal{M}_{\mathrm{vac}}=S^1/\mathbb{Z}_2\\\begin{cases}\alpha_1\neq 0:~\mathcal{N}=0\\\alpha_1=0:~\mathcal{N}=3\end{cases}\\H_0=\mathrm{SU}(2)_D\times \mathrm{U}(1)\end{array}}\xrightarrow{r\to0}\boxed{\begin{array}{c}\mathcal{M}_{\mathrm{vac}}=\mathrm{pt}=\mathcal{O}\\\mathcal{N}=3\\H_0=\mathcal{G}\end{array}}
\end{align}

\subsubsection*{Type (ii): $g_2>2g_1$}
\begin{align}
\label{eq:T2_vac}  &\boxed{\begin{array}{c}(\alpha_1,\alpha_2,\alpha_3)~\text{generic}\\\mathcal{M}_{\mathrm{vac}}=T^3/S_4\\\mathcal{N}=0\\H_0=\mathbf{1}\end{array}}\begin{array}{c}\\\supset\\\\\subseteq\end{array}\boxed{\begin{array}{c}(\alpha_2+\alpha_3,\alpha_2,\alpha_3)\\\mathcal{M}_{\mathrm{vac}}=T^2/K_4\\\mathcal{N}=1;~ (\alpha_2,\alpha_3\neq 0)\\H_0=\mathbf{1}\end{array}}\begin{array}{c}\\\supset\\\\\subset\end{array}\boxed{\begin{array}{c}(\alpha_2,\alpha_2,0)\\\mathcal{M}_{\mathrm{vac}}=S^1/\mathbb{Z}_2\\\mathcal{N}=2;~(\alpha_2\neq 0)\\H_0=\mathrm{U}(1)_D\end{array}}\begin{array}{c}\\\not\supseteq\\\\\subset\end{array}
  \notag \\
  &\boxed{\begin{array}{c}(\alpha_1,\alpha_1,\alpha_1)\\\mathcal{M}_{\mathrm{vac}}=S^1/\mathbb{Z}_2\\\begin{cases}\alpha_1\neq 0:~\mathcal{N}=0\\\alpha_1= 0:~\mathcal{N}=3\end{cases}\\H_0=\mathrm{SO}(3)_D\end{array}}\xrightarrow{r\to0}\boxed{\begin{array}{c}\mathcal{M}_{\mathrm{vac}}=\mathrm{pt}=\mathcal{O}\\\mathcal{N}=3\\H_0=\mathcal{G}\end{array}}
\end{align}
As we commented in~\eqref{eq:SU3_nAb_embedd}, Type (i) vacua are associated with
the embedding
$\mathrm{SU}(2)\subset \mathrm{SU}(3)$ which has a $\mathrm{U}(1)$ commutant. Namely,
one takes the diagonal combination of this $\mathrm{SU}(2)$ subgroup with
the
$\mathrm{SO}(3)$ factor in the gauge group (taking also into account the
$\mathrm{U}(1)$ commutant) in order to arrive at
(see~\eqref{eq:gauge_subgroups})
\begin{align}
H_0^{(1)}=\mathrm{SU}(2)_D\times \mathrm{U}(1)
\end{align}
This is the residual gauge symmetry of the $S^1/\mathbb{Z}_2$ vacua (first box
on second line of~\eqref{eq:T1_vac}). The gauge groups of all other vacua
in the
Type (i) chain are subgroups of this one. Similarly, the only other non-abelian
subgroup of $\mathrm{SU}(3)$ is $\mathrm{SO}(3)$. The embedding $\mathrm{SO}(3)\subset
\mathrm{SU}(3)$ has no commutant, so in this case one arrives at
\begin{align}
H_0^{(1)}=\mathrm{SO}(3)_D
\end{align}
which is the residual gauge group of highest rank for Type (ii) vacua in~\eqref{eq:T2_vac}.

Moreover, let us remark that the singlets with respect to these maximal
subgroups $H_0^{(1)}$ are the unique ones given in~\eqref{eq:coset1}
and~\eqref{eq:coset2} (with the appropriate specification of phases shown
in~\eqref{eq:T1_vac} and~\eqref{eq:T2_vac}). To argue this,
as a first step it is useful to remind the branching rules of the adjoint
representation $\mathbf{8}$ of $\mathrm{SU}(3)$ with respect to its only two
non-abelian subgroups $\mathrm{SU}(2)$ and $\mathrm{SO}(3)$
\begin{align}
\mathbf{8}\Big\rvert_{\mathrm{SU}(3)}&\rightarrow \left(\mathbf{1}\oplus2\times\mathbf{2}\oplus\mathbf{3}\right)\Big\rvert_{\mathrm{SU}(2)}
\\
\mathbf{8}\Big\rvert_{\mathrm{SU}(3)}&\rightarrow \left(\mathbf{3}\oplus\mathbf{5}\right)\Big\rvert_{\mathrm{SO}(3)}
\end{align}
Recall that the scalar fields parameterizing the scalar manifold $\mathcal{M}_S$
transform in the $(\mathbf{3},\mathbf{8}+\mathbf{1})$ representation under
$\mathcal{G}=\mathrm{SO}(3)\times \mathrm{SU}(3)$. So combining the above decomposition with the adjoint representation $\mathbf{3}$ of
$\mathrm{SO}(3)$ and restricting to the diagonal subgroups results in
\begin{align}
  (\mathbf{3},2\times\mathbf{1}\oplus2\times\mathbf{2}\oplus\mathbf{3})\Big\rvert_{\mathrm{SO}(3)\times\mathrm{SU}(2)}&\rightarrow (\mathbf{1}\oplus 2\times\mathbf{2}\oplus 3\times\mathbf{3}\oplus 2\times\mathbf{4}\oplus\mathbf{5})\Big\rvert_{\mathrm{SU}(2)_{\mathrm{D}}} \label{eq:branching_su2u1}\\
  (\mathbf{3},\mathbf{1}\oplus\mathbf{3}\oplus\mathbf{5})\Big\rvert_{\mathrm{SO}(3)\times\mathrm{SO}(3)}&\rightarrow (\mathbf{1}\oplus 3\times\mathbf{3}\oplus 2\times\mathbf{5}\oplus\mathbf{7})\Big\rvert_{\mathrm{SO}(3)_{\mathrm{D}}} \label{eq:branching_so3}
\end{align}
We see that in both cases there is a unique
singlet as we claimed.

Having analyzed the residual supersymmetry  of our distinguished
subset of vacua, we move on to
calculating mass spectra in each of these vacua. In the next section we present
a general algorithm for construction of mass matrices for fields of all spins.
Then we apply these techniques and compute the spectra in all supersymmetric points and show that they organize into
$\mathrm{OSp}(\mathcal{N}|4)$ supermultiplets, for
$\mathcal{N}=1,2,3$~\footnote{We computed mass spectra also for $AdS_4$
vacua
  that break supersymmetry completely to $\mathrm{SO}(3,2)$. However, we are not
going to present these results in this paper.}.



\section{Organizing supergravity fields into \texorpdfstring{$\mathrm{OSp(\mathcal{N}|4)}$}{OSp(N|4)} supermultiplets}\label{sec:supermult}
Here we will show results for vacua that preserve $\mathcal{N}=1,2,3$
supersymmetry. There are however $\mathcal{N}=0$ vacua, which completely break
supersymmetry and the mass spectrum of supergravity excitations around these
vacua has been computed as well. However it is not particularly illuminating and for this
reason it  will not be presented in this paper.
\subsubsection*{General comments on $\mathrm{OSp(\mathcal{N}|4)}$ supermultiplets}
To describe supermultiplets we will follow the notation
of~\cite{Cordova:2016emh}. The particular case of $\mathrm{OSp}(\mathcal{N}|4)$
supermultiplets relevant in this paper was also studied earlier in~\cite{Fabbri:1999ay}.

We briefly summarize just the necessary conventions and definitions of~\cite{Cordova:2016emh}
useful in our special case. For details, the reader is kindly asked to consult
the original paper. Supermultiplets of $\mathrm{OSp}(\mathcal{N}|4)$ will
be classified by Dynkin
labels of its maximal compact subgroup $\mathrm{SO}(\mathcal{N})_R\times \mathrm{SO}(3)_J \times
\mathrm{SO}(2)_\Delta$. The first factor represents the R--symmetry, the second the
(Wick rotated) Lorentz transformations in three dimensions and finally the last
factor is generated by the dilatation operator $D$. At the level of algebras, we
use for the first two factors the isomorphism
$\mathfrak{so}(3)\simeq\mathfrak{su}(2)$, whenever available (always for the spin
part and for the R-symmetry if $\mathcal{N}=3$). In such a situation, $R$ and $J$ are understood as
$\mathfrak{su}(2)$ weights and the authors of~\cite{Cordova:2016emh} work
in conventions common in math literature
where they belong to non-negative integers, $R,J\in\mathbb{Z}_{\ge0}$. For
$\mathcal{N}=2$, the $\mathrm{SO}(2)\simeq\mathrm{U}(1)$ R-charge takes
values in real
numbers, $R\in\mathbb{R}$. Finally, if $\mathcal{N}=1$, there is no R-symmetry and
states are labeled just by spin and scaling dimension.

Then a
supermultiplet will be denoted by its lowest weight state
\begin{align}
X[J]^{(R)}_\Delta,\quad\mathrm{where}\;X=L,A_1,A_2,B_1,B_2
\end{align}
from which the complete supermultiplet is constructed by raising operators.
As explained above $R$ is the R--symmetry charge, $J$ the spin and $\Delta$ the
scaling dimension. The letter $X$ specifies the type of the supermultiplet: $L$
stands for a long supermultiplet, $A$ for a short supermultiplet at the
threshold (i.e. its scaling dimension $\Delta_A$ can be continuously approached from
above), while $B$ represents an isolated short multiplet (i.e. its scaling
dimension $\Delta_B<\Delta_A$ is separated by a gap).

From supergravity computations at the classical
level~\footnote{For $\mathcal{N}=3$ the scaling dimension and hence the
mass is
  a function of quantized quantities only -- the spin and the R-charge. It cannot
receive any corrections and is thus exact.} one obtains not directly the scaling dimensions, but rather
masses of the particles (here we refer to the uncorrected mass; the $AdS_4$ mass
is then obtained by combining this uncorrected mass with curvature
contributions). It is thus useful to build a dictionary between the uncorrected
masses and the scaling dimensions $\Delta$ or equivalently energies $E_0$, depending
on whether we are using a gauge theory or gravity language. For particles
of
various spin it takes the form
\begin{align}\label{eq:mass_energy}
\begin{array}{|c|c|}
  \hline
\textrm{spin} & \Delta\equiv E_0  \\ \hline
0 & \frac{1}{2}\left( 3\pm \sqrt{9+4m^2} \right)  \\ \hline
1 & \frac{1}{2}\left( 3\pm \sqrt{1+4m^2}\right)  \\ \hline
\frac{1}{2},\;\frac{3}{2} & \frac{1}{2}\left( 3+2\vert m\vert \right) \\ \hline
\end{array}
\end{align}
\subsection{\texorpdfstring{$\mathcal{N}=3$}{N=3} vacua}

\subsubsection{ \texorpdfstring{$\mathrm{OSp}(3|4)$}{Osp(3|4)} supermultiplets}\label{sec:N3_supermultiplets}
The R-symmetry Lie algebra is $\mathfrak{so}(3)\simeq\mathfrak{su}(2)$. To label the
states we will use the Dynkin label $(R)$ of $\mathfrak{su}(2)$. We work in
conventions common in the math literature, namely $(R)\in \mathbb{Z}$. So
$(1)$
and $(2)$ denote the fundamental and the adjoint representation of
$\mathfrak{su}(2)$. The remaining labels of states in a supermultiplet are the
spin and the scaling dimension.

In Appendix~\ref{super3}  we list only those $\mathrm{OSp}(3|4)$ supermultiplets that will be
necessary to encompass the supergravity excitations in $\mathcal{N}=3$ vacua
discussed in this paper (in the tables the R-symmetry representation is
denoted by its dimension, i.e. $\mathbf{2}$ for the fundamental):

\subsubsection{\texorpdfstring{$\mathcal{N}=3$}{N=3} vacuum preserving \texorpdfstring{$H_0=\mathcal{G}$}{H=Ggauge}}
The mass spectrum in this isolated $\mathcal{N}=3$ maximally symmetric vacuum is
summarized in Table~\ref{tab:N3_SO3SU3_mass}.
A quick consistency check employs the Goldstone theorem. There are $11$ unbroken
gauge generators and no broken ones in this vacuum. Therefore we expect no
massive vector fields and $11+1$ massless ones. The additional vector comes from
a completely
decoupled massless vector supermultiplet -- the Betti multiplet. This
supermultiplet will be present in all the following spectra. Later, it will be
included without further comments. The supergravity excitations can be assembled into the following supermultiplets of
$\mathrm{OSp}(3\vert4)$
\begin{align}\label{eq:spec_so3su3}
  \mathrm{Spec}=\underbrace{A_1[0]^{(0)}_{\frac{3}{2}}}_{\substack{\text{massless graviton} \\ \text{multiplet}}}\oplus \underbrace{9\times B_1[0]^{(2)}_1}_{\substack{\text{massless vector} \\ \text{multiplets}}},
\end{align}
as can be easily checked by comparing Table~\ref{tab:N3_SO3SU3_mass} with
the
field content of the supermultiplets, which was summarized in the previous section.

\subsubsection{\texorpdfstring{$\mathcal{N}=3$}{N=3} vacuum preserving \texorpdfstring{$H_0=\mathrm{SU}(2)_D\times
  \mathrm{U}(1)\subset \mathcal{G}$}{H=SU(2) x U(1) < Ggauge}}
The spectrum at the single $\mathcal{N}=3$ supersymmetric point (lying on
$S^1/\mathbb{Z}_2$ manifold of vacua, spanned by $\alpha_1=\alpha_2=\alpha_3$, invariant under the same subgroup $H_0$ of the gauge
group) is shown in Table~\ref{tab:N3_SU2U1_mass}.
Inspection of the supermultiplet tables presented in Appendix~\ref{super3} leads to the conclusion that the
spectrum given in Table~\ref{tab:N3_SU2U1_mass} is organized into the following
supermultiplets
\begin{align}
  \mathrm{Spec}=\underbrace{A_1[1]^{(0)}_{\frac{3}{2}}}_{\substack{\text{massless graviton} \\ \text{multiplet}}}\oplus \underbrace{B_1[0]^{(4)}_2 \oplus 2\times B_1[0]^{(3)}_{\frac{3}{2}}}_{\text{massive vector multiplets}} \oplus \underbrace{2\times B_1[0]^{(2)}_1}_{\substack{\text{massless vector} \\ \text{multiplets}}}.
\end{align}
A consistency check is provided by Goldstone theorem. The gauge symmetry
breaking pattern in this vacuum tells that there are $7$ broken generators and
$4$ unbroken ones. Hence the number of massive vector fields is $7$ and that of the
massless ones is $4+1$, in agreement with the above tables.

\subsubsection{\texorpdfstring{$\mathcal{N}=3$}{N=3} vacuum preserving \texorpdfstring{$H_0=\mathrm{SO}(3)_D\subset \mathcal{G}$}{H=SO(3)D < Ggauge}}
As in the previous case, the vacuum manifold that is invariant under the
subgroup $H_0=\mathrm{SO}(3)_{D}$ is $S^1/\mathbb{Z}_2$, spanned by $\alpha_1=\alpha_2=\alpha_3$.
Again, there exists a single supersymmetric point on this circle of vacua
which
preserves $\mathcal{N}=3$ supersymmetry. The spectrum at this special
vacuum consists of states listed in Table~\ref{tab:N3_SO3_mass}.
Comparison with the supermultiplet tables results in a unique grouping of
the states in Table~\ref{tab:N3_SO3_mass} into $\mathrm{OSp}(3\vert4)$
supermultiplets
\begin{align}\label{eq:spec_so3D}
  \mathrm{Spec}=\underbrace{A_1[1]^{(0)}_{\frac{3}{2}}}_{\substack{\text{massless graviton} \\ \text{multiplet}}}\oplus \underbrace{B_1[0]^{(6)}_3 \oplus B_1[0]^{(4)}_2}_{\text{massive vector multiplets}} \oplus \underbrace{B_1[0]^{(2)}_1}_{\substack{\text{massless vector} \\ \text{multiplet}}}.
\end{align}
Goldstone theorem serves as a check of consistency. There are $3$ unbroken and
$8$ broken gauge generators in this vacuum and hence $8$ massive and $3+1$
massless vector fields. Looking at the tables we see that this is
in fact true.

\subsection{\texorpdfstring{$\mathcal{N}=2$}{N=2} vacua}

\subsubsection{\texorpdfstring{$\mathrm{OSp}(2|4)$}{Osp(2|4)} supermultiplets}
We have a $\mathrm{SO}(2)\simeq\mathrm{U}(1)$ R-symmetry and thus states of the
$\mathrm{OSp}(2|4)$ supermultiplets are labeled by the $\mathrm{U}(1)$ R-charge $R\in\mathbb{R}$,
spin and scaling dimension. There are two independent supercharges with R-charge
$(+1)$ and $(-1)$, respectively. The shortening condition for a supermultiplet
can occur for each of them independently. Therefore we have four different types
of shortening conditions: long-long, long-short, short-long and short-short.

The next topic we need to discuss, is what happens when the scaling dimension of
a long multiplet hits the unitarity bound. In such a situation it splits into a
sum of (partially) short multiplets. But the content of states is the same on
both sides of the relation. For instance on the example of a long massive
gravitino multiplet --
$L\bar{L}[1]^{(R)}_{\Delta>|R|+\frac{3}{2}}$. When its scaling dimension hits the
unitarity bound, it splits into a short massive gravitino multiplet and a
short
massive vector multiplet. In equations\footnote{When changing the $R$--symmetry sign we also need to exchange the role of the left and right components of the supermultiplet, those corresponding to the two independent supercharges.}
\begin{align}
&R>0:\;L\bar{L}[1]^{(R)}_{\Delta>R+\frac{3}{2}}\xrightarrow{\Delta\to R+\frac{3}{2}}L\bar{A}_1[1]^{(R)}_{R+\frac{3}{2}}\oplus L\bar{A}_2[0]^{(R+1)}_{R+2} \\
  &R<0:\;L\bar{L}[1]^{(R)}_{\Delta>-R+\frac{3}{2}}\xrightarrow{\Delta\to -R+\frac{3}{2}}A_1\bar{L}[1]^{(R)}_{-R+\frac{3}{2}}\oplus A_2\bar{L}[0]^{(R-1)}_{-R+2}
\end{align}
Analogously, a long massive vector multiplet can split into a short massless
vector multiplet and a conjugate pair of $\tfrac{1}{2}$-hypermultiplets (forming
a full hypermultiplet).

In order to decide whether the scaling dimension is
above the unitarity bound or it has been reached, one needs to compute independently the R-charge and the scaling dimension. We know that the $\mathcal{N}=2$ vacua spontaneously breaks the $R$--symmetry to ${\rm U}(1)
\leftarrow {\rm SO}(3)$, hence we can infer the $R$--charges content from
the breaking pattern of $R$--symmetry representations present in the corresponding $\mathcal{N}=3$ vacua.
Taking all these comments into account, we find a unique way to organize the spectra in ${\rm OSp}(2|4)$ supermultiplets. In Appendix~\ref{super2} we list the relevant ones.

\subsubsection{\texorpdfstring{$\mathcal{N}=2$}{N=2} vacuum preserving
  \texorpdfstring{$H_0=\mathrm{U}(1)_{D}\times\mathrm{U}(1)\subset\mathrm{SU}(2)\times\mathrm{U}(1)\subset \mathcal{G}$}{H=U(1)D x U(1) < SU(2) x U(1) < Ggauge}}
The gauge symmetry breaking pattern in this vacuum takes the form
$\mathcal{G}=\mathrm{SO}(3)\times\mathrm{SU}(3)\rightarrow
H_0=\mathrm{U}(1)_D\times
\mathrm{U}(1)$. According to Goldstone theorem the $12$ vector fields split into $9$ massive ones and $2+1$ massless (two gauging $H_0$ and one belonging to the Betti multiplet).
The supergravity mass spectrum displayed in Table~\ref{tab:N2_U1U1_mass} can be
arranged into the following supermultiplets
\begin{align}
  \mathrm{Spec}=&\underbrace{A_1\bar{A}_1[2]^{(0)}_{3}}_{\substack{\text{massless graviton}\\ \text{multiplet}}}\oplus \underbrace{L\bar{L}[1]^{(0)}_{|m_G^{(1)}| + \frac{1}{2}}}_{\substack{\text{long massive}\\ \text{gravitino multiplet}}} \oplus \underbrace{L\bar{L}[0]^{(0)}_2}_{\substack{\text{long massive} \\ \text{vector multiplet}}} \oplus 2 \times \underbrace{A_2\bar{A}_2[0]^{(0)}_1}_{\substack{\text{massless vector}\\ \text{multiplets}}}  \notag \\
                & \oplus \left(\underbrace{2\times L\bar{A}_2[0]^{(\frac{1}{2})}_{\frac{3}{2}}}_{\substack{\text{short massive}\\ \text{vector multiplets}}}  \underbrace{\oplus 2\times L\bar{B}_1[0]^{(1)}_1\oplus 2\times L\bar{B}_1[0]^{(\frac{3}{2})}_{\frac{3}{2}}\oplus L\bar{B}_1[0]^{(2)}_{2}}_{\tfrac{1}{2}-\text{hypermultiplets}}\right) \oplus ~ \left({R \rightarrow -R}\right)
\end{align}
Where 
\begin{equation}
    m_G^{(1)}= \frac{\sqrt{g_1^4+g_2^4-2 g_1^2 g_2^2 \cos{2 \alpha _2}}}{g_2^2-g_1^2}
\end{equation}
is the mass of the single massive gravitino. 

\subsubsection{\texorpdfstring{$\mathcal{N}=2$}{N=2} vacuum preserving
  \texorpdfstring{$H_0=\mathrm{U}(1)_{D}\subset\mathrm{SO}(3)_D\subset \mathcal{G}$}{H=U(1)D < SO(3)D < Ggauge}}
The gauge symmetry $\mathcal{G}=\mathrm{SO}(3)\times \mathrm{SU}(3)$ is
partially spontaneously broken to $H_0=\mathrm{U}(1)_D$. We conclude that out of the $12$ vector fields $10$ become
massive, while $1+1$ (one belonging to the Betti multiplet) remain massless.

This agrees with the supergravity mass spectrum shown in
Table~\ref{tab:N2_U1_mass}, which can be organized in a supermultiplet
structure given below
\begin{align}
\mathrm{Spec}=&\underbrace{A_1\bar{A}_1[2]^{(0)}_{3}}_{\substack{\text{massless graviton}\\ \text{multiplet}}}\oplus\underbrace{L\bar{L}[1]^{(0)}_{|m_G^{(2)}|+\frac{1}{2}}}_{\substack{\text{long massive}\\ \text{gravitino multiplet}}} \oplus \underbrace{L\bar{L}[0]^{(0)}_2}_{\substack{\text{long
massive} \\ \text{vector multiplet}}}  \oplus \underbrace{A_2\bar{A}_2[0]^{(0)}_1}_{\substack{\text{massless vector}\\ \text{multiplet}}} \oplus  \underbrace{L\bar{L}[0]^{(0)}_3}_{\substack{\text{long massive} \\ \text{vector multiplet}}} \notag \\
              & \oplus \left( \underbrace{ L\bar{L}[0]^{(1)}_3}_{\substack{\text{long massive}\\ \text{vector multiplet}}} \oplus \underbrace{ L\bar{A}_2[0]^{(2)}_3}_{\substack{\text{short massive}\\ \text{vector multiplet}}} \oplus \underbrace{ L\bar{B}_1[0]^{(3)}_3\oplus L\bar{B}_1[0]^{(2)}_2\oplus L\bar{B}_1[0]^{(1)}_1}_{\tfrac{1}{2}-\text{hypermultiplets}}\right) \oplus \left( R \rightarrow -R \right)
\end{align}
Where 
\begin{equation}
    m_G^{(2)}= \frac{\sqrt{16 g_1^4+g_2^4-8 g_1^2 g_2^2 \cos{2 \alpha _2}}}{g_2^2-g_1^2}
\end{equation}
is the mass of the single massive gravitino. 

\subsection{\texorpdfstring{$\mathcal{N}=1$}{N=1} vacua}

\subsubsection{ \texorpdfstring{$\mathrm{OSp}(1|4)$}{Osp(1|4)} supermultiplets}
Since the R-symmetry is trivial, states of irreducible
representations of $\mathrm{OSp(1|4)}$ are labeled just by spin and scaling
dimension. In Appendix~\ref{super1} we list only six supermultiplets that will be needed, four long and
two short ones.

\subsubsection{\texorpdfstring{$\mathcal{N}=1$}{N=1} vacuum preserving
  \texorpdfstring{$H_0=\mathrm{U}(1)\subset\mathrm{U}(1)_{\mathrm{D}}\times \mathrm{U}(1)\subset \mathrm{SU}(2)_\mathrm{D}\times\mathrm{U}(1)\subset \mathcal{G}$}{H=U(1) < U(1)D x U(1) < SU(2)D x U(1) < Ggauge}}
The gauge symmetry $\mathcal{G}=\mathrm{SO}(3)\times \mathrm{SU}(3)$ in this
vacuum is partially spontaneously broken to $H_0=\mathrm{U}(1)$. Thus there are
$\mathrm{dim}(\mathcal{G})-\mathrm{dim}H_0=10$ broken generators and Goldstone
theorem implies in this situation that the total $12$ vector fields split
into
10 massive and $1+1$ massless ones (one in the Betti multiplet). Indeed, the
above reasoning complemented by the computation of the mass spectrum within
supergravity, reported in Table~\ref{tab:N1_U1_mass} leads to a unique $\mathcal{N}=1$ supermultiplet
spectrum in $AdS_4$ (i.e.
$\mathrm{OSp}(1|4)$)
\begin{align}\label{eq:N1_U1_smult}
\mathrm{Spec}=&\underbrace{A_1[3]_{\frac{5}{2}}}_{\substack{\text{massless}\\\text{graviton}\\ \text{multiplet}}}\oplus\underbrace{L[2]_{\Delta^{(1)}_{G1}}\oplus L[2]_{\Delta^{(1)}_{G2}}}_{\substack{\text{massive gravitino}\\ \text{multiplets}}}\oplus\underbrace{2\times A_1[1]_{\frac{3}{2}}}_{\substack{\text{massless vector}\\ \text{multiplets}}}\oplus\underbrace{4\times L[1]_2\oplus L[1]_{\Delta^{(1)}_{V1}}\oplus L[1]_{\Delta^{(1)}_{V2}}}_{\substack{\text{massive vector multiplets}}} \notag \\
              &\oplus \underbrace{L^\prime[0]_3\oplus 2\times L^\prime[0]_2\oplus 8\times L^\prime[0]_{\frac{3}{2}}\oplus L^\prime[0]_{\Delta^{(1)}_{H1}}\oplus L^\prime[0]_{\Delta^{(1)}_{H2}}\oplus 6\times L^\prime[0]_1}_{\text{matter multiplets}}
\end{align}

When comparing the supermultiplet spectrum~\eqref{eq:N1_U1_smult} to the mass
spectrum of supergravity presented in Table~\ref{tab:N1_U1_mass}, Higgs
phenomenon has to be taken into account. Namely, the longitudinal modes of
massive vectors (gravitini) are massless scalars (spin-$\tfrac{1}{2}$ fermions).
The scaling dimensions (energies) appearing in Table~\ref{tab:N1_U1_mass}
are
expressed in terms of the parameters of the supergravity theory as follows
\begin{align}
\label{scaldim}
\Delta^{(1)}_{G1}=&\Delta^{(1)}_{H1}= 1+\frac{\sqrt{g_1^4+g_2^4-2g_1^2g_2^2\cos(2\alpha_2)}}{g_2^2-g_1^2} \\
  \Delta^{(1)}_{G2}=&\Delta^{(1)}_{H2}=1+\frac{\sqrt{g_1^4+g_2^4-2g_1^2g_2^2\cos(2\alpha_3)}}{g_2^2-g_1^2} \\
  \Delta^{(1)}_{V1}=&1+\frac{\sqrt{ \beta^{(1)}_1-4 \sqrt{\beta^{(1)}_2}}}{2(g_2^2-g_1^2)} \\
   \Delta^{(1)}_{V2}=&1+\frac{\sqrt{ \beta^{(1)}_1+4 \sqrt{\beta^{(1)}_2}}}{2(g_2^2-g_1^2)} \\
\beta^{(1)}_1& = 5g_1^4+5 g_2^4-2g_1^2 g_2^2 (4 \cos (2 \alpha_2)+4 \cos (2 \alpha_3)-3) \\
\beta^{(1)}_2& =g_1^8+2g_1^6 g_2^2+10g_1^4 g_2^4+2g_1^2 g_2^6+g_2^8 + \notag \\
& 8g_1^4 g_2^4 \cos (2 (\alpha_2+\alpha_3))+2g_1^2 g_2^2 \left(g_1^2+g_2^2\right)^2 (\cos (2 (\alpha_2 -\alpha_3))-2 \cos (2 \alpha_2)-2 \cos (2 \alpha_3))
\end{align}

\subsubsection{\texorpdfstring{$\mathcal{N}=1$}{N=1} vacuum preserving
  \texorpdfstring{$H_0=\{\mathbf{1}\}\subset\mathrm{U}(1)_{\mathrm{D}}\subset\mathrm{SO}(3)_\mathrm{D}\subset \mathcal{G}$}{H=1 < U(1)D < SO(3)D < Ggauge}}
In this vacuum we observe a complete spontaneous symmetry breaking
$\mathcal{G}=\mathrm{SO}(3)\times \mathrm{SU}(3)\to H_0=\{\mathbf{1}\}$. Hence
Goldstone theorem dictates that there are $11$ massive vector fields and just a
single massless vector in the Betti multiplet.

The mass spectrum of supergravity fields summarized in
Table~\ref{tab:N1_Id_mass} is organized into the following $\mathrm{OSp}(1|4)$
supermultiplets
\begin{align}
  \mathrm{Spec}=&\underbrace{A_1[3]_{\frac{5}{2}}}_{\substack{\text{massless}\\\text{graviton}\\ \text{multiplet}}}\oplus\underbrace{L[2]_{\Delta^{(2)}_{G1}}\oplus L[2]_{\Delta^{(2)}_{G2}}}_{\substack{\text{massive gravitino}\\ \text{multiplets}}}\oplus\underbrace{A_1[1]_{\frac{3}{2}}}_{\substack{\text{massless}\\ \text{vector}\\ \text{multiplets}}}\oplus\underbrace{5\times L[1]_{\frac{7}{2}}\oplus L[1]_{\Delta^{(2)}_{V1}}\oplus L[1]_{\Delta^{(2)}_{V2}}}_{\substack{\text{massive vector multiplets}}} \notag \\
                &\oplus \underbrace{3\times L^\prime[0]_4\oplus 8\times L^\prime[0]_3\oplus 2\times L^\prime[0]_2\oplus  L^\prime[0]_{\Delta^{(2)}_{H1}}\oplus L^\prime[0]_{\Delta^{(2)}_{H2}}\oplus 3\times L^\prime[0]_1}_{\text{matter multiplets}}
\end{align}

The values of scaling dimensions determining the supergravity mass spectrum
presented in Table~\ref{tab:N1_Id_mass} take the form of the ones in the corresponding type (i) vacua with the replacement $g_1 \rightarrow 2 g_1$.

\section{Domain wall solutions}\label{sec:DW}
In the previous section we have studied the (super-) conformal-multiplet arrangement of the fields on the new $AdS_4$ vacua. In this section we will show that the latter can be interpreted as fixed points of RG-flows triggered by  relevant operators which pertain to the CFT dual to the central vacuum.\\
In order to do this, we consider a $(3+1)$-dimensional bulk space-time, parametrized by the coordinates $x^\mu=\left(x^i,y\right)$, and use the standard domain-wall (DW) ansatz for the metric, which has the usual form
\begin{align}
ds^2\,&=\,\,e^{2 A(y)}\,ds^{2}_{1,2}\,-\,d y^2\,=\,\,e^{2A(y)}\,dx^{i}\,\eta_{i j}\,dx^{j}\,-\,d y^2\,\quad,\quad \eta_{i j}\,=\,\left(+,-,-\right)\quad,\label{DW}\\
{\phi}^r\,&=\,{\phi}^r\left(y\right)\quad,\quad i,j\,=\,0,1,2\quad,
\end{align}
where $ds^{2}_{1,2}$  defines the flat Minkowski metric in three dimensions, $A(y)$ is the scale factor, $y$ is the coordinate transverse to the wall, and all scalar fields $\phi(y)$ depend only on the transverse coordinate $y$ \footnote{From now on, we will omit the $y$-dependence of the scalar fields and the scale factor in the DW metric. }.\\
From the ${\rm AdS/CFT}$ point of view, the domain wall ansatz  corresponds to an RG flow between the UV and IR fixed points described by the asymptotic regions $y\to\pm\infty$.\\
Let us be more explict by considering the consistent truncation described in Section \ref{vacuasect}, generated by the three complex scalar fields $z_1,z_2,z_3$. We recall that solutions of the truncated theory are solutions of the complete theory and that all fields in the DW solution
are functions of the transverse coordinate $y$ only. 
From the coset metric \eqref{metrixSTU} and the ansatz in (\ref{DW}) one can obtain, after consistently setting all fermions and vector fields to zero, the effective lagrangian density \footnote{Here, primes denote derivatives respect to the $y$ direction}
\begin{equation}
{\Scr L}\,=\,-e^{3 A}\sum_i^3\left[\,3 A''\,+\,6 A'{}^2\,+\, \left( r'_i \right)^2\,+\,\frac{1}{4}\sinh\left(2r_i\right)^2 \left(\alpha'_i\right)^2\,+\,V(r_i,\alpha_i)\,\right]\label{laggen}\,\,,
\end{equation}
where the potential for Type (i) and Type (ii) models was given in~\eqref{potgen} and~\eqref{potgen2}, respectively.\\
We leave the details of the DW solutions in appendix \ref{DWapp}. Here we focus on the main properties and their possible interpretation in the dual picture. In particular we search for configurations in which the radii $r_i$ are equal to the same field $r$. Then the phases $\alpha_i$ do not depend on $y$. Therefore, the constant values of the phases $\alpha_i$ select the critical point at the end of the flow (IR fixed point) as in Table (\ref{eq:T1_vac}) (or (\ref{eq:T2_vac}) for Type (ii) vacuum), the starting point being the central $\mathcal{N}=3$ vacuum (UV fixed point). The "shape" of the domain wall is implicitly governed by the field $r(y)$ through the warping function $A(y(r))$. For the sake of simplicity let us consider the Type (i) consistent truncation \eqref{eq:coset1} (Type (ii) consistent truncation gives the same results after substituting $g_1\rightarrow\,2g_1$), which provides the vacuum at the origin and the one described by \eqref{eq:rstar}.  In this case we obtain the DW solution, whose explicit expression is given in eq. \eqref{eq:domaini1} Appendix \ref{DWapp}. It is useful to perform the following change of coordinates in order to study the behavior near the fixed points of the flow:

\begin{equation}
x^i\mapsto\left(g_1^2-\varepsilon g_2^2\right)\, x^i\quad,\quad r\,=\,r(y)\quad,\quad \varepsilon\,=\,\left\{\begin{array}{cc}

                         0 & r \rightarrow 0 \\

                         1 & r \rightarrow r^\star

                       \end{array}\right.
\end{equation}
where $r(y)$ is the solution for $r$ in the DW background. Actually, it is enough to know the expression for the inverse relation $y(r)$
given by \eqref{eq:domaini2}. Then the DW metric becomes
\begin{equation}
  ds^2\, =\, \frac{1}{4} \left(\,\frac{(g_1 \text{csch}(r)-g_2 \text{sech}(r))^2  }{g_1^4}dx^i dx_i\,-\,\frac{ \text{csch}^2(r) \text{sech}^4(r)}{(g_1-g_2 \tanh (r))^2}\,dr^2\,\right)\,\,.
\end{equation}

Now, we consider the limit $r \rightarrow 0$ to obtain

\begin{equation}
  ds^2\,\sim\,ds^2_{\text{UV}}\, = \,\frac{1}{4 r^2 g_1^2} \left( -dr^2 + dx^i dx_i \right)
\end{equation}

which is the metric for an $AdS_4$ space with radius
\begin{equation}
R^2\,=\,-\frac{3}{\Lambda}\,=\,\frac{1}{4{g_1}^2}\,\,,
\end{equation}
in agreement with the value of $\Lambda$ at $r=0$ in \eqref{eq:cosmi}. This expression provides directly the asymptotic behavior of $r$ near the conformal boundary. Indeed, in this particular case the metric is in the usual Poincar\'e coordinates with radial direction $z$. Hece, we have $r \sim z$ and $\Delta_r = 1$. 
On the other side, expanding $ds^2$ near $r \rightarrow r^\star$ we get

\begin{equation}
  ds^2\,\sim\,ds^2_{\text{IR}}\, = \, R^2 \left( u^2 dx^j dx_j - \frac{du^2}{u^2} \right)
\end{equation}
where $u=(r-r^\star)$ and
\begin{equation}
R^2\,=\,-\frac{3}{V(r^\star)}\,=\,\frac{{g_2}^2-{g_1}^2}{4{g_1}^2{g_2}^2}\,\,,
\end{equation}
as expected from \eqref{eq:cosmi}. The relation with Poincar\'e coordinates is given by $u=\frac{1}{z}$. So that $(r-r^\star) \sim z^{-1}$ and $\Delta_u = -1$.
\newline
The interpretation as an RG-flow, is the following. When we switch on the
$r$ source (the combination $\delta r_1+ \delta r_2+ \delta r_3$) at the origin we introduce a relevant deformation, indeed the scaling dimension of the operator coupled to $r$ will be $\Delta_{{\Scr O}_r}|_{0}=2$. This triggers an RG-flow that eventually ends at $r=r^\star$ where the operator becomes irrelevant, indeed $\Delta_{{\Scr O}_r}|_{r^\star}=4$. We
are flowing from the ${\Scr N}=3$ $SCFT_3$ dual to the $AdS_4$ background at $r=0$ (the UV region) to a $CFT_3$ dual to the $AdS_4$ background
at $r=r^\star$ (in the IR region). In general the IR three-dimensional dual theory will not be superconformal. For particular values of
$\alpha_i$ the IR critical point will correspond to a $SCFT_3$ with different amount of supersymmetries, in agreement with the classification given in \ref{eq:T1_vac}.\\
As a check for our interpretation we compute the scalar spectrum of the truncation near $r=0$ and $r=r^*$ and we obtain the masses $(-2, -2, -2, -2, -2, -2)$ and $(4, -2, -2, 0, 0, 0)$ respectively. The latter correspond to the combinations $(\delta r_1 + \delta r_2 + \delta r_3, \delta r_2-\delta r_1, \delta r_3- \delta r_1, \delta \alpha_1, \delta
\alpha_2, \delta \alpha_3)$. Another relevant check of the interpretation of $r=0$ as the UV critical point and $r=r^\star$ as the IR one is provided by the holographic c-theorem \cite{Myers:2010tj,Myers:2010xs}. Following these works we compute
\begin{equation}
  a(y)=A'(y)^{-2}\,\,,
\end{equation}
where
\begin{equation}
A'(y )= -2 g_2 \sinh ^3(r(y ))+2 g_1 \cosh ^3(r(y ))\,\,.
\end{equation}
It follows that $a(y(r))$ is monotonically decreasing as a function of $r \in [0,r^\star]$, consistently with the holographic c-theorem $a_{UV} \geq a_{IR}$.
%

\section*{Conclusions}
\addcontentsline{toc}{section}{\protect\numberline{}Conclusions}
In this work, after providing a review of the embedding tensor formulation of $D=4$ extended gauged supergravity specialized to the $\mathcal{N}=3$ case, we have focused on a particular model with gauge group $\mathcal{G}={\rm SO}(3)\times {\rm SU}(3)$ and studied its vacua. We find, aside from the  $\mathcal{N}=3$ vacuum at the origin of the scalar manifold preserving the whole gauge group, one 3-torus of vacua for $g_1\le g_2\le 2 g_1$ and two 3-tori of vacua for $g_2> 2 g_1$. Each of these manifolds contains, aside from a known isolated $\mathcal{N}=3$ vacuum \cite{Karndumri:2016fix}, a line of $\mathcal{N}=2$ vacua, a surface of $\mathcal{N}=1$ vacua and the remaining stable, non-supersymmetric vacua, all of which to our knowledge were overlooked in the literature. These vacua were found in particular consistent truncations of the model described by three complex scalar fields $z_i=r_i\,e^{i\,\alpha_i}$. The three angular coordinates $\alpha_i$ parametrizing the 3-tori of vacua, are flat directions of the scalar potential and thus are reasonably expected to correspond to exactly marginal deformations in the dual CFT at the boundary. Therefore, within each of these 3-dimensional loci, vacua with different amounts of supersymmetry, or no-supersymmetry at all, are connected through flat directions of the scalar potential, indicating in the dual CFT picture a possible (partial) supersymmetry breaking triggered by exactly marginal operators. This is reminiscent of a similar property displayed by a class of vacua recently found within gauged maximal four-dimensional supergravity in \cite{Giambrone:2021wsm}, describing Type IIB S-fold backgrounds \cite{Hull:2004in} and generalizing earlier supersymmetric solutions of the same kind \cite{Inverso:2016eet,Assel:2018vtq,Giambrone:2021zvp,Guarino:2021kyp,Bobev:2021yya,Arav:2021gra}. The vacua studied in \cite{Giambrone:2021wsm} define a locus parametrized by two compact axionic deformations $\chi_1,\,\chi_2$ which are flat directions of the scalar potential at the corresponding critical points. For generic values of  $\chi_1,\,\chi_2$ the solutions are non-supersymmetric, for $\chi_1=-\chi_2$ they feature $\mathcal{N}=2$ supersymmetry and for $\chi_1=\chi_2=0$ supersymmetry is enhanced to $\mathcal{N}=4$. These two flat deformations are conjectured in~\cite{Giambrone:2021wsm} to define marginal deformations of the $\mathcal{N}=4$ S-fold CFT \cite{Assel:2018vtq} dual to the solution found in \cite{Inverso:2016eet}, and thus to span a non-supersymmetric conformal manifold of the dual CFT.\par
It would be interesting to uplift the web of vacua discussed here to Type II superstring theory or to $D=11$ supergravity. One could try to embed a suitable truncation of the model studied here (containing the 3-tori of vacua) into maximal supergravity, and then use exceptional field theory techniques \cite{Hohm:2013pua,Hohm:2013uia,Hohm:2014qga} in order to uplift them to string or M-theory. Another possibility is that the truncation of the model describing the new vacua studied here does not fit in a maximal supergavity. In this case one should work with less supersymmetric consistent truncations possibly implementing the analysis of \cite{Cassani:2019vcl}. In particular one could try to obtain a subsector of our model capturing the new solutions and the central one as a compactification of string or M-theory by means of a suitable $G_S$-structure manifold. There is also the possibility that no consistent truncation can describe our solutions. If the uplift of the whole new family of vacua is possible, assessing perturbative stability of the corresponding $\mathcal{N}=0$ backgrounds would in principle require the computation of the corresponding Kaluza-Klein spectrum in order to check if the $D=4$ scalar-modes have all squared masses exceeding the BF bound. However, borrowing an argument used in \cite{Giambrone:2021wsm} that non-supersymmetric vacua connected to stable supersymmetric ones by  continuous parameters are expected to be pertutbatively stable, we anticipate perturbative stability of the $\mathcal{N}=0$ vacua.
The ten or eleven dimensional backgrounds, if found, would then provide further holographic evidence in favor of the existence of non-supersymmetric conformal manifolds.\par
Another outcome of our analysis is the construction of the RG flows connecting the ${\rm AdS}_4\times N^{0,1,0}$ vacuum at the origin to any of the vacua in the 3-tori, generalizing the flow found in  \cite{Karndumri:2016fix} to solutions connecting the origin to $\mathcal{N}=2$, $\mathcal{N}=1$ and $\mathcal{N}=0$ IR fixed points.
Understanding these new flows in the dual CFT picture and the uplift of their fixed points is also a subject of future investigation. 

\subsection*{Acknowledgements}
A. G. and M. T. are very grateful to Gregoire Josse and  Emanuel Malek for fruitful discussions. A. G. also wishes to thank Davide Cassani for  some useful  remarks about consistent truncations during the early stages of preparation of the present work. P. V. would like to acknowledge hospitality of the Arnold--Regge Center, Turin, where the majority of this work was done during his postdoc.

\begin{appendices}

\section{Ward Identity}
\label{ward}
A particular case of eq. \eqref{quadrato} is the following one
\begin{equation}
  \mathbb{T}_{\underline{\Lambda}\, \underline{R}}^{~~~~~\underline{\Pi}} \mathbb{T}^{\underline{\Sigma}~~\underline{\Delta}}_{~~\underline{\Pi}} - \mathbb{T}^{\underline{\Sigma}~~\underline{\Pi}}_{~~\underline{R}} \mathbb{T}_{\underline{\Lambda} \, \underline{\Pi}}^{~~~~~\underline{\Delta}} + \mathbb{T}_{\underline{\Lambda}~~\underline{\Pi}}^{~~\underline{\Sigma}} \mathbb{T}^{\underline{\Pi}~~\underline{\Delta}}_{~~\underline{R}} = 0
\end{equation}
Terms like $\mathbb{T}_{M}^{~~\underline{\Lambda} \, \underline{\Sigma}}$ and $\mathbb{T}_{M \, \underline{\Lambda} \, \underline{\Sigma}}$ do not appear because ${\Scr
R}$ is block-diagonal.
We further restrict to
\begin{eqnarray}
  \mathbb{T}^{A D \underline{\Pi}} \mathbb{T}_{B\underline{\Pi}C} - \mathbb{T}_B^{~~D\underline{\Pi}} \mathbb{T}^A_{~~\underline{\Pi} C} + \mathbb{T}^A_{~~B\underline{\Pi}} \mathbb{T}^{\underline{\Pi} D}_{~~~~~C} &=& \nonumber\\
  \mathbb{T}^{A D}_{~~~~E} \mathbb{T}^{~~E}_{B~~C} - \mathbb{T}_{B~~E}^{~~D} \mathbb{T}^{AE}_{~~~~C} + \mathbb{T}^{A~~E}_{~~B} \mathbb{T}_{E~~C}^{~~D} && \nonumber\\
 + \,  \mathbb{T}^{A D}_{~~~~I} \mathbb{T}^{~~I}_{B~~C} - \mathbb{T}_{B~~I}^{~~D} \mathbb{T}^{AI}_{~~~~C} + \mathbb{T}^{A~~I}_{~~B} \mathbb{T}_{I~~C}^{~~D} &=& 0
\end{eqnarray}
Now we recall
\begin{equation}
  \mathbb{T}_{\underline{\Lambda} A}^{~~~~B} = - \mathbb{T}_{\underline{\Lambda}~~ A}^{~~~B} ~~~{\rm and}~~~ \mathbb{T}_{\underline{\Lambda} A}^{~~~~I} = \mathbb{T}_{\underline{\Lambda}~~ A}^{~~~I}
\end{equation}
to obtain
\begin{eqnarray}
  Q^{A D}_{~~~~B C} \equiv -\mathbb{T}^{A D}_{~~~~E} \mathbb{T}^{~~~~E}_{BC} + \mathbb{T}_{BE}^{~~~~D} \mathbb{T}^{AE}_{~~~~C} + \mathbb{T}^{AE}_{~~~~B} \mathbb{T}_{EC}^{~~~~D} && \nonumber\\
 + \,  \mathbb{T}^{A D}_{~~~~I} \mathbb{T}^{~~~~I}_{BC} - \mathbb{T}_{BI}^{~~~~D} \mathbb{T}^{AI}_{~~~~C} - \mathbb{T}^{AI}_{~~~~B} \mathbb{T}_{IC}^{~~~~D}  &=& 0
\end{eqnarray}
The following decomposition holds true\footnote{$S_{AB}=S_{BA}$}
\begin{eqnarray}
  \mathbb{T}^{A D}_{~~~~E} &=& \frac{1}{2}(\epsilon^{ADB} S_{B E} + \delta_E^{\left[A \right. } N^{\left. D \right]}) \nonumber\\
  \mathbb{T}^{A D}_{~~~~I} &=& \epsilon^{B A D} T_{B I}
\end{eqnarray}
In terms of $S_{AB}$ and $N^A$, we compute\footnote{${S}^{A B} = ({S}_{AB})^*$}
\begin{eqnarray}
  - \epsilon_{\overline{A} A D} \epsilon^{\overline{B} B C}  {T}^{A D}_{~~~~E} {T}^{~~~~E}_{BC}&=& -({S} {S}^* )_{\overline{A}}^{~~\overline{B}}
-\frac{1}{2}(\epsilon_{\overline{A} E D} {N}^D {S}^{\overline{B} E} + \epsilon^{\overline{B} E D } {N}_D {S}_{\overline{A} E}) \nonumber\\
  && -\frac{1}{4}(\delta_{\overline{A}}^{~\overline{B}} \mathbb{N}_A \mathbb{N}^A-\mathbb{N}_{\overline{A}} \mathbb{N}^{\overline{B}}) \nonumber\\
  \epsilon_{\overline{A} A D} \epsilon^{\overline{B} B C}  \mathbb{T}_{BE}^{~~~~D} \mathbb{T}^{AE}_{~~~~C} &=& \frac{1}{4}(\delta_{\overline{A}}^{~\overline{B}}{\rm Tr}\{{S} {S}^*\} -({S} {S}^* )_{\overline{A}}^{~~\overline{B}} + \frac{1}{2}(\epsilon_{\overline{A} E D} {N}^D {S}^{\overline{B} E} + \epsilon^{\overline{B} E D } {N}_D {S}_{\overline{A} E})  \nonumber\\
  && +\frac{1}{4} \delta_{\overline{A}}^{~\overline{B}} {N}_A {N}^A - \frac{3}{4}{N}_{\overline{A}} {N}^{\overline{B}})\nonumber\\
  \epsilon_{\overline{A} A D} \epsilon^{\overline{B} B C}\mathbb{T}^{A D}_{~~~~I} \mathbb{T}^{~~~~I}_{BC} &=& 4 T_{\overline{A} I} T^{\overline{B} I} \nonumber\\
   \epsilon_{\overline{A} A D} \epsilon^{\overline{B} B C}\mathbb{T}^{AE}_{~~~~B} \mathbb{T}_{EC}^{~~~~D} &=& \epsilon_{\overline{A} A D} \epsilon^{\overline{B} B C} \mathbb{T}_{BE}^{~~~~D} \mathbb{T}^{AE}_{~~~~C} \nonumber\\
  -\epsilon_{\overline{A} A D} \epsilon^{\overline{B} B C} \mathbb{T}_{BI}^{~~~~D} \mathbb{T}^{AI}_{~~~~C} &=& - \epsilon_{\overline{A} A D} \epsilon^{\overline{B} B C} \mathbb{T}^{AI}_{~~~~B} \mathbb{T}_{IC}^{~~~~D}
\end{eqnarray}
It is easy to verify the last two equations. Indeed, eq.(\ref{linea}) implies
\begin{equation}
  \mathbb{T}_{\underline{\Lambda} \underline{\Sigma}}^{~~~~\underline{\Delta}}=-\mathbb{T}_{\underline{\Sigma} \underline{\Lambda}}^{~~~~\underline{\Delta}}
\end{equation}
We get rid of terms of the form ${S} \cdot {N}$ thanks to
\begin{eqnarray}
  Q^{AD}_{~~~~BD}=0 \Leftrightarrow \epsilon^{A E \overline{C}} {S}_{\overline{C} B} N_E &=& -\frac{1}{2}\delta^{~A}_B {N}_C {N}^C + \frac{1}{2}{N}^A {N}_B \nonumber\\
                  && + \, 4 \, \delta^{~A}_B T_{C I} T^{C I} - \, 4 \,  T_{B I} T^{A I} \nonumber\\
                  && - \, 4 \, \mathbb{T}_{I B}^{~~~~D} \mathbb{T}^{I A}_{~~~~D} + \, 4 \, \mathbb{T}^{I A}_{~~~~B} \mathbb{T}_{I D}^{~~~~D}
\end{eqnarray}
Finally, we obtain\footnote{Actually, to get \eqref{eq:fermionic_shifts} we must redefine $S_{AB}=-2\mathbb{S}_{AB}$, $N^D=2\mathbb{N}^D$, $T_{CI}=\frac{1}{2}\mathbb{N}_{CI}$.}
\begin{equation}
  \epsilon_{\overline{A} A D} \epsilon^{\overline{B} B C} Q^{A D}_{~~~~B C}-\frac{32}{3} Q^{A D}_{~~~~ A D} \delta^{~\overline{B}}_{\overline{A}} = 0  \Leftrightarrow  \mathbb{N}_A \mathbb{N}^B + \mathbb{N}_{A I} \mathbb{N}^{B I} + \mathbb{N}_{IC}^{~~~~B} \mathbb{N}^{IC}_{~~~~A} - \, 12 \,
\mathbb{S}_{AC} \mathbb{S}^{BC} = \delta_{\overline{A}}^{~\overline{B}} V
\end{equation}

\section{Fermion Shift Tensors and Mass Matrices from \texorpdfstring{$\mathbb{T}$}{T}-tensor}
\label{SM}

We present a systematic way to identify the interesting components of the
$\mathbb{T}$ tensor involved in the definitions of fermionic shifts and mass matrices.

\subsection{Fermionic shifts}
In order to identify fermionic shifts inside $\mathbb{T}$ we consider what their $H$ representation should correspond to. This task is easy since we know that they enter fermionic supersymmetry transformations with parameter $\epsilon_A \in ( {\bf 3},{ \bf 1} )_{+\frac{1}{2}}$. Indeed, we have
\begin{eqnarray}
 ({\bf 3},{\bf 1})_{+\frac{1}{2}} \ni & \langle \delta \psi_{A\mu} \rangle = \langle\nabla_\mu \epsilon_A + {\rm i}\mathbb{S}_{AB} \gamma_\mu \epsilon^B \rangle & \Rightarrow \mathbb{S}_{AB} \in ({\bf 6}, {\bf 1})_{+1} \nonumber\\
 ({\bf 1},{\bf 1})_{+\frac{3}{2}}  \ni & \langle \delta \chi_\bullet \rangle  = \langle\mathbb{N}^D \epsilon_D\rangle & \Rightarrow \mathbb{N}^D \in (\overline{\bf 3},{\bf 1 })_{+1}\nonumber\\
 ({\bf 3},{\bf n})_{\frac{n+6}{2n}} \ni& \langle \delta \lambda_{IA}\rangle = \langle\mathbb{N}_{IA}^{~~~B} \epsilon_B\rangle  & \Rightarrow \mathbb{N}_{IA}^{~~~B} \in ({\bf 8 }+{\bf 1},{\bf n})_{+\frac{3}{n}}\nonumber\\
  ({\bf 1},{\bf n})_{\frac{3n+6}{2n}} \ni & \langle\delta \lambda_I\rangle=\langle\mathbb{N}_{IA}\epsilon^A\rangle & \Rightarrow \mathbb{N}_{IA} \in ({\bf 3},{\bf n})_{\frac{2n+3}{n}} \nonumber\\
\end{eqnarray}

We see that the wanted components of $\mathbb{T}$, possibly projected with $\mathcal{G} \subset H$-invariant tensors, must have one or two $R$-symmetry indices and no more than one matter index. The independent choices, obtained from $\mathbb{T}^{AB}_{~~~~C}$, $\mathbb{T}^{IA}_{~~~~B}$, $\mathbb{T}^{AB}_{~~~~I}$ up to complex conjugation, are
\begin{eqnarray}
  &\epsilon_{AB(D}\mathbb{T}^{A B}_{~~~~C)} \in ({\bf 6} ,{\bf 1})_{+1} ~~~~& \epsilon_{AB[D}\mathbb{T}^{A B}_{~~~~C]} \Leftrightarrow \mathbb{T}^{EB}_{~~~~B}  \in ({\bf \overline{3}},{\bf 1})_{+1} \nonumber\\
  & \mathbb{T}_{IA}^{~~~~B} \in ({\bf 8}+{\bf 1},{\bf n})_{+\frac{3}{n}} & \epsilon_{CAB}\mathbb{T}^{AB}_{~~~~I} \in ({\bf 3},{\bf n})_{\frac{2n+3}{n}}
\end{eqnarray}
This are exactly the needed representation in the definition of fermionic
shifts.

\subsection{Fermionic Mass Matrices}
Now we move to $\mathbb{M}_{\mathcal{IJ}}$. We play the same game as before. In this case we discover their representations from the possible $\overline{\lambda}^{\mathcal{I}} \mathbb{M}_{\mathcal{IJ}} \lambda^{\mathcal{J}} \in \left({\bf 1},{\bf 1}\right)_{0}$ interactions\footnote{One could find the
needed components for $\mathbb{N}_{\mathcal{I}}^A$ and $\mathbb{S}_{AB}$ looking for a gravitino-gravitino and gravitino-fermions mass terms.} which are of the following form
\begin{eqnarray}
  & \overline{\chi}_{\bullet} \mathbb{M}^{\bullet \bullet} \chi_{\bullet}
& \Rightarrow \mathbb{M}^{\bullet \bullet} \in \left({\bf 1},{\bf 1}\right)_{-3} \label{m}\\
  & \overline{\chi}_{\bullet}  \mathbb{M}^\bullet_I \lambda^I & \Rightarrow \mathbb{M}^\bullet_I \in \left({\bf 1},{\bf n}\right)_{\frac{3}{n}} \label{mi} \\
  & \overline{\chi}_{\bullet} \mathbb{M}^{\bullet, IA} \lambda_{IA} & \Rightarrow \mathbb{M}^{IA} \in \left({\bf \overline{3}},{\bf \overline{n}}\right)_{-\frac{2n+3}{n}} \label{mia}\\
  & \overline{\lambda}^I \mathbb{M}_{IJ} \lambda^J & \Rightarrow \mathbb{M}_{IJ} \in \left({\bf 1},{\bf \frac{1}{2}n(n+1)}\right)_{\frac{3(n+2)}{n}} \label{mij}\\
  & \overline{\lambda}^I \mathbb{M}_I^{~~AJ} \lambda_{AJ} & \Rightarrow \mathbb{M}_{I}^{~~AJ} \in \left({\bf \overline{3}},{\bf n \times \overline{n}}\right)_{+1} \label{miaj}\\
  & \overline{\lambda}_{AI} \mathbb{M}^{AI | BJ} \lambda_{BJ} & \Rightarrow \mathbb{M}^{AI | BJ} \in \left({\bf 3},{\bf \frac{1}{2}\overline{n}(\overline{n}-1)}\right)_{-\frac{n+6}{n}} \label{maibj}
\end{eqnarray}
We can easily convince ourselves that the only components of $\mathbb{T}$, up to identifications, matching these representations are
\begin{equation}
\mathbb{T}^{IJ}_{~~~J}  \in \left({\bf 1},{\bf \overline{n}}\right)_{- \frac{3}{n}}\,\,;\,\,\, \mathbb{T}^{AJ}_{~~~~I} \in  \left({\bf \overline{3}},{\bf n \times \overline{n}}\right)_{+1} \,\,;\,\,\, \epsilon^{ABC} \mathbb{T}^{IJ}_{~~~C} \in \left({\bf 3},{\bf
\frac{1}{2}\overline{n}(\overline{n}-1)}\right)_{-\frac{n+6}{n}} \,\,;\,\,\, \epsilon_{CAB}\mathbb{T}^{AB}_{~~~~I} \in \left({\bf 3},{\bf n}\right)_{\frac{2n+3}{n}} \nonumber
\end{equation}
These are the only ones entering gradient flow equations. Then, $\mathbb{M}_{IJ}$ and $\mathbb{M}_{\bullet \bullet}$ are consistently vanishing.\par
The precise relations between the mass matrices and the corresponding components of the $\mathbb{T}$-tensor is given in Appendix~\ref{D}.

\section{The Gradient Flow equations}\label{D}
We consider here the different projections of eq. (\ref{GF}) into $H$-covariant components:
\begin{eqnarray}
  \mathcal{D}\mathbb{N}^A &=& \frac{1}{2}{\Scr P}^E_{~~I} \mathbb{N}^{IA}_{~~~E} + \frac{1}{2}\epsilon^{EAC}\mathbb{N}_{CI}  {\Scr P}^I_{~~E}+\frac{1}{2}  {\Scr P}^A_{~~I}\mathbb{N}^{IE}_{~~~E} \nonumber\\
  \mathcal{D}\mathbb{N}_{CI}&=&2\epsilon_{ABC}  {\Scr P}^B_{~~J}\mathbb{T}^{AJ}_{~~~I} - 2 \mathbb{S}_{CD}  {\Scr P}^D_{~~I}+\epsilon_{CDB}\mathbb{N}^B  {\Scr P}^D_{~~I} \nonumber\\
  \mathcal{D}\mathbb{N}^{IA}_{~~~B}&=&-2   {\Scr P}^I_{~~C} \epsilon^{CAD}\mathbb{S}_{DB}+  {\Scr P}^I_{~~B}\mathbb{N}^A \nonumber\\
                          &&+   {\Scr P}^J_{~~D}\left(-2\delta^D_{B}\mathbb{T}^{IA}_{~~~J}+\mathbb{T}^{ID}_{~~~J}\delta^A_B \right)+  {\Scr P}^D_{~~J}\left( 2\delta^A_D \mathbb{T}^{IJ}_{~~~B} -\mathbb{T}^{IJ}_{~~~D} \delta^A_B \right) \nonumber\\
\mathcal{D}\mathbb{S}_{BE}&=&-\frac{1}{2}\epsilon_{AD(B}\mathbb{N}^{ID}_{~~~E)}  {\Scr P}^A_{~~I}-\frac{1}{2}  {\Scr P}^I_{(E}\mathbb{N}_{B)I} \label{gflow}
\end{eqnarray}
On the other hand, using the general form of the gradient flow equations required by the supersymmetry of the gauged Lagrangian, see \cite{Trigiante:2016mnt}, specialized to the $\mathcal{N}=3$ models, we find:
\begin{eqnarray}
  \mathcal{D}\mathbb{N}^A &=&   {\Scr P}^E_{~~I} \mathbb{T}^{AI}_{~~~E} + \mathbb{T}^{AE}_{~~~I}  {\Scr P}^I_{~~E} -2   {\Scr P}^A_{~~I}\mathbb{M}^{I}_{~~ \bullet} -2   {\Scr P}^I_{~~F}\mathbb{M}_{\bullet IE}\epsilon^{EAF}\nonumber\\
  \mathcal{D}\mathbb{N}_{CI}&=&- 2 \mathbb{S}_{CD}  {\Scr P}^D_{~~I} -2 \mathbb{M}_{IJ}   {\Scr P}^J_{~~C} -2 \mathbb{M}^{~~JA}_{I}   {\Scr P}^B_{~~J} \epsilon_{ACB}\nonumber\\
  \mathcal{D}\mathbb{N}^{IA}_{~~~B}&=&-2   {\Scr P}^I_{~~C} \epsilon^{CAD}\mathbb{S}_{DB}+  {\Scr P}^J_{~~E}\mathbb{T}^{IE}_{~~~J}\delta^A_B +   {\Scr P}^E_{~~J}\mathbb{T}^{IJ}_{~~~E}\delta^A_B
\nonumber\\
                          &&-2   {\Scr P}^J_{~~B}\mathbb{M}^{IA}_{~~~J} -2\mathbb{M}^{IA|JC}   {\Scr P}^D_{~~J} \epsilon_{CBD} \nonumber\\
\mathcal{D}\mathbb{S}_{BE}&=&-\frac{1}{2}\epsilon_{AD(B}\mathbb{N}^{ID}_{~~~E)}  {\Scr P}^A_{~~I}-\frac{1}{2}  {\Scr P}^I_{(E}\mathbb{N}_{B)I} \label{gflow-report}
\end{eqnarray}
Direct comparison between \eqref{gflow} and \eqref{gflow-report} suggest the following identifications
\begin{equation}
  \mathbb{T}^{IE}_{~~~E}=2 \mathbb{M}^I_{~~\bullet}, ~~~~ \mathbb{T}^{IA}_{~~~J}+\frac{1}{2}\delta^I_J \mathbb{N}^A=\mathbb{M}^{AI}_{~~~J}, ~~~~
\mathbb{T}^{IJ}_{~~~A}=-\frac{1}{2}\epsilon_{ABC}\mathbb{M}^{IB|JC}, ~~~~ \mathbb{T}^{AB}_{~~~~I} = 2 \epsilon^{ABC}\mathbb{M}_{\bullet IC}
\end{equation}
and
\begin{equation}
  \mathbb{M}_{IJ}=0\,.
\end{equation}
The latter condition is consistent with the discussion of Appendix \ref{SM}, where it is also shown that the mass matrix $\mathbb{M_{\bullet \bullet}}$, which does not enter the above gradient flow equations, is in fact vanishing.
\section{Gauge Generators}\label{E}
The ${\rm SO}(3)\times {\rm SU}(3)$ generators $\hat{t}_\ell,\,\hat{t}_m$ in the fundamental representations of the respective groups read:
\begin{align}
&\hat{t}_{\ell=1}=J_1=\left(\begin{matrix} 0 & 0& 0\cr 0 & 0& 1\cr 0 & -1& 0\end{matrix}\right)\,\,;\,\,\,\,
\hat{t}_{\ell=2}=J_2=\left(\begin{matrix} 0 & 0& -1\cr 0 & 0& 0\cr 1 & 0& 0\end{matrix}\right)\,\,;\,\,\,\,
\hat{t}_{\ell=2}=J_3=\left(\begin{matrix} 0 & 1& 0\cr -1 & 0& 0\cr 0 & 0& 0\end{matrix}\right)\,,\nonumber\\
&\hat{t}_{m=3+I}=\frac{i}{2}\,\lambda_I\,,\,\,\,\,\,I=1,\dots, 8\,,\nonumber
\end{align}
where
\begin{align}
\lambda_1&=\left(
\begin{array}{ccc}
 0 & 1 & 0 \\
 1 & 0 & 0 \\
 0 & 0 & 0 \\
\end{array}
\right)\,\,;\,\,\,\,
\lambda_2=\left(
\begin{array}{ccc}
 0 & -i & 0 \\
 i & 0 & 0 \\
 0 & 0 & 0 \\
\end{array}
\right)\,\,;\,\,\,\,
\lambda_3=\left(
\begin{array}{ccc}
 1 & 0 & 0 \\
 0 & -1 & 0 \\
 0 & 0 & 0 \\
\end{array}
\right)\,,\nonumber\\
\lambda_4&=\left(
\begin{array}{ccc}
 0 & 0 & 1 \\
 0 & 0 & 0 \\
 1 & 0 & 0 \\
\end{array}
\right)\,\,;\,\,\,\,
\lambda_5=\left(
\begin{array}{ccc}
 0 & 0 & -i \\
 0 & 0 & 0 \\
 i & 0 & 0 \\
\end{array}
\right)\,\,;\,\,\,\,
\lambda_6=\left(
\begin{array}{ccc}
 0 & 0 & 0 \\
 0 & 0 & 1 \\
 0 & 1 & 0 \\
\end{array}
\right)\,,\nonumber\\
\lambda_7&=\left(
\begin{array}{ccc}
 0 & 0 & 0 \\
 0 & 0 & -i \\
 0 & i & 0 \\
\end{array}
\right)\,\,;\,\,\,\,
\lambda_8=\frac{1}{\sqrt{3}}\,\left(
\begin{array}{ccc}
 1 & 0 & 0 \\
 0 & 1 & 0 \\
 0 & 0 & -2 \\
\end{array}
\right)\,.
\end{align}
\section{Solving for the DW solutions}
\label{DWapp}
Computed on the DW metric (\ref{DW}), the components of the Ricci tensor read
\begin{align}
R_{ij}\,&=\,e^{2 A}\,\left[\,3(A')^2\,+\,A''\,\right]\eta_{ij}\,\,,\\
R_{y y}\,&=\,-3\,\left[\,(A')^2\,+\,A''\,\right]\,\,,
\end{align}
where the $'$ denotes the derivative with respect to the transverse coordinate $y$ and the Ricci scalar is
\begin{align}
R\,&=\,6\,\left[\,2(A')^2\,+\,A''\,\right]\,\,.\end{align}
The Euler-Lagrange equations of motion for (\ref{laggen}) are
\begin{align}
&e^{3 A}\left[\,2r''_i\,+6 A' r'_i\,-\,\frac{1}{2}\sinh\left(4r_i\right)\alpha'_i{}^2-\partial_{r_i}V(r_i,\alpha_i)\,\right]\,=\,0\,\,,\label{eq1}\\
&e^{3 A}\sinh\left(2r_i\right)\,\left[\,4\cosh\left(2r_i\right) r'_i \alpha'_i\,+\,\sinh\left(2r_i\right)\left(\,3 A' \alpha'_i\,+\,\,\alpha''_i\right)\,\right]\,=\,0\,\,,\label{eq2}
\end{align}
while Einstein equations read
\begin{align}
&e^{2 A}\left[\,A''\,+\,3A'{}^2\,+\,V(r_i,\alpha_i)\,\right]\,=\,0\,\,,\\
&\,3A''\,+\,3A'{}^2\,+\,V(r_i,\alpha_i)\,+\,\sum_i^3 \left(\,2r'_i{}^2\,+\,\frac{1}{2}\sinh\left(2r_i\right)^2 \alpha'_i{}^2\,\right)\,=\,0\label{eqein}\,\,.
\end{align}
The critical points of the potentials~\eqref{potgen} and~\eqref{potgen2}, that we choose as end--points of the RG-flow, consist of the origin $\mathcal{O}$ and other vacua at fixed radii
\begin{align}
&\text{Type (i):}\quad\quad r_1=r_2=r_3=r_{\text{vac}}=\frac{1}{2}\log\left(\frac{g_2+g_1}{g_2-g_1}\right)\quad,\quad g_2>g_1\,\,\,,\\
&\text{Type (ii):}\quad\quad r_1=r_2=r_3=r_{\text{vac}}=\frac{1}{2}\log\left(\frac{g_2+2g_1}{g_2-2g_1}\right)\quad,\quad g_2>2g_1\,\,\,.
\end{align}
When imposing that the moduli of all $z_i$ are equal, \eqref{eq1} leads to the conclusion that $\alpha_i$ have to be
constant. In fact, sending all $r_i$ to the same value $r$, and combining the three equations in \eqref{eq1}, one obtains
\begin{align}
e^{3 A}\,\sinh\left(4r\right)\left(\,\alpha'_1{}^2\,-\,\alpha'_2{}^2\,\right)\,=\,0\,\,,\\
e^{3 A}\,\sinh\left(4r\right)\left(\,\alpha'_1{}^2\,-\,\alpha'_3{}^2\,\right)\,=\,0\,\,,\\
e^{3 A}\,\sinh\left(4r\right)\left(\,\alpha'_2{}^2\,-\,\alpha'_3{}^2\,\right)\,=\,0\,\,.
\end{align}

\subsection{The solution}\label{DWsol}
Setting all $\alpha_i$ to constant values along the flow, the equations reduce to the EOM for the field $r$ and the Einstein equations, which read
\begin{align}
&r''\,+\,3A'r'\,-\,\frac{1}{6}\partial_r V(r)\,=\,0\,\,\,, \label{sc1su2u1} \\
&A''\,+\,3\left(A'\right)^2\,+\,V(r)\,=\,0\,\,\,,\label{einst1su2u1}\\
&3\left[A''\,+\,\left(A'\right)^2\,+\,2\left(r'\right)^2\,\right]\,+\,V(r)\,=\,0\label{einst2su2u1}.
\end{align}
$V\left(r\right)$ being the potential given in (\ref{vacuumi}) for Type (i) solution or (\ref{vacuumii}) for Type (ii). The last two equations can be combined into the following constraint
\begin{align}
3\left(A'\right)^2\,-\,3\left(r'\right)^2\,+\,V(r)\,=\,0 \label{constr}\,\,\,.
\end{align}
Now, this system of equations can be obtained from an effective action of the form
\begin{align}
{\Scr L}_\text{eff}\,&=\,e^{3 A}\left[\,3\left(A'\right)^2\,-\,3\left(r'\right)^2\,-\,V(r)\,\right]\,=\\ \nonumber
&=\,\frac{1}{2}G_{ij} {\Phi'}^i {\Phi'}^j\,-\,{\Scr V}(\Phi)\,\,\,,
 \label{effactsu2u1}
\end{align}
with $\Phi^i\,=\,\left(\,A,r\,\right)$, ${\Scr V}(\Phi)\,=\,e^{3 A}\,V(r)$ and $G_{ij}\,=\,6\,e^{3 A}\,\text{diag}\left(\,1\,,\,-1\,\right)$ .\\
The Hamiltonian corresponding to the above Lagrangian is defined via the Legendre transform
\begin{align}
H\,&=\,\Pi_i {\Phi'}^i\,-\,{\Scr L}_\text{eff}\,=\,\frac{1}{2}G^{ij} \Pi_i \Pi_j\,+\,{\Scr V}(\Phi)\,\,,
\end{align}
where
\begin{equation}
\Pi_i\,=\,\frac{\delta{\Scr L}_\text{eff}}{\delta {\Phi'}^i}\,=\,G_{ij}{\Phi'}^j\,\label{canmom1}
\end{equation}
are the usual canonical momenta.
Then we can recast the second-order field equations in the form of first order ones by considering the Hamilton-Jacobi problem, namely by writing
\begin{equation}
\Pi_i\,=\,
\frac{\delta W(\Phi)}{\delta {\Phi}^i}\label{canmom2}\,\,\,,
\end{equation}
where $W(\Phi)$ is the Hamilton's characteristic function, solution to the Hamilton-Jacobi equation:
\begin{align}
H\,&=\,\frac{1}{2}G^{ij} \partial_i W \partial_j W\,+\,{\Scr V}(\Phi)\,.
\end{align}


The characteristic function $W(\Phi)$ can be expressed in terms of a $\alpha_i$-independent ``superpotential'' $\mathcal{W}_0 $, defined in \eqref{Wsup0}, as follows
\begin{align}
W\left(A,r\right)=2\,e^{3 A}\,\mathcal{W}_0\left(r\right),
\end{align}
Note that this ''superpotential" also describes non-supersymmetric flows. Again this is related to the fact that $\alpha_i$, which connect supersymmetric vacua to non-supersymmetric ones, are constants of motion along the flow.
In terms of the superpotential $\mathcal{W}_0(r)$, the scalar potential is
defined through the ''superpotential equation''
\begin{equation}
V\left(r\right)\,=\,{\frac{1}{3}}\left(\partial_r \mathcal{W}_0(r)\right)^2\,-\,\,{3}\,\mathcal{W}_0(r)^2\,\,,
\end{equation}
which holds both for Type (i) and Type (ii) vacuum. Now, from (\ref{canmom1}) and (\ref{canmom2}) we obtain
\begin{equation}
\Phi^{\prime i}=G^{ij}\frac{\partial W}{\partial\Phi^j}\,,
\end{equation}
so that the general form of the first order equations is
\begin{align}
\text{Type (i)}\quad\,\,\,:\,\,\,\quad A'(y)\,&=\,\mathcal{W}_0(r)\,=\,2\,\Big|\, \left[\,g_1 \cosh^3(r)\,-\,g_2 \sinh^3(r)\,\right]\,\Big|\,\,,\\ \nonumber
r'(y)\,&=\,-\sinh(2r)\left[\,g_1 \cosh(r)\,-\,g_2 \sinh(r)\,\right]\,\,\,,\\ \nonumber
\\
\text{Type (ii)}\quad\,\,\,:\,\,\,\quad A'(y)\,&=\mathcal{W}_0(r)\,=\,\Big|\,2\,g_1 \cosh^3(r)\,-\,g_2 \sinh^3(r)\,\Big|\,\,,\\
\nonumber
r'(y)\,&=\,-\frac{1}{2}\sinh(2r)\left[\,2\,g_1 \cosh(r)\,-\,g_2 \sinh(r)\,\right]. \nonumber
\end{align}
These equations can be easily integrated to give
\begin{align}
\text{Type (i)}\quad\,\,\,:\,\,\,\quad A(y)\,=&\,c_1\,+\,\ln\left[\,\frac{g_1 \cosh(r)\,-\,g_2\sinh(r)}{\sinh(2r)}\,\right]\label{eq:domaini1}\,\,\,,\\ \nonumber
y\,=&\,c_2\,-\frac{1}{2g_1g_2}\Bigg(\,2g_1\arctan\left[\tanh\left(\frac{r}{2}\right)\right]\,+\,g_2\ln\left[\tanh\left(\frac{r}{2}\right)\right]\,+\\
&+\,2\sqrt{{g_2}^2-{g_1}^2}\,\tanh^{(-1)}\left[\frac{g_2-g_1\tanh(\frac{r}{2})}{\sqrt{{g_2}^2-{g_1}^2}}\right]\Bigg)\label{eq:domaini2}\,\,\,,\\ \nonumber
\\
\text{Type (ii)}\quad\,\,\,:\,\,\,\quad A(y)\,=&\,c_1\,+\,\ln\left[\,\frac{2\,g_1 \cosh(r)\,-\,g_2\sinh(r)}{\sinh(2r)}\,\right]\label{eq:domainii1}\,\,\,,\\ \nonumber
y\,=&\,c_2\,-\frac{1}{2g_1g_2}\Bigg(\,4g_1\arctan\left[\tanh\left(\frac{r}{2}\right)\right]\,+\,g_2\ln\left[\tanh\left(\frac{r}{2}\right)\right]\,+\\
&+\,2\sqrt{{g_2}^2-4{g_1}^2}\,\tanh^{(-1)}\left[\frac{g_2-2g_1\tanh(\frac{r}{2})}{\sqrt{{g_2}^2-4{g_1}^2}}\right]\Bigg)\label{eq:domainii2}. \nonumber
\end{align}
$c_1$ and $c_2$ are integration constants that can be set to zero by a shift of $x^i$ coordinates.

\section{Relevant Supermultiplets}\label{app:supermult}
\subsection{\texorpdfstring{$\mathcal{N}=3$}{N=3}}\label{super3}
\begin{small}
\begin{align*}
&
\begin{tabular}{|c|c|c|c|}
\hline
\multicolumn{4}{|c|}{$\mathbf{A_1[1]^{(0)}_{\frac{3}{2}}}$:\;{massless graviton multiplet}} \\
\hline
spin & $\Delta\equiv E_0$ & $\mathfrak{su}(2)_R$ irrep & $m^2$ \\ \hline
$\frac{1}{2}$ & $\frac{3}{2}$ & $\mathbf{1}$ & $0$ \\ \hline
$1$ & $2$ & $\mathbf{3}$ & $0$ \\ \hline
$\frac{3}{2}$ & $\frac{5}{2}$ & $\mathbf{3}$ & $1$ \\ \hline
$2$ & $3$ & $\mathbf{1}$ & $0$ \\ \hline
\end{tabular}
&&
\begin{tabular}{|c|c|c|c|}
\hline
\multicolumn{4}{|c|}{$\mathbf{B_1[0]^{(2)}_1}$:\;{massless vector multiplet}} \\
\hline
spin & $\Delta\equiv E_0$ & $\mathfrak{su}(2)_R$ irrep & $m^2$ \\ \hline
\multirow{2}{*}{$0$} & $1$ & $\mathbf{3}$ & $-2$ \\
& $2$ & $\mathbf{3}$ & $-2$ \\ \hline
\multirow{2}{*}{$\frac{1}{2}$} & $\frac{3}{2}$ & $\mathbf{1}$ & $0$ \\
& $\frac{3}{2}$ & $\mathbf{3}$ & $0$ \\ \hline
$1$ & $2$ & $\mathbf{1}$ & $0$ \\ \hline
\end{tabular}
\\
&
\begin{tabular}{|c|c|c|c|}
  \hline
\multicolumn{4}{|c|}{$\mathbf{B_1[0]^{(3)}_{\frac{3}{2}}}$:\;{masssive vector multiplet}} \\
  \hline
spin & $\Delta\equiv E_0$ & $\mathfrak{su}(2)_R$ irrep & $m^2$ \\ \hline
\multirow{3}{*}{ $0$ } & $\frac{3}{2}$ & $\mathbf{4}$ & $-\frac{9}{4}$ \\
& $\frac{5}{2}$ & $\mathbf{4}$ & $-\frac{5}{4}$ \\
& $\frac{5}{2}$ & $\mathbf{2}$ & $-\frac{5}{4}$ \\ \hline
\multirow{3}{*}{$\frac{1}{2}$} & $2$ & $\mathbf{2}$ & $\frac{1}{4}$ \\
& $2$ & $\mathbf{4}$ & $\frac{1}{4}$ \\
& $3$ & $\mathbf{2}$ & $\frac{9}{4}$ \\ \hline
$1$ & $\frac{5}{2}$ & $\mathbf{2}$ & $\frac{3}{4}$ \\ \hline
\end{tabular}
&&
\begin{tabular}{|c|c|c|c|}
     \hline
     \multicolumn{4}{|c|}{$\mathbf{B_1[0]^{(4)}_{2}}$:\;{masssive vector multiplet}} \\
     \hline
     spin & $\Delta\equiv E_0$ & $\mathfrak{su}(2)_R$ irrep & $m^2$ \\ \hline
     \multirow{5}{*}{ $0$ } & $2$ & $\mathbf{5}$ & $-2$ \\
          & $3$ & $\mathbf{1}$ & $0$ \\
          & $3$ & $\mathbf{3}$ & $0$ \\
          & $3$ & $\mathbf{5}$ & $0$ \\
          & $4$ & $\mathbf{1}$ & $4$ \\ \hline
     \multirow{4}{*}{$\frac{1}{2}$} & $\frac{5}{2}$ & $\mathbf{3}$ & $1$ \\
          & $\frac{5}{2}$ & $\mathbf{5}$ & $1$ \\
          & $\frac{7}{2}$ & $\mathbf{1}$ & $4$ \\
          & $\frac{7}{2}$ & $\mathbf{3}$ & $4$ \\ \hline
     $1$ & $3$ & $\mathbf{3}$ & $2$ \\ \hline
\end{tabular}
\\
&
\begin{tabular}{|c|c|c|c|}
\hline
\multicolumn{4}{|c|}{$\mathbf{B_1[0]^{(6)}_{3}}$:\;{masssive vector multiplet}} \\
       \hline
       spin & $\Delta\equiv E_0$ & $\mathfrak{su}(2)_R$ irrep & $m^2$ \\ \hline
       \multirow{5}{*}{ $0$ } & $3$ & $\mathbf{7}$ & $0$ \\
            & $4$ & $\mathbf{3}$ & $4$ \\
            & $4$ & $\mathbf{5}$ & $4$ \\
            & $4$ & $\mathbf{7}$ & $4$ \\
            & $5$ & $\mathbf{3}$ & $10$ \\ \hline
       \multirow{4}{*}{$\frac{1}{2}$} & $\frac{7}{2}$ & $\mathbf{5}$ & $4$ \\
            & $\frac{7}{2}$ & $\mathbf{7}$ & $4$ \\
            & $\frac{9}{2}$ & $\mathbf{3}$ & $9$ \\
            & $\frac{9}{2}$ & $\mathbf{5}$ & $9$ \\ \hline
       $1$ & $4$ & $\mathbf{5}$ & $6$ \\ \hline
     \end{tabular}
\end{align*}
\end{small}

\subsection{\texorpdfstring{$\mathcal{N}=2$}{N=2}}\label{super2}
\begin{small}
\begin{align*}
&
\begin{tabular}{|c|c|c|c|}
\hline
\multicolumn{4}{|c|}{$\mathbf{A_1\bar{A}_1[2]^{(0)}_3}$:\;{massless graviton multiplet}} \\
\hline
spin & $\Delta\equiv E_0$ & $R$ & $m^2$ \\ \hline
$1$ & $2$ & $0$ & 0 \\ \hline
\multirow{2}{*}{$\frac{3}{2}$} & $\frac{5}{2}$ & $-1$ & $1$ \\
& $\frac{5}{2}$  & $+1$ & $1$ \\ \hline
$2$ & $3$ & $0$ & $0$ \\ \hline
\end{tabular}
&&
\begin{tabular}{|c|c|c|c|}
\hline
\multicolumn{4}{|c|}{$\mathbf{A_2\bar{A}_2[0]^{(0)}_1}$:\;{massless vector multiplet}} \\
\hline
spin & $\Delta\equiv E_0$ & $R$ & $m^2$ \\ \hline
\multirow{2}{*}{$0$} & $1$ & $0$ & $-2$ \\
& $2$ & $0$ & $-2$ \\ \hline
\multirow{2}{*}{$\frac{1}{2}$} & $\frac{3}{2}$ & $-1$ & $0$ \\
& $\frac{3}{2}$ & $+1$ & $0$ \\ \hline
$1$ & $2$ & $0$ & $0$ \\ \hline
\end{tabular}
\\
&
\begin{tabular}{|c|c|c|c|}
  \hline
\multicolumn{4}{|c|}{$\mathbf{L\bar{L}[1]^{(0)}_{\Delta_0}}$:\;{long masssive gravitino multiplet}} \\
  \hline
spin & $\Delta\equiv E_0$ & $R$ & $m^2$ \\ \hline
  \multirow{4}{*}{$0$} & $\Delta_0 +\frac{1}{2}$ & $-1$ & $(\Delta_0 +\frac{1}{2})(\Delta_0 - \frac{5}{2})$ \\
& $\Delta_0 +\frac{3}{2}$ & $-1$ & $\Delta_0^2 - \frac{9}{4}$ \\
& $\Delta_0 +\frac{1}{2}$ & $+1$ & $(\Delta_0 +\frac{1}{2})(\Delta_0 - \frac{5}{2})$ \\
& $\Delta_0 +\frac{3}{2}$ & $+1$ & $\Delta_0^2 - \frac{9}{4}$ \\  \hline
\multirow{6}{*}{$\frac{1}{2}$} & $\Delta_0$ & $0$ & $(\Delta_0-\frac{3}{2})^2$ \\
     & $\Delta_0 + 1$ & $0$ & $(\Delta_0-\frac{1}{2})^2$ \\
 & $\Delta_0 + 1$ & $0$ & $(\Delta_0-\frac{1}{2})^2$ \\
 & $\Delta_0 + 2$ & $0$ & $(\Delta_0+\frac{1}{2})^2$ \\
   & $\Delta_0 + 1$ & $-2$ & $(\Delta_0-\frac{1}{2})^2$ \\
& $\Delta_0 + 1$ & $+2$ & $(\Delta_0-\frac{1}{2})^2$ \\ \hline
\multirow{4}{*}{$1$} & $\Delta_0+\frac{1}{2}$ & $-1$ & $(\Delta_0 -\frac{1}{2})(\Delta_0 - \frac{3}{2})$ \\
     & $\Delta_0+\frac{3}{2}$ & $-1$ & $\Delta_0^2 - \frac{1}{4}$ \\
     & $\Delta_0+\frac{1}{2}$ & $+1$ & $(\Delta_0 - \frac{1}{2})(\Delta_0
- \frac{3}{2})$ \\
& $\Delta_0+\frac{3}{2}$ & $+1$ & $\Delta_0^2 - \frac{1}{4}$ \\ \hline
$\frac{3}{2}$ & $\Delta_0 + 1 $ & $0$ & $(\Delta_0-\frac{1}{2})^2$ \\ \hline
\end{tabular}
  &&
 \begin{tabular}{|c|c|c|c|}
  \hline
\multicolumn{4}{|c|}{$\mathbf{L\bar{L}[0]^{(0)}_{\Delta_0>1}}$:\;{long masssive vector multiplet}} \\
  \hline
spin & $\Delta\equiv E_0$ & $R$ & $m^2$ \\ \hline
  \multirow{5}{*}{$0$} & $\Delta_0$ & $0$ & $\Delta_0 (\Delta_0 - 3)$ \\
& $\Delta_0 +1$ & $0$ & $(\Delta_0 +1)(\Delta_0 - 2)$ \\
     & $\Delta_0 +2$ & $0$ & $(\Delta_0 +2)(\Delta_0 -1)$ \\
     & $\Delta_0 +1$ & $-2$ & $(\Delta_0 +1)(\Delta_0 - 2)$ \\
& $\Delta_0 +1$ & $+2$ & $(\Delta_0 +1)(\Delta_0 - 2)$ \\  \hline
\multirow{4}{*}{$\frac{1}{2}$} & $\Delta_0+\frac{1}{2}$ & $-1$ & $(\Delta_0-1)^2$ \\
 & $\Delta_0+\frac{3}{2}$ & $-1$ & $\Delta_0^2$ \\
   & $\Delta_0+\frac{1}{2}$ & $+1$ & $(\Delta_0-1)^2$ \\
& $\Delta_0+\frac{3}{2}$ & $+1$ & $\Delta_0^2$ \\ \hline
$1$ & $\Delta_0+1$ & $0$ & $\Delta_0(\Delta_0 - 1)$  \\ \hline
\end{tabular}
  \\
  &
\begin{tabular}{|c|c|c|c|}
\hline
\multicolumn{4}{|c|}{$\mathbf{L\bar{A}_2[0]^{(R>0)}_{R+1}}$:\;{short masssive vector multiplet}} \\
\hline
spin & $\Delta\equiv E_0$ & $R$ & $m^2$ \\ \hline
\multirow{3}{*}{$0$} & $R+1$ & $R$ & $(R+1)(R-2)$ \\
& $R+2$ & $R-2$ & $(R+2)(R-1)$ \\
& $R+2$ & $R$ & $(R+2)(R-1)$ \\ \hline
\multirow{3}{*}{$\frac{3}{2}$} & $R+\frac{3}{2}$ & $R-1$ & $R^2$ \\
& $R+\frac{3}{2}$ & $R+1$ & $R^2$ \\
& $R+\frac{5}{2}$ & $R-1$ & $(R+1)^2$ \\ \hline
$1$ & $R+2$ & $R$ & $R(R+1)$ \\ \hline
\end{tabular}
&&
\begin{tabular}{|c|c|c|c|}
\hline
\multicolumn{4}{|c|}{$\mathbf{L\bar{B}_1[0]^{(R)}_{R}}$:\;{$\tfrac{1}{2}$- hypermultiplet}} \\
\hline
spin & $\Delta\equiv E_0$ & $R$ & $m^2$ \\ \hline
\multirow{2}{*}{$0$} & $R$ & $R$ & $R(R-3)$ \\
& $R+1$ & $R-2$ & $(R+1)(R-2)$ \\ \hline
$\frac{1}{2}$ & $R+\frac{1}{2}$ & $R-1$ & $(R-1)^2$ \\ \hline
\end{tabular}
\end{align*}
\end{small}

\subsection{\texorpdfstring{$\mathcal{N}=1$}{N=1}}\label{super1}
\begin{small}
\begin{align*}
&
\begin{tabular}{|c|c|c|}
\hline
  \multicolumn{3}{|c|}{$\mathbf{A_1[3]_{\frac{5}{2}}}$:\;{massless gravity multiplet} } \\
\hline
spin & $\Delta\equiv E_0$ & $m^2$ \\ \hline
$\frac{3}{2}$ & $\frac{5}{2}$ & $1$ \\ \hline
$2$ & $3$ & $0$ \\
\hline
\end{tabular}
&&
\begin{tabular}{|c|c|c|}
  \hline
  \multicolumn{3}{|c|}{$\mathbf{L[2]_{\Delta_0>2}}$:\;{massive gravitino multiplet} } \\
  \hline
  spin & $\Delta\equiv E_0$ & $m^2$ \\ \hline
  $\frac{1}{2}$ & $\Delta_0+\frac{1}{2}$ & $(\Delta_0-1)^2$ \\ \hline
  \multirow{2}{*}{$1$} & $\Delta_0$ & $(\Delta_0-1)(\Delta_0-2)$ \\
  & $\Delta_0+1$ & $\Delta_0(\Delta_0-1)$ \\ \hline
$\frac{3}{2}$ & $\Delta_0+\frac{1}{2}$ & $(\Delta_0-1)^2$ \\
  \hline
\end{tabular}
  \\
&
\begin{tabular}{|c|c|c|}
\hline
\multicolumn{3}{|c|}{$\mathbf{A_1[1]_{\frac{3}{2}}}$:\;{massless vector multiplet} } \\
\hline
spin & $\Delta\equiv E_0$ & $m^2$ \\ \hline
$\frac{1}{2}$ & $\frac{3}{2}$ & $0$ \\ \hline
$1$ & $2$ & $0$ \\
\hline
\end{tabular}
&&
\begin{tabular}{|c|c|c|}
\hline
\multicolumn{3}{|c|}{$\mathbf{L[1]_{\Delta_0>\frac{3}{2}}}$:\;{massive vector multiplet} } \\
\hline
spin & $\Delta\equiv E_0$ & $m^2$ \\ \hline
$0$ & $\Delta_0+\frac{1}{2}$ & $(\Delta_0+\frac{1}{2})(\Delta_0-\frac{5}{2})$ \\ \hline
\multirow{2}{*}{$\frac{1}{2}$} & $\Delta_0$ & $(\Delta_0-\frac{3}{2})^2$ \\
     & $\Delta_0+1$ & $(\Delta_0-\frac{1}{2})^2$ \\ \hline
$1$ & $\Delta_0+\frac{1}{2}$ & $(\Delta_0-\frac{1}{2})(\Delta_0-\frac{3}{2})$ \\
\hline
\end{tabular}
  \\
&
\begin{tabular}{|c|c|c|}
\hline
\multicolumn{3}{|c|}{$\mathbf{L^{\prime}[0]_{\Delta_0>\frac{1}{2}}}$:\;{matter multiplet} } \\
\hline
spin & $\Delta\equiv E_0$ & $m^2$ \\ \hline
\multirow{2}{*}{$0$} & $\Delta_0$ & $\Delta_0(\Delta_0-3)$ \\
& $\Delta_0+1$ & $(\Delta_0+1)(\Delta_0-2)$ \\ \hline
$\frac{1}{2}$ & $\Delta_0+\frac{1}{2}$ & $(\Delta_0-1)^2$ \\
\hline
\end{tabular}
\end{align*}
\end{small}

\section{Supergravity spectra in various vacua}\label{app:spectra}
The tables below do not contain Goldstone bosons and Goldstinos. Recall that $\Delta^{(1)}_{G1/2}=\Delta^{(1)}_{H1/2}$. The same holds true for $\Delta^{(2)}_{G1/2}$. The expression of the latter and $\Delta^{(2)}_{V1/2}$ is obtained from the corresponding quantities with the superscript $(1)$ and by replacing $g_1$ with $2g_1$.
\subsection{\texorpdfstring{$\mathcal{N}=3$}{N=3}}
\begin{samepage}
\begin{footnotesize}
\begin{table}[ht!]
\centering
\begin{tabular}{|c|c|c|c|}
\hline
spin & $m^2$ & $\Delta\equiv E_0$ & multiplicity \\ \hline
0 & -2 & \{1,2\} & 54=27+27 \\ \hline
$\frac{1}{2}$ & 0 & $\frac{3}{2}$ & 37 \\ \hline
1 & 0 & 2 & 12 \\ \hline
$\frac{3}{2}$ & 1 & $\frac{5}{2}$ & 3 \\ \hline
2 & 0 & 3 & 1 \\ \hline
\end{tabular}
\caption{Mass spectrum in the $\mathcal{N}=3$ vacuum preserving the full gauge group
  $\mathrm{SO}(3)\times\mathrm{SU}(3)$}
\label{tab:N3_SO3SU3_mass}
\end{table}
\pagebreak
\begin{table}[h!]
\centering
\begin{tabular}{|c|c|c|c|}
  \hline
spin & $m^2$ & $\Delta\equiv E_0$ & multiplicity \\ \hline
\multirow{5}{*}{0} & 4 & 4 & 1 \\
 & 0 & 3 & 9 \\
 & $-\frac{5}{4}$ & $\frac{5}{2}$ & 12 \\
 & -2 & 1,2 & 17 \\
 & $-\frac{9}{4}$ & $\frac{3}{2}$ & 8 \\[3pt] \hline
\multirow{5}{*}{$\frac{1}{2}$} & 4 & $\frac{7}{2}$ & 4 \\
 & $\frac{9}{4}$ & 3 & 4 \\
 & 1 & $\frac{5}{2}$ & 8 \\
 & $\frac{1}{4}$ & 2 & 12 \\
 & 0 & $\frac{3}{2}$ & 9 \\[3pt] \hline
\multirow{3}{*}{1} & 2 & 3 & 3 \\
 & $\frac{3}{4}$ & $\frac{5}{2}$ & 4 \\
 & 0 & 2 & 5 \\ \hline
  $\frac{3}{2}$ & 1 & $\frac{5}{2}$ & 3 \\ \hline
2 & 0 & 3 & 1 \\ \hline
\end{tabular}
\caption{Mass spectrum in the single $\mathcal{N}=3$ vacuum invariant under the
  subgroup $\mathrm{SU}(2)\times \mathrm{U}(1)$ of the gauge group.}
\label{tab:N3_SU2U1_mass}
\end{table}
\pagebreak
\begin{table}[h!]
\centering
\begin{tabular}{|c|c|c|c|}
\hline
spin & $m^2$ & $\Delta\equiv E_0$ & multiplicity \\ \hline
\multirow{4}{*}{0} & 10 & 5 & 3 \\
 & 4 & 4 & 16 \\
 & 0 & 3 & 16 \\
 & -2 & 1,2 & 11 \\ \hline
\multirow{4}{*}{$\frac{1}{2}$} & 9 & $\frac{9}{2}$ & 8 \\
 & 4 & $\frac{7}{2}$ & 16 \\
 & 1 & $\frac{5}{2}$ & 8 \\
 & 0 & $\frac{3}{2}$ & 5 \\ \hline
\multirow{3}{*}{1} & 6 & 4 & 5 \\
 & 2 & 3 & 3 \\
 & 0 & 4 & 4 \\ \hline
$\frac{3}{2}$ & 1 & $\frac{5}{2}$ & 3 \\ \hline
2 & 0 & 3 & 1 \\ \hline
\end{tabular}
\caption{Mass spectrum in a $\mathcal{N}=3$ vacuum invariant under the subgroup
 $\mathrm{SO}(3)_{D}$ of the gauge group.}
\label{tab:N3_SO3_mass}
\end{table}
\end{footnotesize}
\end{samepage}
\pagebreak

\begin{samepage}
\subsection{\texorpdfstring{$\mathcal{N}=2$}{N=2}}
\begin{footnotesize}
\begin{table}[h!]
\centering
\begin{tabular}{|c|c|c|c|}
\hline
spin & $m^2$ & $\Delta\equiv E_0$ & multiplicity \\ \hline
\multirow{7}{*}{$0$}
& $-\frac{9}{4}$ & $\frac{3}{2}$ & $8$\\
& $-2$ &$\{1,2\}$ & $15=6+9$ \\
& $-\frac{5}{4}$ & $\frac{5}{2}$ & $12$ \\
& $0$ &$3$& $5$ \\
& $4$ &$4$& $1$ \\
& $(\delta+2)(\delta-1)$ & $\delta+2$ & $2$ \\
& $\delta(\delta+3)$  & $\delta+3$ & $2$ \\
\hline
\multirow{8}{*}{$\frac{1}{2}$}
&$0$ & $\frac{3}{2}$& $8$\\
& $\frac{1}{4}$ & $2$ & $12$ \\
& $\frac{9}{4}$ & $3$ & $4$ \\
&$1$ &$\frac{5}{2}$ & $4$\\
&$4$ &$\frac{7}{2}$ & $2$\\
& $\delta^2$ & $\delta+\frac{3}{2}$ & $1$ \\
& $(\delta+1)^2$ & $\delta+\frac{5}{2}$ & $4$ \\
& $(\delta+2)^2$ & $\delta+\frac{7}{2}$ & $1$ \\
\hline
\multirow{5}{*}{$1$} & $0$ & $2$ & $3$ \\
& $\frac{3}{4}$ & $\frac{5}{2}$ & $4$ \\
& $2$ & $3$ & $1$ \\
& $\delta(\delta+1)$ & $\delta+2$ & $2$ \\
& $(\delta+1)(\delta+2)$ & $\delta+3$ & $2$ \\
\hline
  \multirow{2}{*}{$\frac{3}{2}$} & $1$ & $\frac{5}{2}$ & $2$ \\
     &$(\delta+1)^2$ &$\delta+\frac{5}{2}$ & $1$\\
\hline
$2$ & $0$ & $3$ & $1$ \\ \hline
\end{tabular}
\caption{Mass spectrum in the $\mathcal{N}=2$ vacuum invariant under a
  $\mathrm{U}(1)_D\times \mathrm{U}(1)$ subgroup of the gauge group. Here, $\delta=|m_G^{(1)}|-1$.}
\label{tab:N2_U1U1_mass}
\end{table}
\pagebreak

\begin{table}[h!]
\centering
\begin{tabular}{|c|c|c|c|}
\hline
spin & $m^2$ & $\Delta\equiv E_0$ & multiplicity \\ \hline
\multirow{6}{*}{$0$}
& $-2$ &$\{1,2\}$ & $9=4+5$ \\
& $0$ &$3$& $12$ \\
& $4$ &$4$& $16$ \\
& $10$ & $5$ & $3$ \\
& $(\delta+2)(\delta-1)$ & $\delta+2$ & $2$ \\
& $\delta(\delta+3)$  & $\delta+3$ & $2$ \\
\hline
\multirow{7}{*}{$\frac{1}{2}$}
&$0$ & $\frac{3}{2}$& $4$\\
&$1$ &$\frac{5}{2}$ & $4$\\
&$4$ &$\frac{7}{2}$ & $14$\\
& $9$ & $\frac{9}{2}$ & $8$ \\
& $\delta^2$ & $\delta+\frac{3}{2}$ & $1$ \\
& $(\delta+1)^2$ & $\delta+\frac{5}{2}$ & $4$ \\
& $(\delta+2)^2$ & $\delta+\frac{7}{2}$ & $1$ \\
\hline
\multirow{4}{*}{$1$} & $0$ & $2$ & $2$ \\
& $2$ & $3$ & $1$ \\
& $6$ & $4$ & $5$ \\
& $\delta(\delta+1)$ & $\delta+2$ & $2$ \\
& $(\delta+1)(\delta+2)$ & $\delta+3$ & $2$ \\
\hline
  \multirow{2}{*}{$\frac{3}{2}$} & $1$ & $\frac{5}{2}$ & $2$ \\
     &$(\delta+1)^2$ &$\delta+\frac{5}{2}$ & $1$\\
\hline
$2$ & $0$ & $3$ & $1$ \\ \hline
\end{tabular}
\caption{Mass spectrum in the $\mathcal{N}=2$ vacuum invariant under a
  $\mathrm{U}(1)_D$ subgroup of the gauge group. Here, $\delta=|m_G^{(2)}|-1$}
\label{tab:N2_U1_mass}
\end{table}
\end{footnotesize}
\end{samepage}
\subsection{\texorpdfstring{$\mathcal{N}=1$}{N=1}}
\tiny{
\begin{table}[t!]
\centering
\begin{tabular}{|c|c|c|c|}
\hline
spin & $m^2$ & $\Delta\equiv E_0$ & multiplicity \\ \hline
\multirow{11}{*}{$0$} & $-\frac{9}{4}$ & $\frac{3}{2}$& $8$ \\
& $-2$ &$\{1,2\}$ & $14=12+2$ \\
& $-\frac{5}{4}$ &$\frac{5}{2}$& $12$ \\
& $0$ &$3$& $3$ \\
& $4$ &$4$& 1 \\
     & $(\Delta^{(1)}_{V1}+\frac{1}{2})(\Delta^{(1)}_{V1}-\frac{5}{2})$ & $\Delta^{(1)}_{V1}+\frac{1}{2}$ & $1$ \\
     & $(\Delta^{(1)}_{V2}+\frac{1}{2})(\Delta^{(1)}_{V2}-\frac{5}{2})$ & $\Delta^{(1)}_{V2}+\frac{1}{2}$ & $1$ \\
     & $\Delta^{(1)}_{G1}(\Delta^{(1)}_{G1}-3)$ & $\Delta^{(1)}_{G1}$ & $1$ \\

     & $(\Delta^{(1)}_{G1}+1)(\Delta^{(1)}_{G1}-2)$ & $\Delta^{(1)}_{G1}+1$ & $1$ \\

     & $\Delta^{(1)}_{G2}(\Delta^{(1)}_{G2}-3)$ & $\Delta^{(1)}_{G2}$ & $1$ \\

     & $(\Delta^{(1)}_{G2}+1)(\Delta^{(1)}_{G2}-2)$ & $\Delta^{(1)}_{G2}+1$ & $1$ \\
\hline
\multirow{11}{*}{$\frac{1}{2}$}
&$0$ & $\frac{3}{2}$& $8$\\
&$\frac{1}{4}$ &$2$ & $12$\\
&$1$ &$\frac{5}{2}$ &$2$ \\
& $\frac{9}{4}$&$3$ & $4$\\
&$4$ &$\frac{7}{2}$ & $1$\\
& $(\Delta^{(1)}_{G1}-1)^2$&$\Delta^{(1)}_{G1}+\frac{1}{2}$ & $2$ \\
& $(\Delta^{(1)}_{G2}-1)^2$&$\Delta^{(1)}_{G2}+\frac{1}{2}$ & $2$ \\
& $(\Delta^{(1)}_{V1}-\frac{3}{2})^2$& $\Delta^{(1)}_{V1}$& $1$\\
& $(\Delta^{(1)}_{V1}-\frac{1}{2})^2$ & $\Delta^{(1)}_{V1}+1$ & $1$ \\
& $(\Delta^{(1)}_{V2}-\frac{3}{2})^2$& $\Delta^{(1)}_{V2}$& $1$\\
& $(\Delta^{(1)}_{V2}-\frac{1}{2})^2$ & $\Delta^{(1)}_{V2}+1$ & $1$ \\
\hline
\multirow{8}{*}{$1$} & $0$ & $2$ & $2$ \\
& $\frac{3}{4}$ & $\frac{5}{2}$ & $4$ \\
&$(\Delta^{(1)}_{G1}-1)(\Delta^{(1)}_{G1}-2)$ & $\Delta^{(1)}_{G1}$& $1$\\
&$\Delta^{(1)}_{G1}(\Delta^{(1)}_{G1}-1)$ &$\Delta^{(1)}_{G1}+1$ & $1$\\
&$(\Delta^{(1)}_{G2}-1)(\Delta^{(1)}_{G2}-2)$ & $\Delta^{(1)}_{G2}$& $1$\\
&$\Delta^{(1)}_{G2}(\Delta^{(1)}_{G2}-1)$ &$\Delta^{(1)}_{G2}+1$ & $1$\\
&$(\Delta^{(1)}_{V1}-\frac{1}{2})(\Delta^{(1)}_{V1}-\frac{3}{2})$ &$\Delta^{(1)}_{V1+\frac{1}{2}}$ & $1$\\
&$(\Delta^{(1)}_{V2}-\frac{1}{2})(\Delta^{(1)}_{V2}-\frac{3}{2})$ &$\Delta^{(1)}_{V2+\frac{1}{2}}$ & $1$\\
\hline
  \multirow{3}{*}{$\frac{3}{2}$} & $1$ & $\frac{5}{2}$ & $1$ \\
     &$(\Delta^{(1)}_{G1}-1)^2$ &$\Delta^{(1)}_{G1}+\frac{1}{2}$ & $1$\\
     &$(\Delta^{(1)}_{G2}-1)^2$ &$\Delta^{(1)}_{G2}+\frac{1}{2}$ & $1$\\
\hline
$2$ & $0$ & $3$ & 1 \\ \hline
\end{tabular}
\caption{Mass spectrum in the $\mathcal{N}=1$ vacuum invariant under a
  $\mathrm{U}(1)$ subgroup of the gauge group.}
\label{tab:N1_U1_mass}
\vspace{-65pt}
\end{table}
}

\begin{footnotesize}
\begin{table}[t!]
  \centering
\begin{tabular}{|c|c|c|c|}
\hline
spin & $m^2$ & $\Delta\equiv E_0$ & multiplicity \\ \hline
  \multirow{10}{*}{$0$} & $-2$ & $\{1,2\}$& $6+2=8$ \\
     & $0$ &$3$& $10$ \\
     & $4$ &$4$& $16$ \\
     & $10$ &$5$& $3$ \\
     & $(\Delta^{(2)}_{V1}+\frac{1}{2})(\Delta^{(2)}_{V1}-\frac{5}{2})$ & $\Delta^{(2)}_{V1}+\frac{1}{2}$ & $1$ \\
     & $(\Delta^{(2)}_{V2}+\frac{1}{2})(\Delta^{(2)}_{V2}-\frac{5}{2})$ & $\Delta^{(2)}_{V2}+\frac{1}{2}$ & $1$ \\
     & $\Delta^{(2)}_{G1}(\Delta^{(2)}_{G1}-3)$ & $\Delta^{(2)}_{G1}$ & $1$ \\

     & $(\Delta^{(2)}_{G1}+1)(\Delta^{(2)}_{G1}-2)$ & $\Delta^{(2)}_{G1}+1$ & $1$ \\

     & $\Delta^{(2)}_{G2}(\Delta^{(2)}_{G2}-3)$ & $\Delta^{(2)}_{G2}$ & $1$ \\

     & $(\Delta^{(2)}_{G2}+1)(\Delta^{(2)}_{G2}-2)$ & $\Delta^{(2)}_{G2}+1$ & $1$ \\
  \hline
\multirow{10}{*}{$\frac{1}{2}$}
&$0$ & $\frac{3}{2}$& $4$\\
&$1$ &$ \frac{5}{2}$ & $2$\\
&$4$ &$\frac{7}{2}$ & $13$\\
&$9$ & $\frac{9}{2}$ & $8$ \\
& $(\Delta^{(2)}_{G1}-1)^2$&$\Delta^{(2)}_{G1}+\frac{1}{2}$ & $2$ \\
& $(\Delta^{(2)}_{G2}-1)^2$&$\Delta^{(2)}_{G2}+\frac{1}{2}$ & $2$ \\
& $(\Delta^{(2)}_{V1}-\frac{3}{2})^2$& $\Delta^{(2)}_{V1}$& $1$\\
& $(\Delta^{(2)}_{V1}-\frac{1}{2})^2$ & $\Delta^{(2)}_{V1}+1$ & $1$ \\
& $(\Delta^{(2)}_{V2}-\frac{3}{2})^2$& $\Delta^{(2)}_{V2}$& $1$\\
& $(\Delta^{(2)}_{V2}-\frac{1}{2})^2$ & $\Delta^{(2)}_{V2}+1$ & $1$ \\
\hline
\multirow{8}{*}{$1$} & $0$ & $2$ & $1$ \\
& $6$ & $4$ & $5$ \\
&$(\Delta^{(2)}_{G1}-1)(\Delta^{(2)}_{G1}-2)$ & $\Delta^{(2)}_{G1}$& $1$\\
&$\Delta^{(2)}_{G1}(\Delta^{(2)}_{G1}-1)$ &$\Delta^{(2)}_{G1}+1$ & $1$\\
&$(\Delta^{(2)}_{G2}-1)(\Delta^{(2)}_{G2}-2)$ & $\Delta^{(2)}_{G2}$& $1$\\
&$\Delta^{(2)}_{G2}(\Delta^{(2)}_{G2}-1)$ &$\Delta^{(2)}_{G2}+1$ & $1$\\
&$(\Delta^{(2)}_{V1}-\frac{1}{2})(\Delta^{(2)}_{V1}-\frac{3}{2})$ &$\Delta^{(2)}_{V1+\frac{1}{2}}$ & $1$\\
&$(\Delta^{(2)}_{V2}-\frac{1}{2})(\Delta^{(2)}_{V2}-\frac{3}{2})$ &$\Delta^{(2)}_{V2+\frac{1}{2}}$ & $1$\\
\hline
  \multirow{3}{*}{$\frac{3}{2}$} & $1$ & $\frac{5}{2}$ & $1$ \\
     &$(\Delta^{(2)}_{G1}-1)^2$ &$\Delta^{(2)}_{G1}+\frac{1}{2}$ & $1$\\
     &$(\Delta^{(2)}_{G2}-1)^2$ &$\Delta^{(2)}_{G2}+\frac{1}{2}$ & $1$\\
\hline
$2$ & $0$ & $3$ & 1 \\ \hline
\end{tabular}
  \caption{Mass spectrum of $\mathcal{N}=1$ vacuum with completely broken gauge symmetry.}
  \label{tab:N1_Id_mass}
    \vspace{-43pt}
\end{table}
\end{footnotesize}
\end{appendices}

\clearpage
\bibliography{biblio_N3D4SUGRA2}
  \bibliographystyle{jheppub}
\end{document}